\documentclass{article}

\usepackage{listings}
\usepackage{hyperref}

\makeatletter
\AtBeginDocument{%
  \def\doi#1{\url{https://doi.org/#1}}}
\makeatother
\usepackage{multicol}
\usepackage{amsmath}
\usepackage{amsthm}

\usepackage{amssymb,amsfonts}
\usepackage{latexsym}
\usepackage{subcaption}
\usepackage[export]{adjustbox}
\usepackage{fancybox}
\usepackage{tikz}
\usepackage{tikzit}

\tikzstyle{new edge style 0}=[-]

\usepackage{makecell}
\usepackage{lipsum,adjustbox}

\usepackage{pgfplots}
\pgfplotsset{compat=newest}
\usetikzlibrary{decorations.pathreplacing}
\usepackage{setspace}
\usepackage{algorithm}
\usepackage[noend]{algpseudocode}
\usepackage[framemethod=tikz]{mdframed}
\usepackage{xspace}
\usepackage{framed}
\usepackage{tabularx}

\usepackage{xintexpr}

\newcounter{thm}
\newtheorem{experiment}[thm]{Experiment}

\newtheorem{invariant}[thm]{Invariant}

\newtheorem{construction}[thm]{Construction}

\newenvironment{proofof}[1]{\begin{trivlist} \item {\bf Proof
#1:~~}}
  {\qed\end{trivlist}}
\renewenvironment{proofof}[1]{\par\medskip\noindent{\bf Proof of #1: \ }}{\hfill$\Box$\par\medskip}
\newcommand{\namedref}[2]{\hyperref[#2]{#1~\ref*{#2}}}
\newcommand{\thmlab}[1]{\label{thm:#1}}
\newcommand{\thmref}[1]{\namedref{Theorem}{thm:#1}}

\newcommand{\seclab}[1]{\label{sec:#1}}
\newcommand{\secref}[1]{\namedref{Section}{sec:#1}}
\newcommand{\applab}[1]{\label{app:#1}}
\newcommand{\appref}[1]{\namedref{Appendix}{app:#1}}

\newcommand{\figlab}[1]{\label{fig:#1}}
\newcommand{\figref}[1]{\namedref{Figure}{fig:#1}}

\newcommand{\tablelab}[1]{\label{tab:#1}}
\newcommand{\tableref}[1]{\namedref{Table}{tab:#1}}
\newcommand{\deflab}[1]{\label{def:#1}}
\newcommand{\defref}[1]{\namedref{Definition}{def:#1}}
\newcommand{\explab}[1]{\label{exp:#1}}
\newcommand{\expref}[1]{\namedref{Experiment}{exp:#1}}
\newcommand{\eqnlab}[1]{\label{eq:#1}}

\newcommand{\constref}[1]{\namedref{\textbf{Construction}}{const:#1}}
\newcommand{\constlab}[1]{\label{const:#1}}

\newenvironment{remindertheorem}[1]{\medskip \noindent {\bf Reminder of  #1.  }\em}{}

\def \A    {\mdef{\mathcal{A}}}
\def \D    {\mdef{\mathcal{D}}}
\def \C    {\mdef{\mathcal{C}}}
\def \B    {\mdef{\mathcal{B}}}
\def \M   {\mdef{\mathcal{M}}}
\def \U   {\mdef{\mathcal{U}}}

\def \Enc    {\mdef{\mathsf{Enc}}}
\def \Dec    {\mdef{\mathsf{Dec}}}
\def \FHE    {\mdef{\mathsf{FHE}}}

\def \Cond   {\mdef{\mathsf{C}}}
\def \KG   {\mdef{\mathsf{KeyGen}}}
\def \Setup   {\mdef{\mathsf{Setup}}}
\def \Eval   {\mdef{\mathsf{Eval}}}

\def \Sim   {\mdef{\mathtt{Sim}}}
\def \CE   {\mdef{\mathtt{CE}}}
\def \AE   {\mdef{\mathtt{AE}}}
\def \PKDF   {\mdef{\mathtt{PKDF}}}

\def \Setup    {\mdef{\mathsf{Setup}}}

\def \Pwd    {\mdef{\mathcal{PWD}}}
\def \Id    {\mdef{\mathcal{ID}}}
\def \pwd   {\mdef{\mathtt{pwd}}}
\def \salt   {\mdef{\mathtt{salt}}}

\def \id   {\mdef{\mathtt{Id}}}

\def \QSgin    {\mdef{\mathsf{RegisterQuery}}}

\def \Lgin    {\mdef{\mathsf{Login}}}
\def \QLgin    {\mdef{\mathsf{LoginQuery}}}

\def \ValidLginQ {\mdef{\mathtt{ValidLginQuery}}}
\def \ToInt  {\mdef{\mathsf{ToInt}}}
\def \ToOrig  {\mdef{\mathsf{ToInt}^{-1}}}
\def \Pad  {\mdef{\mathtt{Pad}}}

\def \Pail {\mdef{\mathtt{P}}}
\def \SS {\mdef{\mathtt{SS}}}
\def \PtoCMul {\mdef{\mathtt{PlainToCtxMul}}}
\def \Add {\mdef{\mathtt{Add}}}

\def \lcm {\mdef{\mathsf{lcm}}}

\newcommand\Mbb[1]{\mathbb{#1}}

\def \Capslock {\mdef{\mathsf{CAPSLOCK}}}

\def \RanEnc   {\mdef{\mathsf{REnc}}}
\def \RanDec   {\mdef{\mathsf{RDec}}}

\def \Auth   {\mdef{\mathsf{Auth}}}

\def \Ham  {\mdef{\mathtt{Ham}}}
\def \ED  {\mdef{\mathtt{ED}}}

\def \apnd  {\mdef{\mathtt{Append}}}

\def \Init {\mdef{\mathsf{Initialization}}}
\def \Sgin {\mdef{\mathsf{RegisterNewUser}}}
\def \Lgin {\mdef{\mathsf{Login}}}
\def \EvType {\mdef{\mathsf{EventType}}}
\def \Register {\mdef{\mathsf{Register}}}
\def \Login {\mdef{\mathsf{Login}}}

\def \acpt {\mdef{\mathtt{acpt}}}
\def \rjct {\mdef{\mathtt{rjct}}}
\def \isReg {\mdef{\mathtt{isRegistered}}}
\def \PullPwd {\mdef{\mathtt{PullPwd}}}
\def \view {\mdef{\mathtt{view}}}

\def \Share   {\mdef{\mathsf{ShareGen}}}
\def \recover   {\mdef{\mathsf{SecretRecover}}}
\def \ValidShr   {\mdef{\mathsf{ValidShare}}}
\def \InterPol   {\mdef{\mathsf{InterPol}}}
\def \shrs   {\mdef{\mathsf{shares}}}

\def \Experiment   {\mdef{\mathtt{Experiment}}}

\newcommand{\InvertCase}{\mathsf{InvertCase}}

\def \iif   {\mdef{\mathtt{if~}}}

\def \ffor   {\mdef{\mathtt{for~}}}

\def \ooutput   {\mdef{\mathtt{output~}}}

\def \ccompute   {\mdef{\mathtt{compute~}}}
\def \sset   {\mdef{\mathtt{set~}}}
\def \ssample   {\mdef{\mathtt{sample~}}}
\def \pparse   {\mdef{\mathtt{parse~}}}
\def \aand   {\mdef{\mathtt{~AND~}}}

\def \run   {\mdef{\mathtt{run~}}}
\def \rreturn   {\mdef{\mathtt{return~}}}
\def \condit   {\mdef{\mathtt{Cond}}}

\newcommand{\ldb}{\mathopen{\lbrack\!\lbrack}} 
\newcommand{\rdb}{\mathclose{\rbrack\!\rbrack}}

\newcommand\norm[1]{\left\lVert#1\right\rVert}

\newcommand{\eps}{\epsilon}

\newcommand{\mdef}[1]{{\ensuremath{#1}}\xspace}  

\DeclareMathOperator*{\poly}{poly}

\newcommand{\set}[1]{\mdef{\left\{#1\right\}}}                        

\newcommand{\ignore}[1]{}

\newif\ifnotes\notestrue 
\ifnotes
\newcommand{\samson}[1]{\textcolor{purple}{{\bf (Samson:} {#1}{\bf ) }} \marginpar{\tiny\bf
             \begin{minipage}[t]{0.5in}
               \raggedright S:
            \end{minipage}}}       
            
\newcommand{\jeremiah}[1]{\textcolor{red}{{\bf (Jeremiah:} {#1}{\bf ) }} \marginpar{\tiny\bf
             \begin{minipage}[t]{0.5in}
               \raggedright J:
            \end{minipage}}    }

\else
\newcommand{\samson}[1]{}
\newcommand{\jeremiah}[1]{}
\fi

\hypersetup{
     colorlinks   = true,
     citecolor    = blue,
		 linkcolor		= red
}

\makeatletter
\renewcommand*{\@fnsymbol}[1]{\textcolor{mahogany}{\ensuremath{\ifcase#1\or *\or \dagger\or \ddagger\or
 \mathsection\or \triangledown\or \mathparagraph\or \|\or **\or \dagger\dagger
   \or \ddagger\ddagger \else\@ctrerr\fi}}}
\makeatother

\definecolor{electricpurple}{rgb}{0.75, 0.0, 1.0}
\definecolor{fluorescentpink}{rgb}{1.0, 0.08, 0.58}
\definecolor{mahogany}{rgb}{0.75, 0.25, 0.0}
\definecolor{darkblue}{rgb}{0.0, 0.0, 0.55}
\definecolor{darkpastelgreen}{rgb}{0.01, 0.75, 0.24}
\definecolor{darkgreen}{rgb}{0.0, 0.2, 0.13}
\definecolor{darkgoldenrod}{rgb}{0.72, 0.53, 0.04}
\definecolor{darkred}{rgb}{0.55, 0.0, 0.0}
\hypersetup{
     colorlinks   = true,
     citecolor    = electricpurple,
		 linkcolor		= blue
}
\newcommand{\fullversion}[2]{#1}
\usepackage{fullpage}

\newcounter{def}
\newtheorem{definition}[def]{Definition}
\newtheorem{theorem}[thm]{Theorem}
\begin{document}

\title{Conditional Encryption with Applications to Secure Personalized Password Typo Correction}

\author{Mohammad Hassan Ameri \and Jeremiah Blocki}
\date{Purdue University\\ West Lafayette, IN USA}


\maketitle
\begin{abstract}
We introduce the notion of a conditional encryption scheme as an extension of public key encryption. In addition to the standard public key algorithms ($\KG$, $\Enc$, $\Dec$) for key generation, encryption and decryption, a conditional encryption scheme for a binary predicate $P$ adds a new conditional encryption algorithm $\Cond\Enc$. The conditional encryption algorithm $c=\Cond\Enc_{pk}(c_1,m_2,m_3)$ takes as input the public encryption key $pk$, a ciphertext $c_1 = \Enc_{pk}(m_1)$ for an unknown message $m_1$, a control message $m_2$ and a payload message $m_3$ and outputs a conditional ciphertext $c$. Intuitively, if $P(m_1,m_2)=1$ then the conditional ciphertext $c$ should decrypt to the payload message $m_3$. On the other hand if $P(m_1,m_2) = 0$ then the ciphertext should not leak {\em any} information about the control message $m_2$ or the payload message $m_3$ even if the attacker already has the secret decryption key $sk$. We formalize the notion of conditional encryption secrecy and provide concretely efficient constructions for a set of predicates relevant to password typo correction. Our practical constructions utilize the Paillier partially homomorphic encryption scheme as well as Shamir Secret Sharing. We prove that our constructions are secure and demonstrate how to use conditional encryption to improve the security of personalized password typo correction systems such as TypTop. We implement a C++ library for our practically efficient conditional encryption schemes and evaluate the performance empirically. We also update the implementation of TypTop to utilize conditional encryption for enhanced security guarantees and evaluate the performance of the updated implementation.  

\end{abstract}



{\bf Full Version:} This document provides the full version of the article published at CCS 2024 under the same title -- see \url{https://doi.org/10.1145/3658644.3690374}. 

\sloppy

\section{Introduction}
Traditionally, public key encryption allows any party who has the public key $pk$ to encrypt a message $m$ and obtain a ciphertext $c$ which can only be decrypted by a party who possesses the corresponding secret key $sk$. The implicit assumption is that anyone who posses the secret key $sk$ is a trusted party. However, there are some applications where the party encrypting a message may only want to conditionally reveal that message to the party who possesses the secret key if certain conditions hold. For example, consider the problem of having an authentication server maintain an (encrypted) cache of incorrect login attempts for each user. Such a cache might be used to design a (personalized) typo tolerant password authentication scheme \cite{SP:CAAJR16,CCS:CWPCR17} and/or to help identify malicious login attempts. In the TypTop \cite{CCS:CWPCR17} system the sever generates a public/private key pair $(sk_u,pk_u)$ for each user account $u$ and encrypts the secret key $sk_u $ with a symmetric encryption key $K_u = \PKDF(s_u, \pwd_u)$ derived from the user's password $\pwd_u$ and a random salt value $s_u$. The server then stores the resulting ciphertext $c_u = \Enc_{K_u}(sk_u)$ and public key $pk_u$ along with the salt value $s_u$. Whenever the user logs in with an incorrect password $\pwd'$ the authentication server uses the public key $pk_u$ to generate and store the ciphertext $c' = \Enc_{pk_u}(\pwd')$. Later if the user logs in with a correct password $\pwd$ we can re-derive the symmetric key $K_u = \PKDF(s_u,\pwd)$, use the symmetric key to recover the secret key $sk_u = \Dec_{K_u}(c_u)$ and then use the secret key $sk_u$ to decrypt each password $\pwd' = \Dec_{sk_u}(c')$ in our encrypted cache. The TypTop system can then examine each particular password $\pwd'$ in the cache to determine whether or not this password is an ``acceptable typo" that should be accepted during future login attempts. An online password cracker will not be able to peek inside the encrypted vault unless he has already guessed the correct password and derived $K_u$. However, the above approach still has a security drawback in that an offline attacker who manages to crack the user's password $\pwd_u$ will be able to access {\em any} password stored in the encrypted vault. 

While the TypTop \cite{CCS:CWPCR17} system maintains a cache of {\em all} incorrect login attempts, only the passwords that are ``sufficiently close" to the original password are considered as candidates to be added to a list of ``acceptable typos" for future login attempts. Thus, in the password typo tolerant application, one only needs to store incorrect login attempts that are plausibly typos of the user's original password e.g., the Hamming Distance between the two passwords is at most $2$ or the password was incorrectly capitalized because the CAPSLOCK key was not turned off. If the user mistakenly logs into his social media account (12345\_SOCIAL) with his bank password (STR@NG\_BANK\_\#;aym7*5) and the social media site was running TypTop then it would add the bank password (STR@NG\_BANK\_\#; aym7*5) to its encrypted cache even though this password is not close to the social media password and would never be added to the list of ``acceptable typos". The potential presence of additional user passwords in the encrypted cache could increase the incentive for an offline adversary to crack the social media password (12345\_SOCIAL) in order to decrypt the cache which might contain the user's passwords for other accounts e.g., the banking password STR@NG\_BANK\_\#;aym7*5. Ideally, when the authentication server sees an incorrect login attempt it would add the password to the encrypted cache if and only if that login attempt is a plausible typo. However, the authentication server should not store the password $\pwd_u$ in plaintext form so it is difficult to know whether or not the login attempt is a plausible typo a priori. To this end it would be useful to generate a ``conditional" ciphertext $c'$ such that (1) if $\pwd_u$ and $\pwd'$ are sufficiently close then $c'$ decrypts to $\pwd'$; (2) otherwise $c'$ leaks no information about the password $\pwd'$.  
\subsection{Our Contributions} We introduce the notion of a conditional encryption scheme and demonstrate a concrete application to improve the security of personalized typo tolerant systems such as TypTop \cite{CCS:CWPCR17}. We conjecture that Conditional Encryption may find many other MPC applications e.g., securing Triger-Action Platforms (``If-this-than-that" operations) for IoT services \cite{SP:CCWSCF21} or designing Fuzzy Password Authenticated Key Exchange Protocols \cite{EC:DHPRY18,EC:CHKLM05,PKC:RoyXu23,AC:BFHHO23}. Intuitively, a conditional encryption scheme $(\KG,\Enc,\Dec,\Cond\Enc)$ for a binary predicate $P(\cdot,\cdot)$ is a public key encryption scheme with the addition of a new ``conditional encryption" algorithm. The conditional encryption $\Cond\Enc_{pk}(c,m_2,m_3)$ algorithm accepts four inputs: a public encryption key $pk$, a (regular) public key ciphertext $c = \Enc_{pk}(m_1)$ for an {\em unknown} message $m_1$, a control message $m_2$ and a payload message $m_3$ and the output is a ciphertext $c'$. Intuitively, if $P(m_1,m_2)=1$, then control message $m_2$ is related to the unknown and encrypted message $m_1$ (e.g., the Hamming Distance between $m_2$ and $m_1$ is sufficiently small) and the output ciphertext $c'$ should encrypt the payload message $m_3$ i.e., $\Dec_{sk}(c') = m_3$. On the other hand, if $P(m_1,m_2) =0$ then the messages $m_1$ and $m_2$ are unrelated and the ciphertext $c'$ should not reveal {\em any} information about the control message $m_2$ or the payload message $m_3$ --- even if the attacker knows the secret decryption key $sk$. 

\subsubsection{Conditional Encryption Security} We provide formal security definitions for a conditional encryption scheme in the semi-honest settings. If the attacker does not know the secret decryption key $sk$, then we require that the encryption scheme satisfies the traditional notion of real-or-random security for a public key encryption scheme. When the attacker does have the secret key $sk$ and the predicate  does not hold (i.e., $P(m_1,m_2)=0$) we still want to ensure that the ciphertext $c'=\Cond\Enc_{pk}(c,m_2,m_3)$ does not leak any information about $m_2$, $m_1$ or $m_3$. We formalize this ``conditional encryption secrecy" property using a simulator $\Sim(pk)$ who is {\em only} given access to the public key and must generate a ciphertext which is indistinguishable from $c'$ even if the distinguisher $\D$ is given access to the secret key $sk$ as well as the original ciphertext $c$ and the original messages $m_1$, $m_2$ and $m_3$. We elect to follow a concrete security definition instead of asymptotic definitions to provide concrete guidance on selecting the concrete security parameters in practice.


\subsubsection{Efficient Constructions} We next provide {\em efficient} constructions of conditional encryption for the equality predicate and for predicates based on hamming distance\footnote{For the Hamming Distance predicate (e.g., $P(m_1,m_2) = 1$ if and only if $m_1[i] \neq m_2[i]$ for at most $d$ locations $i\leq n$) decrypting a conditional ciphertext predicate requires time proportional to ${ n \choose d}$ where $n$ is the length of the messages $m_1$ and $m_2$ and $d$ is the Hamming Distance that we tolerate. This can be slow when both $d$ and $n$ are large. However, in our target password applications the desired distance parameter $d$ for our Hamming Distance predicate (resp. edit-distance predicate) is relatively small (e.g., $d=2$) as is the parameter $n$ (e.g., $n\leq 32$). Finding efficient constructions for the Hamming Distance (or Edit-Distance) predicate when $n$ and $d$ are both large is left as an open challenge for future research.}, edit distance and CAPSLOCK. 
Our constructions use the Pallier partially homomorphic encryption scheme, secret sharing and authenticated encryption as the constructive building blocks. We also provide a generic composition theorem for OR predicates in the semi-honest setting. In particular, if we have separate constructions of conditional encryption schemes for predicates $P_1(\cdot,\cdot), \ldots, P_k(\cdot,\cdot)$ then we can obtain a conditional encryption scheme for the predicate $P_{OR}(m_1,m_2)\doteq \bigvee_{i=1}^k P_i(m_1,m_2)$ simply by concatenating the conditional ciphertexts generated by the conditional encryption algorithm $\Cond\Enc_i$ for predicate $P_i$. As an application we consider the ``typo predicate" which is the OR of several predicates:  Hamming Distance at most two, Edit-Distance at most one and a CAPSLOCK predicate. This is the same predicate used by Chaterjee et al. in the TypTop personalized typo tolerant password authentication system. In the appendix, we also provide a general construction  of conditional encryption for arbitrary predicates circuit private fully homomorphic encryption (FHE). However, the practicality of this construction is unclear as circuit private FHE is substantially more expensive than Pallier.

\subsubsection{Application to TypTop} We show how to (slightly) modify the TypTop system to improve security using conditional encryption. \fullversion{In the appendix we formally define the notion of ``typo privacy" (see \defref{def:TypoPrivacy}) which ensures that the authentication server never collects ciphertexts of passwords which are not plausible typos of the original password.}{Intuitively, ``typo privacy" ensures that the authentication server never collects ciphertexts of passwords which are not plausible typos of the original password --- see the full version\cite{fullversion} for a formal definition of typo privacy.} While the original TypTop scheme does not satisfy typo privacy, we prove that our modified TypTop construction does satisfy typo privacy and is still efficient. See \secref{ConcretCondEnc} and \secref{sec:TypPredicate} for more details. 

\subsubsection{Implementation and Empirical Evaluation} We provide a C++ implementation for each of our practical conditional encryption schemes (excluding our general construction from circuit private FHE). Our implementations include conditional encryption scheme for the following predicates: CAPSLOCK, Edit Distance One, Hamming distance at most $t$, as well as  OR composition of these predicates (CAPSLOCK or Edit Distance One or Hamming Distance Two). We also modify the C++ implementation of the TypTop system for personalized password typo correction to use conditional encryption for enhanced security (Typo Privacy). We further modified the TypTop system to utilize memory hard functions \cite{STOC:AlwSer15,alwen2017towards,EC:AlwBloPie18,biryukov2016argon2} for key-derivation --- an update recommended by the TypTop authors. Our code is available on Github\footnote{\url{https://github.com/mhassanameri/CondEncCCS24Artifact}} and Zenodo\footnote{\url{https://zenodo.org/records/13744111}}.


We evaluate the performance of our conditional encryption schemes for each predicate. As an example, when we consider the OR predicate for messages of length $n \leq 32$ characters (e.g., almost all passwords\footnote{Over $99.9\%$ of leaked RockYou passwords were less than $30$ characters.}) and instantiate the Pallier Cryptosystem with a $1024$-bit modulus $N$ we observe average running times $158.15$ (ms), $645.945 $ (ms) and $261.44 $ (ms) for the regular encryption $\Enc_{pk}(\cdot)$, conditional encryption $\Cond\Enc_{pk}(\cdot)$ and decryption of a conditional ciphertext $\Cond\Dec_{sk}(\cdot)$, respectively. The size of a regular (resp. conditional) ciphertext was $16 $ KB (resp. $24 $ KB ). The results are summarized in \tableref{CondEncEmp}. We find that modifying the TypTop system does not increase authentication delays although it does increase the storage requirements for the authentication server  by a factor of $\approx 246$. Fortunately, per user storage will not be a limiting factor in most settings. See \secref{sec:TypPredicate} and \tableref{CondTypComp} for more details. 

\subsubsection{Related work} At a high level the notion of conditional encryption might seem similar to other advanced public key primitives such as  identity based encryption \cite{C:BonFra01,EC:SahWat05,EC:Gentry06,CCS:BolGoyKum08,C:BonRagSeg13}, predicate encryption \cite{EC:KatSahWat08,C:GorVaiWee15,TCC:SheShiWat09,C:BCFG17,C:AgrYadYam22}, attribute based encryption \cite{SP:BetSahWat07,C:AgrYadYam22,TCC:Chase07,CCS:GPSW06,CCS:ChaCho09,EC:LewWat11,gorbunov2015attribute,ameri2018key,JC:WanPanChe23}, functional encryption \cite{CCS:NAPWAH14,TCC:GGHZ16,EC:GDGJKLSSZ14,CCS:SahSey10,C:AGVW13,ACNS:ErnMit23} and fully homomorphic encryption\cite{STOC:Gentry09,TOCT:ZGCV14,EC:VGHV10,SP:ViaJatHit21}. However, we stress that the security requirements for conditional encryption are quite distinctive in that we require that secrecy guarantees hold *\textit{even if}* the attacker has the secret decryption key. By contrast, the security definitions for other public key primitives (identity based encryption, predicate encryption, attribute based encryption, functional encryption and fully homomorphic encryption) all assume that the attacker {\em does not} have the secret decryption key.  While we do use the Paillier partially homomorphic encryption scheme to construct conditional encryption schemes for particular predicates, these constructions do not use Pallier as a blackbox. We are also able to construct conditional encryption for general predicates using {\em circuit private} FHE, but the circuit privacy requirement seems to be inherent i.e., there exists regular (non circuit private) FHE schemes for which our conditional encryption construction is explicitly broken. \fullversion{See discussion in \appref{CondEncFromFHE}}{See \appref{CondEncFromFHE} and the full version \cite{fullversion} for additional discussion}.



\subsection{Preliminaries}\seclab{preliminaries}
In this section, we review the notations and cryptographic primitive which will be used in the rest of the paper. \newline 

Given a randomized algorithm $\A$ (e.g., key-generation)  we use $y=A(x;r)$ to denote the deterministic output of $A$ when run on input $x$ with fixed random $r \in \{0,1\}^*$ and we use the random variable  $y \leftarrow \A(x)$ to denote the output of $A(x;r)$ when $r$ is selected randomly.  

Let  $ \Sigma $ denote an alphabet (e.g., ASCII or unicode). Given a string $w \in \Sigma^*$ we use $\norm{w}$ to denote the length of $w$ and for $i \leq \norm{w}$ we use $w[i]$ to denote the $i$th character of $w$. We let $\M_n = \Sigma^{\leq n}$ denote the set of all strings $w$ with length $\norm{w} \leq n$. It will be convenient to assume that all passwords have the same length. Of course most user passwords do not have the same length but if the maximum length of a user password is $n-1$ then we can easily define a $1$ to $1$ function $\Pad:\Sigma^{\leq n-1} \rightarrow \Sigma^n$ and consider $\Pwd = \Sigma^n$ to be the set of all possible user passwords after padding. In practice, we could select $n=30$ as essentially all user passwords are shorter than this (e.g., over $99.9\%$ of leaked RockYou passwords were less than 30 characters.). The symbol ``$\|$'' will be used for concatenation. Thus, $ y = x_1 \| x_2 $ is concatenation of $ x_1 $ and $ x_2 $. \newline

Let $ L = \langle l_1, \ldots, l_{|L|}  \rangle $ be list of $ |L| $ elements. We also define the operation $ L' = \apnd(L, l) $ which adds $ l $ to the list and we have $ L' =  \langle l_1, \ldots, l_{|L|}, l \rangle $. We note that $ l_i $ can be an element in $ \Mbb{Z}_{N^2}, \Sigma^n, \M_n$ or $ \Pwd $, etc.  \newline

For the message $m \in \Sigma^{\leq n}$ we use the notation  $m_{-i} \in \Sigma^{\leq n-1} $ to denote the string $ m$ when the $ i $-th char is deleted and if $i > |m|$ then $m_{-i}=m$.

\subsubsection{Partially homomorphic Encryption}
The  Paillier cryptosystem is a partially homomorphic cryptographic scheme which supports ciphertext addition, plaintext to ciphertext multiplication and subtraction.  Specifically, the public key $pk=(N,g)$ (resp. secret key $sk=(\beta, \mu)$) consists of $N=pq$ where $p,q$ are prime numbers and the number $g=N+1 \in \mathbb{Z}_{N^2}^*$ (resp. $\beta =\lcm(p-1,q-1)$ and $\mu=\varphi(N)^{-1} \mod N$). We note that for all $i \in \mathbb{Z}_N$ we have $g^i = \sum_{j=0}^i {i \choose j} N^j = 1+Ni \mod{N^2}$ so that $g$ has multiplicative order $N$ modulo $N^2$. The secret key $sk=(\beta,\mu)$ consists of two parameters $\beta=\lcm(p-1,q-1)$ and $\mu = \varphi(N)^{-1} \mod N$ is defined to be the multiplicative inverse of $\varphi(N) = (p-1)(q-1)$ modulo $N$. \\ 

The algorithm $\Enc_{pk}(m;r)$ takes as input a message $m \in \mathbb{Z}_N$ and a nonce $r \in \mathbb{Z}_N^*$ and outputs $g^{m}r^N \mod{N^2}$. The function $\Enc_{pk}$ acts as a bijective map from $\mathbb{Z}_N \times \mathbb{Z}_N^* \rightarrow \mathbb{Z}_{N^2}^*$. In particular, for {\em every} $c \in \mathbb{Z}_{N^2}^*$ there is a message $m \in \mathbb{Z}_N$ and a nonce $r \in \mathbb{Z}_N^*$ such that $c=g^m r^N \mod{N^2}$ \cite{EC:Paillier99}. \\


The encryption scheme has several homomorphic properties in particular if $c_1 = g^{m_1} r_1^N \mod{N^2}$ and $c_2= g^{m_2} r_2^N \mod{N^2}$ encrypt message $m_1,m_2 \in \mathbb{Z}_N$ respectively then $c_1c_2 = g^{m_1+m_2} (r_1r_2)^N \mod{N^2}$ encrypts the message $m_1+m_2 \mod N$. Similarly, if $c= g^m r^N \mod {N^2}$ encrypts the message $m$ then $c^k = g^{mk} (r^k)^N \mod {N^2}$ encrypts the message $mk \mod N$. See \appref{App:PailDetails} for a full description of the Paillier encryption scheme. \\


When we apply the Paillier Cryptosystem, our desired message space $\mathcal{M}$ is typically not the set of integers $\mathbb{Z}_N$. Thus, we assume that there is an injective map $ \ToInt: \mathcal{M} \to \Mbb{N} $ and $ \ToOrig: \Mbb{N} \to  \mathcal{M}$. We will also assume that $|\mathcal{M}| \leq N$ and that $\forall m \in \mathcal{M}$ that $0 \leq \ToInt(m) < \left| \mathcal{M}\right| \leq  N$. Given $x \in \mathbb{Z}_N$ we define $\ToOrig (x) = \bot$ if $x$ has no preimage i.e., $\forall m \in \mathcal{M}$ we have $\ToInt(m) \neq x$.
\subsubsection{Secret Sharing (SS)}\seclab{Sec:SecretSharing}
Several of our constructions rely on a primitive called secret sharing. A $(t,n)$-secret sharing scheme consists of two polynomial time algorithms $\mathtt{ShareGen}$ and $\recover$. Intuitively, $\left( \ldb s \rdb_1, \ldots,  \ldb s \rdb_n\right) \gets \mathtt{ShareGen}(n,t,s)$ takes as input a secret $s \in \mathbb{F}$ along with parameters $n,t$ and outputs $n$ shares $\left(\ldb s \rdb_1, \ldots, \ldb s \rdb_n\right) \in \mathbb{F}$. Given any subset $S = \{i_1,\ldots, i_t\} \subseteq [n]$ of $|S| = t$ shares we can recover the secret $s$ using  $$\recover\left(\left(i_1, \ldb s \rdb_{i_1}), \ldots, (i_t, \ldb s \rdb_{i_t}\right)\right) = s$$ However, given any smaller subset $S=\{i_1,\ldots, i_{t-1}\} \subseteq [n]$ of size $|S| \leq t-1$ shares an attacker cannot infer {\em anything} about $s$ from the shares $\ldb s \rdb_{i_1}, \ldots, \ldb s \rdb_{i_{t-1}}$. In particular, we require that  for all secrets $s \in \mathbb{F}$,  all subsets $S = \{i_1,\ldots, i_{t-1}\} \subseteq [n]$ of size $t-1$ the shares  $\ldb s \rdb_{i_1}, \ldots,  \ldb s \rdb_{i_{t-1}}$ can be viewed as uniformly random independent elements in $\mathbb{F}$ unrelated to the secret $s$. The Shamir Secret sharing scheme \cite{CACM:Shamir79} satisfies this requirement. See appendix \appref{Sec:SecretSharing} for more detail about (Shamir) Secret Sharing.

\subsubsection{String Distance and Close Passwords} Given a string $w \in \Sigma^n$ and $i \leq n$ we use $w[i] \in \Sigma$ to denote the $i$th character of $\Sigma$ and given two strings $w_1,w_2 \in \Sigma^n$ we use $\Ham(w_1,w_2) = \left|\{ i | w[i] \neq w[j]\} \right|$ to denote the hamming distance between them. Similarly, given two strings $w_1, w_2 \in \Sigma^*$ we use $\ED(w_1,w_2)$ to denote the edit-distance between them i.e., the minimum number of insertions/deletions to transform $w_1$ into $w_2$ (or vice versa). Note that if $w_1=w_2$ then $\Ham(w_1,w_2)=0$ and $\
\ED(w_1,w_2)=0$. We will often use Hamming/Edit Distance to determine if two passwords $\pwd_1,\pwd_2$ are close e.g., we could define a predicate $P(\pwd_1,\pwd_2)=1$ if $\Ham $ $(\pwd_1, \pwd_2) \leq 2$ or $\ED(\pwd_1, \pwd_2) \leq 1$; otherwise $P(\pwd_1,\pwd_2)=0$. We could also combine Hamming/Edit distance with other common password typos such as CAPSLOCK/SHIFT errors e.g., \allowbreak $P(\pwd_1, \pwd_2)=1$ if $\InvertCase(\pwd_1) = \pwd_2$ or $\Ham(\pwd_1, \pwd_2) \leq 2$ or \-$\ED(\pwd_1, \pwd_2) \leq 1$; otherwise, $P(\pwd_1,\pwd_2)=0$.


\section{Conditional Encryption}\seclab{sec:CondEnc}
In this section, we will introduce the notion of a {\em conditional encryption} scheme. A conditional encryption scheme is similar to a regular public key encryption scheme with the addition of a special algorithm $\Cond\Enc_{pk}$. This algorithm takes three inputs: a ciphertext $c = \Enc_{pk}(m_1;r)\in \mathcal{C}_{\Enc}$ (Where $\mathcal{C}_{\Enc}$ is the ciphertext space of traditional encryption scheme) encrypting some unknown message $m_1 \in \mathcal{M}$ in our message space using random coin $r\in_{R}\{0,1\}^{\lambda}$, a control message $m_2$ and a payload message $m_3$. A conditional encryption scheme is defined with respect to a binary predicate $P:\mathcal{M} \times \mathcal{M} \rightarrow \{0,1\}$. Intuitively, if  $P(m_1,m_2) = 1$ then $\Cond\Enc_{pk}(c,m_2,m_3)\in \mathcal{C}_{\Cond\Enc}$\footnote{Similarly, we define $\mathcal{C}_{\Cond\Enc}$ as the ciphertext space for a conditional encryption scheme.} should produce valid encryption of our payload message $m_3$; otherwise, if $P(m_1,m_2)=0$ the output should reveal {\em no information} about any of the messages $m_1,m_2$ or $m_3$ --- even to an adversary that knows $sk$. 

We now formally define conditional encryption along with its associated security/correctness requirements.

\begin{definition}\deflab{def:ConEncFirtFormalDef}
A conditional encryption scheme $ \Pi $ for a binary predicate $P: \mathcal{M} \times \mathcal{M} \rightarrow \{0,1\}$ consists of four main algorithms $ (\KG,\Enc, \Cond\Enc, \Dec) $ which are described as follows: 

\begin{itemize}
	\item$  (sk, pk)\gets \KG(1^\lambda; r) $:  takes as input the security parameter $ \lambda$ and  random coins $ r\gets_R \{0,1\}^{p(\lambda)}$ and generates a secret key $sk \in \mathcal{SK}$ and the corresponding public key $pk \in \mathcal{PK} $ for our conditional encryption scheme.  
	\item $ (b =0, c) =  \Enc_{pk} (m_1; r)$: takes as input a plaintext message $ m_1 \in \mathcal{M}$ the public key $pk$  random nonce  $ r\gets_R \{0,1\}^{p(\lambda)}$ and outputs a ciphertext $ (b, c)  \in \{0\}\times \mathcal{C} $ encrypting $m_1$. The flag $ b =0$ indicates that $ c $ is output of the regular encryption scheme $ \Enc_{pk} $.  

	\item $ (b= 1, \tilde{c}) = \Cond\Enc_{pk}\left(\left(0,c_{m_1}\right), m_2, m_3; r\right) $: This conditional encryption algorithm takes as input a public key $pk$, a ciphertext  $ (0,c_{m_1})$ with $c_{m_1} \in \mathcal{C}$ corresponding to an unknown message  $ m_1 \in \mathcal{M} $, a control message $ m_2$, a payload message $m_3 \in \mathcal{M}$,  and random nonce  $ r\gets_R \{0,1\}^{p(\lambda)}$  and outputs a ciphertext $ (b, \tilde{c}) \in \{ 1 \} \times \mathcal{C}$. The flag $b=1$ indicates that this ciphertext is the output of the conditional encryption $ \Cond\Enc $ algorithm. Note: When the control message and the payload message are the same $m_2=m_3$ we will sometimes write  $\Cond\Enc_{pk}(c_{m_1}, m_2; r)$ instead of $\Cond\Enc_{pk}(c_{m_1}, m_2, m_2; r)$. If the input ciphertext takes the form $(b=1,c_m)$ then $\Cond\Enc_{pk}\left(\left(1,c_{m_1}\right), m_2, m_3; r\right)  = \bot$ i.e., $(b=0,c_{m_1})$ must be the output of the regular encryption scheme.
	
	\item $ \{m, \bot\} = \Dec_{sk}(c) $: takes as input a ciphertext $ c \in \{ 0, 1\} \times \mathcal{C}$ and the secret key $sk$ and outputs a message $ m \in \mathcal{M} $ or $ \bot $ (indicating failure).  
\end{itemize}
\end{definition}

We require that for any valid pair $(sk,pk)$ produced that $$\Pr\left[\Dec_{sk}\left(\Enc_{pk}\left(m\right)\right)=m\right]=1$$ i.e., perfect correctness for ciphertexts output by the regular encryption algorithm. For correctness of the conditional encryption algorithm we want to ensure that $\Dec_{sk}\left(\Cond\Enc_{pk}\left(c', m_2, m_3\right)\right)=m_3$ whenever $\exists r', m_1$ s.t. $c'=\Enc_{pk}(m_1;r')$ and $P(m_1,m_2)=1$. Intuitively, we can extract our intended payload $ m_3 $ if and only if $ P(m_1, m_2) = 1 $. For conditional encryption we relax our requirement of perfect correctness and instead require that  \[ \Pr\left[\Dec_{sk}\left(\Cond\Enc_{pk}\left(c', m_2, m_3\right)\right)=m_3\right] \geq 1-\epsilon\left(\lambda\right) \] for a negligible function $\epsilon(\cdot)$ whenever \[\exists r', m_1 \text{ s.t. } c'=\Enc_{pk}(m_1;r')\] and $P(m_1,m_2)=1$. We stress that the correctness condition only holds when when $c' \gets \Enc_{pk}(m_1)$ was the output of the regular encryption algorithm. If $c'= (1,\tilde{c})$ was generated by the conditional encryption algorithm then we provide no guarantees that the ciphertext $c''=\Cond\Enc_{pk}(c', m_2,m_3)$ can be decrypted correctly or is even well formed i.e., in all of our constructions $\Cond\Enc_{pk}(c', m_2,m_3)$ will simply output $\bot$ when $c'= (1,\tilde{c})$ is a conditional ciphertext.

\begin{definition}[Correctness]\deflab{CondCorr} We say that $\Pi$ is $1-\epsilon(\cdot)$-correct if the following conditions hold: 
	\begin{itemize}
		\item \textbf{Regular encryption correctness}.  $\forall r_1, r_2 \in  \{0,1\}^{p(\lambda)}, m \in \mathcal{M}$ we have $\Dec_{sk}\left(\Enc_{pk}\left(m; r_2\right)\right) = m$ whenever $\{sk, pk\}\gets \KG(1^\lambda; r_1)$.

		\item (\textbf{Conditional encryption correctness}.) $\forall r_1, r_2 \in_R \{0,1\}^{p(\lambda)}$,  $m_1 \in \mathcal{M}, m_2 ,m_3 \in \mathcal{M}$ such that $P(m_1,m_2)=1$ we have \[\Pr\left[\Dec_{sk}\left( \Cond\Enc_{pk}\left(\Enc_{pk}\left(m_1; r_2\right), m_2, m_3; r_3\right) \right) = m_3\right] \geq 1- \epsilon(\lambda)\] where the randomness is taken over the random coins $r_3$ of $\Cond\Enc$  and we fix $\{sk, pk\}\gets \KG(1^\lambda; r_1)$.

	\end{itemize}
If $\epsilon(\lambda)=0$ we simply say that $\Pi$ is correct. 
\end{definition}

\begin{definition}[Efficiency]\deflab{CondEff}
We say that the conditional encrypt scheme  $ \Pi = (\KG, \Enc, \Cond\Enc, \Dec)$ is efficient if all four algorithms run in probabilistic polynomial time in the security parameter $\lambda$. 
\end{definition}

We can optionally require that our conditional encryption $\Pi$ is error detecting  i.e., whenever $P(m_1,m_2)=0$, the ciphertext $c=\Cond\Enc_{pk}(c_1, m_2, m_3)$ will decrypt to a special symbol $\Dec_{sk}(c)=\bot$ (whp). If $\Pi$ is error detecting this allows us to detect which outputs of the $\Cond\Enc$ algorithm are (in)valid. 

\begin{definition}[Error Detecting]\deflab{ErrorDetecting} We say that $\Pi$ is $1-\epsilon(\cdot)$-error detecting if 
	$\forall r_1, r_2 \in_R \{0,1\}^{p(\lambda)},  m_1 \in \mathcal{M}, m_2 ,m_3 \in \mathcal{M}$ such that $P(m_1, m_2)=0$ we have \[\Pr\left[\Dec_{sk}\left( \Cond\Enc_{pk}\left(\Enc_{pk}(m_1; r_2), m_2, m_3; r_3\right) \right) = \bot\right] \geq 1- \epsilon(\lambda) \] where the randomness is taken over the selection of the random coins $r_3$ of $\Cond\Enc$ and we fix $\{sk, pk\}\gets \KG(1^\lambda; r_1)$. 
 

\end{definition}

We now formally define the security of a conditional encryption scheme. Intuitively, in the security game we ask a distinguisher to distinguish between a simulated ciphertext $\Sim(pk)$ and a conditionally encrypted ciphertext $\Cond\Enc_{pk}(c_1, m_2,m_3)$ --- assume that $c_1 = \Enc_{pk}(m_1,r_1)$ with $P(m_1,m_2)=0$. Clearly, the simulated ciphertext $\Sim(pk)$ cannot leak any information to the adversary as it is generated without knowledge of the control message $m_2$, the payload message $m_3$ or the ciphertext $c_1$.

\begin{definition}[Conditional Encryption Secrecy] \deflab{CondSec}
	We say that conditional encryption scheme $ \Pi = (\KG, \Enc,$ $ \Cond\Enc, \Dec)$ provides $\left(t\left(\cdot\right), t_{\Sim}\left(\cdot\right),\epsilon\left(\cdot\right)\right)$-conditional encryption secrecy  if there exists a simulator $ \Sim $ running in time at most $t_{\Sim}(\lambda)$ such that  for all messages $m_1, m_2, m_3 \in \mathcal{M}$ such that $P(m_1,m_2)=0$, all $\lambda \in \mathbb{N}$, $r,r_1\in \{0,1\}^\lambda$ and all distinguishers $\D$ running in time at most $t(\lambda)$ 

\begin{align}
\Big| &\Pr\left[\D\left(sk, pk, m_1, m_2, m_3, c_1, \Sim\left(pk\right)\right)=1\right] \nonumber\\ 
& - \Pr\left[\D\left(sk, pk, m_1, m_2, m_3, c_1, \Cond\Enc_{pk}\left(c_1, m_2,m_3\right)\right)=1\right] \Big|\nonumber\\
 &\leq \eps\left(t\left(\lambda\right), \lambda\right)   \label{AdvCondEncSemi}
\end{align}
where the randomness is taken over the random coins of the distinguisher and the  conditional encryption algorithm $ \Cond\Enc $. Here, $ (sk, pk) = \KG(1^\lambda; r) $ and $ c_1 =\Enc_{pk}(m_1; r_1) $ denote the public/secret key and the ciphertext computed under the a priori  fixed random strings $ r  $ and $ r_1 $. If $\epsilon(\cdot)=0$ and $t(\cdot)=\infty$ then we say that $\Pi$ has perfect conditional encryption secrecy. 
\end{definition}

The definition of conditional encryption secrecy holds in a semi-honest setting where we assume that the public key $(sk,pk) = \KG(1^\lambda)$ and the ciphertext $c_1 = \Enc_{pk}(m_1,r_1)$ were both generated honestly. We remark that this assumption is reasonable in our password typo vault application because the keys and ciphertexts are generated by the authentication server (a trusted party). However, one can imagine applications where we do not want to assume that $c_1$ and $sk$ were generated honestly. We leave it as an open question to define/construct maliciously secure conditional encryption schemes.



{\noindent \bf Real-Or-Random Security:} We also require that a conditional encryption scheme satisfies the traditional notion of {\em real-or-random} (RoR) security i.e., an attacker who does not have the secret key cannot distinguish between real and random ciphertexts. \fullversion{In \appref{apdx:RealOrRandomProofs}}{In the full version \cite{fullversion}} we extend the traditional definition of {\em real-or-random} (RoR) security to conditional encryption schemes. All of our conditional encryption schemes constructions will satisfy RoR security under the plausible assumption that Pallier encryption itself satisfies RoR security. Because our focus is on conditional encryption secrecy we will defer all RoR security proofs to \fullversion{\appref{apdx:RealOrRandomProofs}}{the full version \cite{fullversion}}.

\section{Concrete Constructions of Conditional Encryption}\seclab{ConcretCondEnc}
In this section we present concrete constructions of conditional encryption for several different binary predicates. As a warm-up we first consider the equality predicate $P_=(m_1,m_2)=1$ if and only if $m_1 = m_2$. As an application we can use this construction to obtain conditional encryption for the CAPSLOCK predicate since $P_{\Capslock} (m_1, m_2) = P_= (m_1,  \InvertCase(m_2))$. We then provide constructions for predicates based on the Hamming Distance (resp. Edit Distance) between $m_1$ and $m_2$. Finally, given conditional encryption schemes for predicates $P_1,\ldots, P_k$ we show how to compose these results to obtain conditional encryption schemes for the OR predicate $P_{OR}(m_1,m_2) \doteq \bigvee_{i=1}^k P_i(m_1,m_2) $. 

\subsection{Conditional Encryption for the Equality Test Predicate $ P_= (x)$}
In this part, we will start off by providing a concrete construction of conditional encryption when the predicate is equality test $ P_{=} $. That means that given a ciphertext of an unknown message $ m_1 $, and the input messages $ m_2 $ and payload $ m_3 $, we compute the encryption of $ m_3 $ if and only if $ m_1 = m_2 $, i.e., $ P_{=} (m_1, m_2)=1 $.

Our construction utilizes the Paillier public key encryption scheme (see \fullversion{\appref{App:PailDetails}}{the full version \cite{fullversion}} for more details about Pallier) which contains three main algorithms $ \Pi_{\Pail} = (\Pail.\KG, \Pail.\Enc, \Pail.\Dec)$. It will also be convenient to let $\ToInt$ denote an injective mapping from our message space $\Sigma^{\leq n}$ to $\mathbb{Z}_{|\Sigma|^{n+1}}$ and use $\ToOrig$ to denote the inverse mapping --- we define $\ToOrig(y)=\bot$ if there is no preimage $m \in \Sigma^{\leq n}$ such that $y = \ToInt(m)$.

Our conditional encryption construction $\Pi$ sets $\Pi.\KG \doteq \Pail.\KG$ and we set $\Pi.\Enc_{pk}(m_1) \doteq \left(0, \Pail.\Enc_{pk}(m_1)\right)$ i.e., to encrypt $ m_1 \in \Sigma^{\leq n} $ we simply compute $ c = \Pail.\Enc_{pk} \left(\ToInt\left(m_1\right)\right) $ and output the ciphertext $ c_{m_1} = (b = 0, c) $. Given a ciphertext $c=(0,c_1)$  with flag $ b=0 $ we define $\Pi.\Dec_{sk}(c=(0,c_1)) \doteq \ToOrig\left(\Pail.\Dec_{sk}(c_1)\right)$. The conditional encryption algorithm $\Pi.\Cond\Enc_{pk}\left(c_{m_1} = (0,c),m_2,m_3\right)$ works as follows:  First we extract $N$ from the public key $pk$ and compute $ \hat{m_2}  = \ToInt(m_2)$ and $ \hat{m_3} =\ToInt(m_3) $ to map them to Paillier's plaintext space. Next we pick random numbers $ R\in_R \mathbb{Z}_N $ and $r \in_R \mathbb{Z}^*_N$ and compute $ c^= = c^R \left(N+1\right)^{-R\hat{m}_2+\hat{m}_3} r^N \mod{N^2}$. Finally, we output   $\Pi.\Cond\Enc_{pk}\left(c_{m_1} = (0,c),m_2,m_3; R, r\right) = (1,c^=)$. Given a conditional ciphertext $(1, c^{=})$ the decryption algorithm will simply output $\Pi.\Dec_{sk}(1,c^=) \doteq \ToOrig(\Pail.\Dec_{sk}(c^=))$. \fullversion{See \constref{Const:EqualityTest} in \appref{AppConstructions} for a more formal description of our construction.}{We refer to the above construction as \constlab{Const:EqualityTest} \constref{Const:EqualityTest}.  }

 Intuitively, we have $c= (N+1)^{\hat{m}_1} r_1^N \mod{N^2}$ for some message $m_1 = \ToOrig(\hat{m}_1)$ and random value $r_1 \in \mathbb{Z}_N^*$. In this case, the final ciphertext $c^=$ can be written as $c^= = \left(N+1\right)^{R(\hat{m}_1-\hat{m}_2)+\hat{m}_3} (rr_1^R)^N \mod{N^2}$. If $\hat{m}_1-\hat{m}_2 = 0$ then $R(\hat{m}_1-\hat{m}_2)=0$ cancels and we are left with a valid encryption of $m_3$. Otherwise, the value  $R(\hat{m}_1-\hat{m}_2)+\hat{m}_3 \mod{N}$ can be viewed as a fresh Paillier ciphertext encrypting a uniformly random integer in $\mathbb{Z}_{N}$. More specifically, let $y=\ToInt(m_2)-\ToInt(m_1)$. As long as $\gcd(N, y)=1$ (due to the selection of $ p,q $ s.t., $\min(p, q) > |\Sigma^{n+1}|$)  the value $R'=R(\hat{m}_1-\hat{m}_2)+\hat{m}_3 \mod{N}$ will also be distributed uniformly in $\Mbb{Z}_N$ i.e., for all $x \in \Mbb{Z}_N$ we have $\Pr[yR = x \mod{N}] = \Pr[R=xy^{-1} \mod{N}]=1/N$ when $R \in \Mbb{Z}_N$ is random and $y^{-1}$ is the multiplicative inverse of $y \mod{N}$. When we pick our Paillier public key ,we can ensure we always have $\gcd(N, y)=1$ by selecting primes $p$ and  $q$ such that $p,q \geq \max_{m_1 \in \Sigma^{n}} \ToInt(m_1)<  |\Sigma^{n+1}|$ and setting $N=pq$.

\newcommand{\ThmEqCorrectness}{	
\constref{Const:EqualityTest}  is a  perfectly correct and $1-\epsilon(\lambda)$-error detecting conditional encryption scheme with $\eps(\lambda) = \dfrac{|\Sigma|^{n+1}}{N} \leq \dfrac{1}{\max\{p,q\}}$.}

\begin{theorem} \thmlab{thm:EqCor}
\ThmEqCorrectness
\end{theorem}
The proof of \thmref{thm:EqCor} can be found in \fullversion{\appref{apdx:MissingProofs}}{the full version \cite{fullversion}}.

\newcommand{\ThmEqTestSecrecy}{	The conditional encryption scheme described in \constref{Const:EqualityTest} provides $ (\infty, t_{\Sim}, 0) $ conditional encryption secrecy in which $ t_{\Sim} = t_{\Pail.\Enc} $ is time of doing one Paillier Encryption. }

\begin{theorem}\thmlab{thm:EqualityTestSecrecy}
\ThmEqTestSecrecy
\end{theorem}
 Intuitively, the simulator $\Sim(pk)$ simply picks random values $r' \in \mathbb{Z}_N^*$ and $R' \in \mathbb{Z}_N$ and outputs $(N+1)^{R'} r'^N \mod{N^2}$ i.e., the Paillier encryption of a uniformly random message. Intuitively, if $\hat{m}_1\neq \hat{m}_2$ and $\mathbf{gcd}(\hat{m}_1-\hat{m}_2,N)=1$ then the value $R(\hat{m}_1\neq \hat{m}_2) + \hat{m}_3$ is uniformly random in $\mathbb{Z}_N$ i.e., statically indistinguishable from $R'$. Similarly, if we fix any $z \in \mathbb{Z}_N^*$ (we will use $z=r_1^R \mod{N}$) and pick $r \in \mathbb{Z}_N^*$ randomly then the value $rz \mod{N}$ is also uniformly random in $\mathbb{Z}_N^*$ i.e., statically indistinguishable from $r'$. The formal proof of \thmref{thm:EqualityTestSecrecy} is available in \fullversion{\appref{apdx:MissingProofs}}{the full version}. We also prove that \constref{Const:EqualityTest} satisfies the traditional notion of Real-or-Random security  --- see \fullversion{\thmref{thm:ROREquality} in \appref{apdx:RealOrRandomProofs}}{the full version \cite{fullversion}}. 

\subsection{CAPSLOCK Predicate}
We can immediately use our conditional encryption scheme for equality test to obtain a construction for the the CAPSLOCK predicate $ P_{\Capslock}$ which is defined as $ P_{\Capslock} (m_1, m_2) =1 $ if and only if $ m_1 = \InvertCase(m_2) $; otherwise $ P_{\Capslock} (m_1, m_2) =0 $. Observe that we can equivalently define \[ P_{\Capslock} (m_1, m_2) = P_= (m_1,  \InvertCase(m_2)) .\]
Thus, our conditional encryption construction $\Pi_{\Capslock}$ for $ P_{\Capslock}$ is exactly as our construction $\Pi_{=}$ for $P_{=}$  with the following modification to the conditional encryption algorithm  $\Pi_{\Capslock}.\Cond\Enc_{pk}$ $ (c_{m_1}, m_2, m_3) = \Pi_{=}.\Cond\Enc_{pk} (c_{m_1}, \InvertCase(m_2), m_3)$
Conditional encryption secrecy and correctness of the construction $\Pi_{CPSLCK}$ follows immediately from \thmref{thm:EqualityTestSecrecy} and \thmref{thm:EqCor} respectively. {RoR security also follows directly from RoR security of \constref{Const:EqualityTest} \fullversion{---  see \thmref{thm:ROREquality} in \appref{apdx:RealOrRandomProofs} for RoR security}{}. 

\subsection{Hamming Distance Predicate}
We now describe our conditional encryption construction for the Hamming Distance predicate  $P_{\ell,\Ham}$ $(m_1,m_2) =1$ if and only if $ \Ham(m_1, m_2 ) \leq \ell $. For ease of exposition, we will describe our construction under the assumption that all messages $m_1,m_2 \in \Sigma^{n}$ have the exact same length. In several of our applications (e.g., password typos) it may not necessarily be the case that all messages have the same length. However, we can easily deal with this issue by defining an injective padding function $\Pad: \Sigma^{\leq m} \rightarrow \Sigma'^n$ where $\Sigma' = \Sigma \cup \{y\}$ extends the alphabet $\Sigma$ by adding a new symbol $y$. In particular, given $m \in \Sigma^{\leq n}$ we define $\Pad(m) = m \| y^{n-|m|}$ to be the length $n$ string from our larger alphabet $\Sigma'$ obtained by padding $m$ with $y$'s. Clearly, the function $\Pad$ is injective so we can define an inverse $\Pad^{-1}$ such that $\Pad^{-1}\left(\Pad\left(m\right)\right)=m$ for any $m \in \Sigma^{\leq m}$. Given two messages $ m, m'\in \Sigma^{\leq n} $ and an integer $ \ell \geq 0 $, we define the binary predicate $ P_{\ell,\Ham,\Pad}(m, m') =1  $ if and only if $ \Ham\left(\Pad\left(m\right), \Pad\left(m'\right)\right) \leq \ell $. Clearly, if we have a conditional encryption scheme for the Hamming Distance predicate $P_{\ell,\Ham}(m_1,m_2)$ with message space $m_1,m_2 \in \Sigma'^{n}$ then we can immediately apply padding to obtain a  conditional encryption scheme for the related predicate $ P_{\ell,\Ham,\Pad}(m_1, m_2)$ with message space $m_1,m_2 \in \Sigma^{\leq n}$.

{\noindent \bf Attempt 1:} As an initial attempt at constructing conditional encryption for our predicate $ P_{\ell,\Ham}$ we can use our equality predicate construction as a blackbox along with the Shamir secret sharing scheme $\SS =  (\Share, \recover) $ and a symmetric key authenticated encryption scheme. The basic idea is to split messages into individual characters $m_1 = \left(m_1\left[1\right],\ldots,m_1\left[n\right]\right)$ and encrypt character by character to obtain $c_1 = \left(c_1\left[1\right],\ldots,c_1\left[n\right]\right) = \Enc_{pk}(m_1)$. The conditional encryption algorithm $\Cond\Enc_{pk}(c_1,m_2,m_3)$ would similarly split the control message $m_2$ up into $n$ individual characters and generate $n$ secret shares $\ldb s\rdb_1,\ldots, \ldb s\rdb_n$ of a fresh symmetric key $K$. We would then use the original conditional encryption scheme for $P_=$ to compute $c_2[i] = \Pi_{=}.\Cond\Enc_{pk}(c_i, m_2[i], \ldb s \rdb_i)$ for each $i\leq n$ using $m_2[i]$ as the control message and $\ldb s \rdb_i$ as the payload message. The final conditional ciphertext would include $c_2[1],\ldots,c_2[n]$ as well as an encryption of $m_3$ using the symmetric key $K$. The decryption algorithm would decrypt each $c_2[i]$ to obtain the share  $\ldb s \rdb_i$. As long as the Hamming Distance predicate holds the decryption algorithm would obtain enough shares to recover $K$ and decrypt $m_3$. 

The problem with this construction is that if the predicate does not hold then an attacker who knows the secret key can still (whp) identify which shares are valid. In particular, it would be trivial for a party who knows the secret key $sk$ to distinguish between the encryption of a valid share  $\ldb s \rdb_i < 2^{\lambda}$ (recovered when $m_1[i]=m_2[i])$ and the encryption of a random element in $\mathbb{Z}_{N}$ (as is the case when $m_2[i] \neq m_1[i]$) since $2^\lambda \ll N$. This would allow the attacker to learn the set $S = \left\{i: m_1\left[i\right] = m_2\left[i\right]\right\}$ of indices $i\leq n$ where $m_2$ matches $m_1$ even if $P_{\ell,\Ham}(m_1,m_2)=0$ --- a clear violation of conditional encryption secrecy!

{\bf The Fix:} To address the above issue we use a randomized encoding to ensure that, when the predicate does not hold, it is impossible to identify which shares are (in)valid. In particular, instead of computing $c_2[i] = \Pi_{=}.\Cond\Enc_{pk}(c_i, m_2[i], \ldb s \rdb_i)$ we instead compute $c_2[i]= \Pi_{=}.\Cond\Enc_{pk}(c_i, m_2[i],x_i)$ where $x_i = \RanEnc(\ldb s \rdb_i)$ is a random encoding of the share  $\ldb s \rdb_i$ as an integer in larger Pallier plaintext space $\mathbb{Z}_N$. In more detail $\RanEnc(x) = a_i 2^\lambda + x$ where the value $a_i \leq \lfloor \frac{N-1-x}{2^\lambda}\rfloor$ is chosen uniformly at random. Intuitively, $x_i = \RanEnc(\ldb s \rdb_i)$ encodes the share $\ldb s \rdb_i$ as a random element in $\mathbb{Z}_{N'}$ subject to the constraint that $\ldb s \rdb_i=\RanEnc(\ldb s \rdb_i) \mod{2^\lambda}$. Since the value of the share $\ldb s \rdb_i$ itself is random this ensures that the attacker cannot distinguish $x_i$ from a random element in $\mathbb{Z}_N$ ---  unless the predicate $ P_{\ell,\Ham}$ holds and we can recover enough correct shares to recover the secret decyrption key. Decrypting a conditional ciphertext will require more work since we do not know a priori which recovered shares are valid and we have to consider all possible subsets. Fortunately, when $\ell$ is constant the number of subsets remains polynomial in $n$. 


In a bit more detail the conditional encryption scheme $\Pi$ works as follows:

\begin{enumerate}
    \item The regular encryption algorithm $\Pi.\Enc_{pk}(m)$ takes as input a message $m =\left(m\left[1\right],\ldots,m\left[n\right]\right)\in \Sigma^{n}$ and encrypts $m$ character by character to obtain a vector of Paillier ciphertexts $c=\left(c\left[1\right],\ldots, c\left[n\right]\right)$ where $ c_i = \Pail.\Enc_{pk}\left(\ToInt(m[i]); r_i\right)$ The regular encryption algorithm outputs $\left(b=0,c\left[1\right],\ldots,c\left[n\right]\right)$ where the flag $b=0$ indicates that this ciphertext was produced by the regular encryption algorithm. 
    \item The conditional encryption algorithm $\Pi.\Cond\Enc(c_1,m_2,m_3)$ takes as input a ciphertext $c_1=\left(b=0, c_1\left[1\right],\ldots,c_1\left[n\right]\right)$ corresponding to some unknown message $m_1 = \left(m_1\left[1\right],\ldots,m_1\left[n\right]\right)$, a control message $m_2 = \left(m_2\left[1\right],\ldots m_2\left[n\right]\right) \in \Sigma^{n}$ and a payload message $m_3$. We first generate a random symmetric key $K \in \{0,1\}^\lambda$ for our authenticated encryption scheme and encrypt the payload message $m_3$ using $K$ to obtain $ c_{AE}  = \Auth\Enc_k( m_3)$. Second we use the Shamir secret sharing scheme to generate $n$ shares $ (\ldb s \rdb_1, \ldots,  \ldb s \rdb_n)\gets \Share(n, n-\ell, K)$ for our secret key $K$. We configure our secret sharing scheme such that $n-\ell$ shares are sufficient to recover $K$, but any subset of $n-\ell-1$ shares information theoretically leaks nothing about $K$. We now follow our equality test construction and compute $c[i] = c_1[i]^{R_i} (N+1)^{-R_i m_2[i] + x_i} r_i^N $ where $x_i = \RanEnc(\ldb s\rdb_i)$ is the random encoding of the share $\ldb s\rdb_i$, $R_i$ is a uniformly random integer in $\mathbb{Z}_N$ and $r_i$ is uniformly random in $\mathbb{Z}_N^*$.  Our final output is $(b=1,c,c_{AE})$ in which $c= (c[1], \ldots, c[n])$.

    \item Given a ciphertext $\left(b=0,c\left[1\right],\ldots,c\left[n\right]\right)$ with $b=0$ the decryption algorithm will simply decrypt character by character to recover $m=\left(m\left[1\right],\ldots,m\left[n\right]\right)$ where $m[i] = \ToInt(x_i)$ and $x_i = \Pail.\Dec_{sk}\left(c[i]\right)$. Given a conditionally encrypted ciphertext $\left(b=1,c\left[1\right],\ldots,c\left[n\right], c_{AE}\right)$ we will first extract shares $\ldb s'\rdb_i  = \RanDec(x_i)$ with $x_i = \Pail.\Dec_{sk}\left(c\left[i\right]\right)$. We will then look through all ${n \choose n-\ell}$ subsets $S \subseteq [n]$ of $n-\ell$ indexes and their corresponding shares to recover a string $$K_{S} = \recover\left(\left(S\left[i\right],\ldb s'\rdb_{S[i]}\right) \forall 0\leq i \leq n-\ell \right)$$ which may or may not be valid. If $\Auth.\Dec_{K_S}(c_{AE}) = \bot$ then we conclude that $K_S$ is invalid and move on to the next subset; otherwise if  $\Auth.\Dec_{K_S}(c_{AE}) = m$, we return $m$. If $\Auth.\Dec_{K_S}(c_{AE}) = \bot$ for all subsets $S \subseteq [n] $ and their $n-\ell$ corresponding shares, then we output $\bot$.  
\end{enumerate}

\fullversion{See \constref{const:ArbHamm} in \appref{AppConstructions} for a formal description of the construction and see \thmref{thm:ArbHamm} in \appref{apdx:MissingProofs} for a proof that \constref{const:ArbHamm}  is $ 1- \eps(\lambda)$-correct and a $1-\epsilon(\lambda)$-error detecting for a negligible function $\eps(\lambda)$. }{We refer to the above construction as \constlab{const:ArbHamm}\constref{const:ArbHamm}. The full version \cite{fullversion} contains proofs that \constref{const:ArbHamm} is $ 1- \eps(\lambda)$-correct and a $1-\epsilon(\lambda)$-error detecting for a negligible function $\eps(\lambda)$.  }

\subsubsection{Correctness of the \constref{const:ArbHamm}}

\newcommand{\ThmHammingDistCorrectness}{	\constref{const:ArbHamm}  is a  $ 1- \eps(\lambda)$-correct and a $1-\epsilon(\lambda)$-error detecting conditional encryption scheme with $\eps(\lambda) = {n \choose \ell} \epsilon_{AE}(\lambda) + 2^{-\lambda}$. Here, \[ \epsilon_{AE}\left(\lambda\right)\doteq \max_{m} \Pr_{K_1,K_2 \in \{0,1\}^\lambda}\left[\Auth.\Dec_{K_1}\left(\Auth.\Enc_{K_2}\left(m\right) \right) \neq \bot \right] \] denotes the negligible probability that the ciphertext $c=\Enc_{K_2}(m)$ is still valid under an unrelated key $K_1$.}


We now prove that  \constref{const:ArbHamm}, satisfies the security definition \defref{CondSec}, i.e., conditional encryption secrecy. We first make a basic statistical observation.

\newcommand{\thmStatDistTwo}{Let $b = ak + r$ where $0\leq r < a$ is the reminder (i.e., $r = b \mod a$). Consider the uniform distributions $\U_b$ which outputs a random value in $\Mbb{Z}_b$ and the distribution $\D_{ak}$ which outputs random values between $0, \cdots, ak$. Then the statistical distance between these two distributions is $\mathtt{SD}(\D_{ak}, \U_b) = \frac{r}{b} \leq \frac{1}{k+1}$.}

\begin{theorem}{\thmlab{thm:StatDistUak}}  
\thmStatDistTwo
\end{theorem}

Intuitively, \thmref{thm:StatDistUak} implies that we cannot distinguish between $\RanEnc(x)$ and a uniformly random $y \in \mathbb{Z}_N$ whenever $0 \leq x < 2^{\lambda}$ is picked randomly. The proof of \thmref{thm:StatDistUak} can be found in \fullversion{\appref{apdx:MissingProofs}}{the full version \cite{fullversion}}. 

\newcommand{\thmSemiHonest}{	[Conditional Encryption Secrecy of \constref{const:ArbHamm}] Assume that our Authenticated encryption scheme $ \Pi_{AE} = (\Auth\Enc, \Auth\Dec) $ is $ \left(t_{AE}, \eps_{AE}\left(t_{AE},\lambda\right)\right) $-secure for any security parameter $\lambda$ and any running time parameter $t_{AE}$. Then for any $t$ and any security parameter $\lambda$ \constref{const:ArbHamm} provides $\left(t, t_{\Sim}, \eps\left(t,\lambda\right)\right)$ conditional encryption secrecy with $ \eps(t,\lambda) \leq \eps_{AE}(t,\lambda) + 2^{-\lambda}$ and $ t_{\Sim}  = n \cdot t_{P.\Enc} + \mathtt{poly}(\lambda)$. }

\begin{theorem} \thmlab{thm:CondSecArbHamm}
 \thmSemiHonest
\end{theorem} 

\thmref{thm:CondSecArbHamm} follows by applying \thmref{thm:StatDistUak}  with $b=N$, $a= \lfloor N/2^{\lambda} \rfloor$ and $k=2^{\lambda}$.
We defer the formal proof of \thmref{thm:CondSecArbHamm} and \thmref{thm:StatDistUak}  to \fullversion{\appref{apdx:MissingProofs}}{the full version \cite{fullversion}} where we also prove that \constref{const:ArbHamm} provides Real-or-Random security\fullversion{ (see \thmref{thm:RORHamming})}{}.


\subsubsection{Efficiency} \label{subsec:Efficiency} The running time of the key generation algorithm $\KG$ is essentially equivalent to Pallier --- with high probability we will have $\min\{p,q\} > \max\left\{2n2^{2\lambda}, \left|\Sigma\right|\right\}$. The running encryption algorithm $\Enc$ is essentially $n \times t_{p}$ where $t_p$ denotes the running time for regular Pallier Encryption and the resulting ciphertext has size $1+n \cdot \lceil \log_2 N^2 \rceil$ (bits). The running time for the conditional encryption algorithm is  essentially $n \times t_{p} + t_{AE} + t_{SS}$  where $t_p$ (resp. $t_{AE}$, $t_{SS}$) denotes the time for one Pallier Encryption (resp. one authenticated encryption/one execution of \Share over a field of size $2^{\lambda}$). The size of a conditionaly encrypted ciphertext is $1+n\lceil \log_2 N^2 \rceil + s_{AE}$ where $s_{AE}$ denotes the length of the authenticated encryption ciphertext. The running time of $\Dec_{sk}$ on a conditionally encypted ciphertext is roughly ${n \choose \ell} (t_{SSrec} + t_{AE})$ where $t_{SSrec}$ (resp. $t_{AE}$) denotes the running time for $\recover$ over a field of size $2^{\lambda}$ (resp. $\Auth.\Dec$). If we incorporate a second secret sharing scheme over a smaller finite field then it is possible to slightly optimize the performance to achieve running time ${n \choose \ell} t_{SSrec}' + O(t_{SSrec} + t_{AE})$ where $t_{SSRec}'$ denotes the execution time for secret share recovery over the {\em smaller} finite field --- see details in Section \ref{subsec:OptimizedHamming}.

\subsection{Edit Distance One}\seclab{Sec:EDOne}

Given two messages $ m, m'\in \Sigma^{ \leq n} $ and an integer $ \ell \geq 0 $, we define the binary predicate $ P_{\ell,\ED}(m, m') =1$ if and only if  $ \ED(m, m') \leq \ell $; 
otherwise, $ P_{\ell,\ED}(m, m') =0 $. In this section, we will construct a conditional encryption scheme for $P_{1,\ED}$ i.e., edit-distance $1$. It would be possible to implement the same general construction for $\ell > 1$. However, the ciphertext sizes would grow proportional to $O\left(n^{\ell}\right)$. Thus, we focus on the $\ell = 1$ case since it is the most useful case for password typo correction (and yields the most efficient construction). We will let $\Pi_{1, \ED} = (\KG, \Enc,\Cond\Enc,\Dec)$ denote our conditional encryption for $P_{1,\ED}$ described below.

Given $m = \left(m\left[1\right],\ldots, m\left[k\right]\right) \in \Sigma^{k}$ we define $$m_{-i} = \left(m\left[1\right],\ldots, m\left[i-1\right],m\left[i+1\right],\ldots, m\left[k\right]\right) \in \Sigma^{k-1}$$ to be the string obtained by deleting the $i$th character from $m$ e.g., if $m=$``bead" then $m_{-2}$=``bad". If $j=0$ or $j>k=|m|$ then we just define $m_{-j} = m$. Observe that $P_{1,\ED}(m,m')=1$ if and only if there exists $j$ such that $m_{-j} = m'$ or such that $m = m'_{-j}$. 

With this observation our construction for $P_{1,\ED}$ will use our construction for $P_{=}$ as a black box. $\KG(1^\lambda)$ works in the exact same way as the conditional encryption scheme for the equality predicate and will generate a key $\left(sk,pk=\left(N,g=N+1\right)\right)$ with $N=pq$ and $\min\{p,q\} \geq |\Sigma|^{n+1}$. Our regular encryption algorithm $\Enc_{pk}(m)$ takes as input $m \in \Sigma^{\leq n}$ and outputs a vector $(0,\tilde{c}_0, \tilde{c}_1,\ldots, \tilde{c}_n)$ where  $\tilde{c}_i = P.\Enc_{pk}\left(\ToInt\left(m_{-i} \right)\right)$is the Pallier encryption of $m_{-i}$ encoded as an integer using the injective mapping $\ToInt:\Sigma^{\leq n} \rightarrow \mathbb{Z}_{|\Sigma|^{n+1}}$ --- $\ToOrig$ is the inverse mapping. The conditional encryption algorithm $\Cond\Enc_{pk}(c_1,m',m'')$ works by running $\Pi_{=}.\Cond\Enc_{pk}$, the conditional encryption algorithm for the equality predicate, on $2n+1$ different inputs to generate $\tilde{c}_0,\tilde{c}_1,\ldots, \tilde{c}_{2n}$ --- if for some $j$ we have $\tilde{c}_j = \bot$ then we simply output $\bot$. First, we parse $c_1 = \left(0, c_1\left[0\right],\ldots, c_1\left[n\right]\right) $ and set $\tilde{c}_{i} = \Pi_{=}.\Cond\Enc_{pk}\left(c_1\left[i\right], m', m''\right)$ for each $0 \leq i \leq n$. Intuitively, if $m' = m_{-i}$ then $\tilde{c}_{i}$ is a Pallier encryption of our payload $m''$; otherwise, $\tilde{c}_{i} = g^{y_i} r_i^N \mod{N^2}$ will be the random Pallier encryption of a uniformly random $y_i \in \mathbb{Z}_N$ under a uniformly random nonce $r_i \in \mathbb{Z}_N^*$. Similarly, we can set $\tilde{c}_{n+i} = \Pi_{=}.\Cond\Enc_{pk}(c_1[0], m_{-i}', m'')$ for each $1 \leq i \leq n$. Intuitively, if $m_{-i}' = m_{-0} = m$ then $\tilde{c}_{n+i}$ is a Pallier encryption of our payload $m''$; otherwise, $\tilde{c}_{i} = g^{y_i} r_i^N \mod{N^2}$ will be the random Pallier encryption of a uniformly random $y_i \in \mathbb{Z}_N$ under a uniformly random nonce $r_i \in \mathbb{Z}_N^*$. The decryption algorithm $\Dec_{sk}$ is defined in the natural way. In particular, $\Dec_{sk}\left(0,c\left[0\right],c\left[1\right],\ldots,c\left[n\right]\right)$ simply decrypts $c[0]$ as $x_0 = P.\Dec_{sk}\left(c\left[0\right]\right)$ using regular Pallier decryption $P.\Dec$ and then outputs $m=\ToOrig(x_0)$.

Similarly, $\Dec_{sk}(1,\tilde{c}_0,\ldots, \tilde{c}_{2n})$ will run our conditional decryption algorithm $P_{=}$ on each individual ciphertext $\tilde{c}_i$ to recover $x_0, x_1,\ldots, x_{2n}$ with $x_i = P.\Dec_{sk}(\tilde{c}_{i})$. If $\min\left\{x_0,\ldots, x_{2n}\right\} > |\Sigma|^{n+1}$ then we output $\bot$; otherwise we can simply return  $\ToOrig\left( \min\{x_0,\ldots, x_{2n} \}\right)$

\newcommand{\thmEDCorrect}{$\Pi_{1, \ED}$ is a $1-\epsilon(\lambda)$ correct conditional encryption scheme for the predicate $P_{1,\ED}$ and $\Pi_{1, \ED}$ is $1-\epsilon(\lambda)$-error detecting with $\eps(\lambda) = \frac{(2n+1)|\Sigma|^{n+1}}{N} \leq \frac{2n+1}{\max\{p,q\}}$. }
\begin{theorem} \thmlab{thm:EDCor}
\thmEDCorrect
\end{theorem}

\newcommand{\thmEDPrivacy}{$\Pi_{1, \ED}$ provides $ (\infty, t_{\Sim}, 0) $ conditional encryption secrecy for the predicate $P_{1,\ED}$. Here, $ t_{\Sim} = (2n+1)t_{\Pail.\Enc} $ is time of doing $(2n+1)$ Paillier Encryptions.  }

\begin{theorem}\thmlab{thm:EDPrivacy}
\thmEDPrivacy
\end{theorem}
\begin{proofof}{\thmref{thm:EDPrivacy}} (Sketch) Assume that $P_{1,\ED}(m,m')=0$ it follows from \thmref{thm:EqualityTestSecrecy} that for any $j \leq n$ that $\Cond\Enc^{=}_{pk}(\Enc_{pk}(m_{-j}),$ $ m',m'')$ outputs $g^{R_j} r_j^N \mod{N^2}$ for a uniformly random $R_j \in \mathbb{Z}_N$ and $r_j \in \mathbb{Z}_N$. Similarly,  for any $j \leq n$ it follows that $\Cond\Enc^{=}(\Enc_{pk}$ $(m), m_{-j}',m''))$ outputs $g^{R_j} r_j^N \mod{N^2}$ for a uniformly random $R_j \in \mathbb{Z}_N$ and $r_j \in \mathbb{Z}_N$. Thus, $\Cond\Enc_{pk}\left(\Enc_{sk}\left(m_{-j}, m',m''\right)\right)$ outputs $(1,\tilde{c}_{0},\ldots, \tilde{c}_{2n})$ where for each $j \leq 2n$ the Pallier Ciphertext $\tilde{c}_i$ is uniformly random in $\mathbb{Z}_{N^2}^*$. 

We define the simulator $ \Sim(pk) $ as follows. The simulator $ \Sim(pk) $ takes as input the Paillier public key $ pk $. For each $0 \leq i \leq 2n+1$ the simulator then selects $ R_i\in_R \mathbb{Z}_{N} $ and $r_i\in_R \mathbb{Z}^*_N$ uniformly at random and then encrypts $ R_s $ as $ C_{\Sim}[i] = \Pail.\Enc_{pk}(R_i; r_i) = g^{R_i} r_i^{N} \mod N^2$ i.e., $C_{\Sim}[i]$ is uniformly random in $\mathbb{Z}_{N^2}^*$. Finally, the simulator outputs $C_{\Sim} =  (1,C_{\Sim,0},\ldots, C_{\Sim,2n})$. 
\end{proofof}

\subsection{OR Composition}\seclab{CompositionAND/OR}
Suppose we have conditional encryption schemes $\Pi_1,\ldots, \Pi_k$ for $k$ different predicates $P_1,\ldots, P_k$ and that each scheme has the same message space. Let $P_{or}(m_1,m_2) = \bigvee_{i=1}^k P_i(m_1,m_2)$ the predicate which is $0$ (false) if and only if all of the predicates are false i.e., $P_i(m_1,m_2)=0$ for all $i \leq k$. We will define a conditional encryption scheme $\Pi_{or} = (\KG, \Enc,\Cond\Enc,\Dec)$ for the predicate $P_{or}$. 

Intuitively, our key generation algorithm $\KG(1^{\lambda})$ runs $(sk_i,pk_i) \gets \KG_i(1^\lambda)$ for each $i$ and outputs $(sk,pk)$ where $sk = (sk_1,\ldots, sk_k)$ and $pk = (pk_1,\ldots, pk_k)$\footnote{As an optimization if $\Pi_i.\KG_i(1^\lambda)$ generates a Pallier key for each $i$ then we can generate one Pallier key $(sk_0,pk_0)$ and set $(sk_i,pk_i)=(sk_0,pk_0)$ for all $1 \leq i \leq k$. }. The algorithm $\Enc_{pk}(m)$ simply generates $c_i = \Pi_i.\Enc_{pk}(m)$ for each $i\leq k$ and outputs $(0,c_1,\ldots, c_k)$. Similarly, the algorithm $\Cond\Enc_{pk}\left(c=\left(0,c_1,\ldots,c_k\right),m',m''\right)$ simply generates $c_i = \Pi_i.\Cond\Enc_{pk}(c_i,m',m'')$ for each $i\leq k$ and outputs $(0,\tilde{c}_1,\ldots, \tilde{c}_k)$ --- if $\tilde{c}_i = \bot$ for any $i \leq k$ then we instead output $\Pi_{or}.\Cond\Enc_{pk}(c,m',m'') = \bot$. Finally, the $\Dec_{sk}(c)$ will run $m_i = \Pi_i.\Dec_{sk}(c)$ to obtain $m_i \in \mathcal{M} \cup \{\bot\}$. If $m_i=\bot$ for all $i\leq k$ then the algorithm outputs $\bot$; otherwise we output $m_j$ where $j$ is largest integer such that $m_j \neq \bot$.

\newcommand{\thmORPrivacy}{Suppose that we are given $ k $ separate conditional encryption schemes $ \Pi_1, \ldots, \Pi_k $ corresponding predicates $ P_1, \ldots, P_k $ and that each $\Pi_i$ provides $(t(\lambda), t_{\Sim,i}(\lambda), $ $\eps_i(t(\lambda),\lambda))$-conditional encryption secrecy. The construction $\Pi_{or}$ provides $ (t'(\lambda), t_{\Sim}'(\lambda), $ $\eps'(t'(\lambda), \lambda)) $-conditional encryption secrecy with $ t'(\lambda) = O\left(t\left(\lambda\right)\right)$, $ t'_{\Sim}(\lambda) \approx \sum_{i = 1}^{k} t_{\Sim_i}(\lambda) $ and $ \eps'\left(t'\left(\lambda\right),\lambda\right) = \sum_{i}^k \eps_i\left(t'\left(\lambda\right),\lambda\right)$. }

\begin{theorem}{\thmlab{ORComp:Security}}
     \thmORPrivacy
\end{theorem} 


The formal proof is available in \fullversion{\appref{apdx:MissingProofs}}{the full version}. Intuitively, the simulator $\Sim_{OR}(pk)$ for $\Pi_{OR}$ will run the simulator $\Sim_i(pk_i)$ for each conditional encryption scheme and concatenate all of the ciphertexts.

\newcommand{\thmORCorrect}{Suppose that we are given $ k $ separate conditional encryption schemes $ \Pi_1, \ldots, \Pi_k $ corresponding to predicates $ P_1, \ldots, P_k $ and that each $\Pi_i$ is $1-\epsilon_i(\lambda)$-correct and $1-\epsilon_i'(\lambda)$-error detecting. Then the construction $\Pi_{or}$ is $1-\epsilon(\lambda)$-correct (resp.$1-\epsilon'(\lambda)$-error detecting) with  $\eps(\lambda) = \sum_{i}^k \eps_i'(\lambda) + \sum_{i}^k \eps_i(\lambda)$ (resp. $\eps'(\lambda) = \sum_{i=1}^\lambda\eps_i'(\lambda)$). }

\begin{theorem}{\thmlab{ORComp:Correct}}
\thmORCorrect
\end{theorem}

The formal proof is available in \fullversion{\appref{apdx:MissingProofs}}{the full version}. We also prove that the suggested construction provides Real-or-Random security as well \fullversion{--- see \thmref{ORComp:ROR} and its corresponding proof in \appref{apdx:RealOrRandomProofs}.}{--- see the full version}

\section{The Typo Predicate: Personalized Typo Correction} \seclab{sec:TypPredicate}

Motivated by the application of password typo correction we now introduce the predicate $P_{typo}(m_1,m_2)=P_{\Capslock}(m_1,m_2) \vee P_{\ell=2,\Ham,\Pad} \vee P_{\ell=1,\ED}$. Chaterjee et al. \cite{CCS:CWPCR17} conducted an empirical study of password typos finding that nearly $78\%$ of legitimate typos fit one of the above three categories i.e., CAPSLOCK error, Hamming Distance $\leq 2$ or a single character insertion/deletion. As application of \thmref{ORComp:Security} we obtain a conditional encryption scheme $\Pi_{typo}$ for the predicate $P_{typo}$ with $\left(t\left(\lambda\right), t_{sim},\epsilon\left(t\left(\lambda\right),\lambda\right)\right)$-security for $\epsilon\left(t\left(\lambda\right),\lambda\right) = 2^{-\lambda} + \epsilon_{AE}\left(t\left(\lambda\right),\lambda\right)$. Correctness and error detection of $\Pi_{typo}$  follow directly from \thmref{ORComp:Correct}.

\subsubsection{Application to Personalized Password Typo Correction} 
We can use our conditional encryption scheme to fix a drawback in the personalized typo correction scheme of Chaterjee et al. \cite{CCS:CWPCR17}. 

\subsubsection{The Security Issue.} Chaterjee et al. \cite{CCS:CWPCR17} proposed to derive a public/secret key pair $(pk_u,sk_u)$ for every user $u$. The public key  $pk_u$ is stored on the authentication server. Any incorrect login attempt $pw' \neq pwd_u$ for this account is encrypted $C_{pw} = \Enc_{pk}(pw)$ and stored in a typo vault. The secret key $sk_u$ is not directly stored on the server, but can be recovered whenever the user logs in with the correct password $pw_u$. In particular, we store $c_{sk_u} = \Auth.\Enc_{K_u}(sk_u)$ where the symmetric key $K_u = \mathtt{KDF}(s_u,pwd_u)$ is derived from the users password $pwd_u$ and a random salt value $s_u$ that is stored on the server in plaintext form. Thus, once the correct password $pw_u$ it is possible to recover $K_u$, then $sk_u$, decrypt all of the password in the vault and identify common typos. The drawback of this approach is that {\em every} incorrect login attempt will appear in the encrypted typo vault. It is not unlikely that the typo vault might include unrelated passwords from the user's other accounts. This could significantly increase the incentives  for a rational offline brute-force attacker to crack the user's password \cite{SP:BloHarZho18}, and simultanously increasing the potential harm to users. 

\subsubsection{The Fix} Our fix is straightforward: replace the regular encryption scheme with our conditional encryption scheme for the predicate $P_{typo}$! In addition to $pk_u$ the authentication server will also store $c_{u}= \Enc_{pk_u}(pwd_u)$ the encryption of the user's password under $pk_u$. Now whenever their is an incorrect login attempt $pw' \neq pwd_u$ we can set $c_{pw'} = \Pi_{typo}.\Cond\Enc_{pk_u}(c_u,pw',pw')$. If $P_{typo}(pwd_u,pw')=1$ then we have $\Pi_{typo}.\Dec_{sk_u}(c_{pw'})=pw'$ so that we can recover $pw'$ later when the user logs into the server with the correct password. However, if $P_{typo}(pwd_u,pw')=0$ then the ciphertext $c_{pw'}$ will be {\em entirely} useless to an offline attacker even if the attacker can recover $pwd_u$ and $sk_u$!  

\subsubsection{Security Proof} In \fullversion{\appref{apdx:TypoPrivacySecDef}}{the full version \cite{fullversion}} we formalize the notion of typo privacy \fullversion{(see \defref{def:TypoPrivacy})}{} for an authentication server that maintains a password typo vault, and we prove that the construction above provides typo privacy\fullversion{ (see \appref{CondTyoTopCompelete} for the details and security proofs.)}{.} We also showed that the TypTop system does not provide ``typo privacy''\fullversion{(see \secref{NoTypoPrivacyFortyptop})}{}. Intuitively, in the typo privacy game, the attacker gets to specify an initial password $pw_u$ for the user. The goal of the attacker is to predict a random bit $b$ selected by the challenger. The adversary may repeatedly either (1) submit a login query $pw'$ to the authentication server, or (2) submit a pair $(pw_0',pw_1')$ of guesses with $P_{typo}(pwd_u,pw_0')=0 = P_{typo}(pwd_u,pw_1')$ to the challenger who will then forward the guess $pw_b'$ to the authentication server. The adversary is allowed to observed the state $\sigma_i$ of the authentication server immediately before $(\sigma_{i-1})$ and immediately after $(\sigma_{i})$ each query $i$. Intuitively, if the conditional encryption scheme is secure then the attacker should not be able to predict the secret bit $b$ since the only update after a type (2) query is to store the new ciphertext $c_b = \Cond\Enc_{pk_u}(c_u,pw_b',pw_b')$. Since $P_{typo}(pwd_u,pw_0')=0 = P_{typo}(pwd_u,pw_1')$ both $c_0 = \Cond\Enc_{pk_u}(c_u,pw_0',pw_0')$ and $c_1 = \Cond\Enc_{pk_u}(c_u,pw_1',pw_1')$ are indistinguishable from a random ciphertext $\Sim(pk)$ generated without knowledge of $pwd_u, pw_1'$ or $pw_0'$.


\section{Implementation and Empirical analysis } \applab{App:Implementation}

In this section, we discuss our implementations of Conditional Encryption for the predicates $P_{=}, P_{\Capslock}, P_{\ell,\Ham,\Pad}$ and $P_{\ell=1,\ED}$ as well as $P_{typo} = P_{\Capslock} \vee P_{\ell =2,\Ham, \Pad} \vee P_{\ell=1,\ED}$. We also implement a modified version TypTop Personalized Typo Correction service to instantiate the Typo Vault using our conditional encryption scheme for the predicate $P_{typo}$. We empirically evaluate the performance of each implementation e.g., running time, ciphertext size etc.



\subsection{Conditional Encryption}

\subsubsection{Implementation} We implemented our conditional encryption schemes in C++. The implementation is available on Github \url{https://github.com/mhassanameri/CondEncCCS24Artifact} and Zenodo \url{https://zenodo.org/records/13744111}. Our implementation includes conditional encryption schemes for the following predicates: CAPSLOCK, Edit Distance One, Hamming Distance at most one, Hamming Distance at most two, as well as the OR of these predicates. We also implemented conditional encryption for a general Hamming Distance predicate for arbitrary distance thresholds $t = \{1, 2, 3, 4 \}$. We defined our message space to be  $\Sigma^{\leq n}$ where $\Sigma$ denotes the set of all ASCII characters and $n \in \{8,16,32,64,128\}$ --- all $x \in \Sigma^{\leq n}$ are first padded to $\Pad(x) \in \left(\Sigma \cup \{y\}\right)^n$ for a special new symbol $y$. 

We used the Pallier Library \cite{PailCPPLib} as our implementation of the Pallier cryptosystem and we used the GMP library \cite{gmpLib} for computation with big integers. We instantiated Pallier with a 1024-bit modulus  (80-bit security),  2048-bit modulus (112-bit security) and 3072-bit modulus (128-bit security)  \cite{polk2015cryptographic}. We remark that for our applications to password typo vaults 80-bit security should be sufficient as it would {\em almost certainly} be easier for an offline attacker to brute-force the user's password and then extract the Pallier secret key directly than to factor a 1024-bit modulus $N$ e.g., see \cite{SP:Bonneau12,blocki2023towards,USENIX:MUSKBCC16}.  \constref{Const:EqualityTest} and $\Pi_{1,\ED}$ (our edit-distance construction) requires that $|\Sigma|^{n+1} \leq \min\{p,q\}$. Thus, when our message length is $n=64$ characters (resp. $n=128$ characters) we must use a 2048-bit (resp. 3072-bit) modulus $N$ to ensure that $\left|\{0,1 \}^8 \right|^n = 2^{8(n+1)} < \min\{p,q\}$ since $\min\{p,q\} < \sqrt{N}$.

 Like TypTop \cite{CCS:CWPCR17}, our implementations of conditional encryption  use CryptoPP \cite{CryptoPP} for Authenticated Encryption and Shamir Secret Sharing. For authenticated encryption, we use AES-GCM with $128$-bit keys and we use Shamir Secret Sharing over a field of size $2^{128}$ to generate shares of the secret symmetric key. Our code is available on Github\footnote{\url{https://github.com/mhassanameri/CondEncCCS24Artifact}} and Zenodo\footnote{\url{https://zenodo.org/records/13744111}}. 


\subsubsection{Optimized Implementation of the Hamming Distance Predicate} \label{subsec:OptimizedHamming} We implemented several versions of our conditional encryption scheme for the Hamming Distance Predicate $P_{\ell,\Ham,\Pad}$ to optimize performance. The (unoptimized) implementation follows \constref{const:ArbHamm} without any optimizations. As noted previously the worst-case running time to decrypt a conditionally encrypted ciphertext is roughly ${n \choose \ell} (t_{SSrec} + t_{AE})$ where $t_{SSrec}$ (resp. $t_{AE}$) denotes the running time for $\recover$ over a field of size $2^{\lambda}$ (resp. $\Auth.\Dec$). 

We can make a simple optimization to speed up the running time of $\Dec$. In particular, we modify $\Cond\Enc$ to generate $n$ shares $\ldb z \rdb_1,$ $\ldots, \ldb z \rdb_n \leftarrow \Share(n,n-\ell, 0)$ of $0$ over a smaller field of size $2^{32} \ll 2^{\lambda}$ {\em in addition to} the $n$ shares $\ldb s \rdb_1,\ldots,\ldb s \rdb_n $ of our secret key $K$. For each character $i$ where $m_1[i]=m_2[i]$ the ciphertext $\tilde{c}_i$ will allow us to extract both shares $\ldb s \rdb_i$ and $\ldb z \rdb_n$. Thus, for each subset $S\subseteq [n]$ of size $|S|$ we can first compute $x_S =  \recover\left( \{ (i,\ldb z \rdb_i) \}_{i \in S} \right)$ by running $\recover$ over our smaller field. Only if $x_S = 0$ do we then proceed to compute $K_S=\recover\left( \{ (i,\ldb s \rdb_i) \}_{i \in S} \right)$ by running $\recover$ over our larger field and then attempt to decrypt our authenticated encryption ciphertext using $K_S$. We will still have to run $\recover$ over our smaller field ${n \choose n-\ell}$ times. However, in expectation we will only need to run $\recover$ over the large field (resp. $\Auth.\Dec$) {\em at most} $1+{n \choose n-\ell}2^{-32}$ times. Our empirical analysis indicates that this optimization significantly speeds up the worst-case running time of our decryption algorithm --- see \figref{fig:HDdifT}. For example, when $n=32$ and $\ell=4$ the  optimized version of $\Dec$ is more than six times faster than the unoptimized version of $\Dec$. 

Our second optimization exploits the simple observation that most user passwords are somewhat short. The goal of finding the subset $S \subseteq [n]$ of $|S| = n-\ell$ correct shares is equivalent to finding the set $C = \{i \in [n]: \Pad(m_1)[i] \neq \Pad(m_2)[i]\} \subseteq [n]$ of corrupted shares. Suppose that we know $m_1,m_2 \in \Sigma^{\leq k}$ are both shorter passwords of length at most $k < n$ and that  $\Ham(\Pad(m_1),\Pad(m_2)) \leq \ell$. In this case there would only be ${k \choose \ell} \ll {n \choose \ell}$ possible choices of $C$ to check. In the breached RockYou password dataset 99\% (resp. 99.9 \%) of passwords were shorter than 15 (resp. 30) characters. Thus, if we expect that most of the inputs are short we can optimize the decryption algorithm by iterating from $k=\ell$ to $n$, iterating over all ${k-1 \choose \ell -1}$ subsets $C' \subseteq [k-1]$ of size $\ell -1$, setting $C=C' \cup {k}$, $S = [n]\setminus C$ and then running $\recover$ with the shares in $S$. In our password typo application we will consider the Hamming Distance predicate with distance parameter $\ell=2$. Examining the password typo dataset collected by Chatterjee et al. \cite{SP:CAAJR16} we observed that in over $80\%$ of the instances where the predicate  $P_{\ell=2,\Ham}$ holds that the Hamming Distance was actually just $1$. If there is only one invalid share then we are guaranteed to find the correct secret after just $n/2$ attempts by first running  $\recover$ with the shares $S=[n]\setminus C$ for each $C \in \{ \{2i-1,2i \} : 1 \leq i \leq n/2 \}$. 

This optimization significantly improved the  conditional decryption algorithm when the padding size is larger like $n \in \set{32, 64, 128}$. As an example, if we consider Hamming distance with $\ell = 4$ and $n = 32$, we observe that  the decryption algorithm takes $14.664$ seconds for our unoptimized implementation, while the average running time (using random RockYou passwords) is reduced to just $205.69$ \textit{milliseconds} when both optimizations are applied.

\subsubsection{Evaluation} 
We evaluated the performance of our implementation of conditional encryption on a Lenovo ThinkStation S30 with a $2.9$ GHz 8-core Intel\textsuperscript{\textregistered} Xeon\textsuperscript{\textregistered} E5-26900x 16 CPU processor and $28$ GB DDR4 RAM memory. \figref{fig:FullEvalCondEnc} and \tableref{CondEncEmp} shows the running time for $ \Cond.\KG$,  $\Enc$ ,$\Cond\Enc$,  $\Cond\Dec$ (we slightly abuse notation and use $\Cond\Dec$ to refer to the decryption algorithm $\Dec$ when the input is a conditional ciphertext) as well as the ciphertext size for the aforementioned predicates. The primary difference between \tableref{CondEncEmp} and \figref{fig:FullEvalCondEnc} are as follows (1) \figref{fig:FullEvalCondEnc} plots the worst-case running time for $\Cond\Dec$ (when the relevant predicates do not hold) while the performance analysis in \tableref{CondEncEmp} is based on empirical user typos i.e., we evaluate the running time $\Cond\Dec$ by selecting random password/typo pairs from the password typo dataset of Chatterjee et al. \cite{SP:CAAJR16} subject to the constraint that the relevant predicate holds. (2) \tableref{CondEncEmp} focuses exclusively on conditional encryption schemes for messages of length at most $n=32$ i.e., the parameter that we use for TypTop.

\begin{table*}\fontsize{9}{10}\selectfont
	\caption{Conditional Encryption: Computation Time and Ciphertext Size ($n =32$, 80-bit security) }\tablelab{CondEncEmp}
	
	\makebox[\textwidth]{\begin{tabular}{|p{ 2in}|ll|ll|c|}
		\hline
		&    \multicolumn{2}{c|}{ $\Enc$}                   &         \multicolumn{2}{c|}{ $\Cond\Enc$}   &   $\Cond\Dec$  \\ \hline
  
		{\bf Predicate:}  & \multicolumn{1}{l|}{Time  (ms)} & $|c|$ &  \multicolumn{1}{l|}{Time  (ms)} & $|c|$ & Time  (ms)   \\ \hline
		
	EdDist One & \multicolumn{1}{l|}{$ 108.68 $} & $8.27$ & \multicolumn{1}{l|}{406.842} & $16.29$ & $ 104.31 $  \\ \hline
		
		 HamDist ($\ell=1, n =32$)	& \multicolumn{1}{l|}{ 85.582 } & $8.01$ & \multicolumn{1}{l|}{ 412.424 } & $8.04$ &  85.644   \\ \hline

    HamDist ($\ell=1, n =32$) OPT	& \multicolumn{1}{l|}{92.384 } & $8.01$ & \multicolumn{1}{l|}{ 445.714} & $8.04$ &  263.626   \\ \hline
   
   HamDist ($\ell=2, n =32$) 	& \multicolumn{1}{l|}{93.88} & $8.01$ & \multicolumn{1}{l|}{445.8} & $8.04$ & 347.953  \\ \hline

   HamDist ($\ell=2, n =32$) OPT 	& \multicolumn{1}{l|}{98.0633} & $8.01$ & \multicolumn{1}{l|}{475.58} & $8.04$ & 264.273  \\ \hline

   HamDist ($\ell=3, n =32$) 	& \multicolumn{1}{l|}{90.1867} & $8.01$ & \multicolumn{1}{l|}{433.63} & $8.04$ & 2268.54  \\ \hline
   
   HamDist ($\ell=3, n =32$) OPT	& \multicolumn{1}{l|}{105.98} & $8.01$ & \multicolumn{1}{l|}{498.75} & $8.04$ & 254.61  \\ \hline

   HamDist ($\ell=4, n =32$)	& \multicolumn{1}{l|}{97.52} & $8.01$ & \multicolumn{1}{l|}{461.79} & $8.04$ & 14664.8  \\ \hline

   HamDist ($\ell=4, n =32$) OPT	& \multicolumn{1}{l|}{98.77} & $8.01$ & \multicolumn{1}{l|}{466.457} & $8.04$ & 205.69  \\ \hline

   CAPSLOCK on	& \multicolumn{1}{l|}{3.0025} & $0.27$ & \multicolumn{1}{l|}{13.26} & $0.29$ & $ 1.01 $  \\ \hline
		
   OR*& \multicolumn{1}{l|}{$ 201.15 $} & $16.54$ & \multicolumn{1}{l|}{$ 900.945 $} & $24.64$ & $ 360 $  \\ \hline
		
 \multicolumn{6}{p{4in}}{ $|c|=$ Ciphertext size (KB)} \\
		
  \multicolumn{6}{p{4in}}{* OR = EditDistOne or HamDistTwo or CAPSLOCKon}   \\
	
    \multicolumn{6}{p{4in}}{** $ P_i $ is the predicate and we define our CondCrypto over this predicate for $i = \{1, 2, 3, 4\}$, which implies 4 different predicates. }   \\

    \multicolumn{6}{p{5in}}{*** For hamming distance (HamDist), $\ell$ represents the threshold value and $n= 32$ is the padding size. Also, OPT means using optimized decryption algorithm.} 
	\end{tabular}}
\end{table*}

 For the Hamming distance, we consider four different thresholds at most one, at most two, at most three and at most four. In  \figref{fig:HDdifT} and \figref{fig:HDdifL} we focus on the worst case running time for $\Cond\Dec$ when the predicate does not hold and we have to iterate over all $n \choose \ell$ possible subsets for secret recovery. \figref{fig:HDdifT} plots the running time of $ \Cond\Dec$ as the input length $n$ varies for different Hamming Distance thresholds $\ell\in \{1,2,3,4\}$. \figref{fig:HDdifL} plots how the running time of $ \Cond\Dec$ is impacted by the Hamming Distance threshold $\ell$. The figure includes separate plots for messages of length $n \in \{8,16,32,64,128\}$. In figures \figref{fig:HDdifT} and \figref{fig:HDdifL} the blue (resp. red) curves highlight the running time of our optimized (resp. non-optimized) implementation. \figref{fig:HDEnc} plots the running time of the encryption and conditional encryption algorithms $\Enc$ and $\Cond\Enc$ and \figref{fig:hCTXsize} plots the size of a regular and conditional ciphertext for the Hamming Distance predicate as the message length varies. As expected we note that the ciphertext size is independent of the threshold $\ell$ and that the size of a conditional ciphertext is approximately equal to the size of a regular ciphertext.    
 

We did similar for the CAPSLOCK predicate and considered the evaluation time over different message lengths $ n =\{ 8,16, 32, 64, 128 \}$ and the average time of each algorithm is presented in \figref{fig:CAPSLOCK}. The running time for each algorithm $ \Enc $, $ \Cond\Enc $ and $ \Dec $ is independent of the message length until we have to increase the size of our Pallier Public key to satisfy the requirement that $|\Sigma|^{n+1} \leq \min\{p,q\}$. This explains the jumps at input length $64$ and $128$.  

\figref{EDOne} plots the running time of $\Enc$, $\Cond\Enc$ and $\Cond\Dec$ for our edit distance one predicate under different padding lengths $ n = \{8, 16, 32, 64, 128 \} $. Similarly, \figref{fig:OR} plots the running time of  of $\Enc$, $\Cond\Enc$ and $\Cond\Dec$ for the OR predicate $P_{typo}$. For $\Cond\Dec$ we report the worst-case running time to decrypt a conditionally encrypted ciphertext i.e., when the predicate does not hold. When eveluating decryption time for the OR predicate $P_{typo}$ we use our optimized implementation of  conditional decryption for the hamming distance predicate.

 \figref{fig:iCTXsize} plots the size of a regular and conditional ciphertexts as the message length increases for each predicate: CAPSLOCK (CAPS), Edit Distance One (ED), Hamming Distance Two (HD) and the OR of the above. Some plots are difficult to see because they are identical to other lines. For example, we first note that the size of a regular ciphertext for the OR predicate is identical to the size of a ED ciphertext. Similarly, the size of a conditional ciphertext is approximately equal for the Hamming Distance (HD) and Edit Distance (ED) predicates. The plots at the bottom of \figref{fig:iCTXsize} are for CAPS as a regular/conditional encryption for this predicate consists of a single Pallier ciphertext.

\subsubsection{Discussion}
Our empirical analysis demonstrates the practicality of our constructions especially for password typos. For example, when $n=32$ the worst-case time to decrypt a conditional ciphertext for the password typo predicate $P_{typo}$ (OR) is less than 250 (ms). While the overhead is higher than traditional encryption schemes, it is important to note that, for our TypTop application, the algorithms $\Cond\Enc$ and $\Dec$ can be evaluated offline and will not delay user authentication. 

 \input{CondEncPerfEval}

\subsection{TypTop with Typo Privacy}
We also implemented a modified version of TypTop system for personalized typo correction \cite{CCS:CWPCR17} as outlined in \secref{sec:TypPredicate}. We  consider two primary modifications to the regular TypTop system. First, we replace the Key Derivation Function (KDF) with Argon2id \cite{biryukov2016argon2} a Memory-Hard Key Derivation Function. This modification was already suggested by the designers of TypTop. Second, we replace the public key encryption scheme with a conditional encryption scheme for the OR predicate. For the purpose of empirical evaluation we implement TypTop with our optimized implementation of conditional encryption and with our unoptimized implementation refering to these as modifications (2A) and (2B) respectively.  

Our code is available at our open source github\footnote{\url{https://github.com/mhassanameri/CondEncCCS24Artifact}} repository. In our analysis, we analyze six versions of the TypTop system: the original system (no modifications), modification (1) only, modification (2A) only, modification (2B) only, modifications (1)+(2A) and finally modifications (1)+(2B). The recommended version is TypTop with modifications (1)+(2A) i.e., using the Memory-Hard Key Derivation Function and the optimized implementation of our conditional encryption scheme for the OR predicate.

In our modified implementation of TypoTop, we assume that the length of each user passwords is at most 32 characters. This is a valid assumption as 99.9 \% of the passwords has length of lower than 32 characters. To support this assumption, as an evidence we extracted this stats from the leaked passwords of LinkedIn and RockYou. More specifically, 99\% of passwords from LinkedIn Frequency Corpus\footnote{Link to the data set: \url{https://figshare.com/articles/dataset/linkedin_files_zip/7350287}} as well as passwords from RockYou\footnote{Link to the RockYou with count dataset: \url{https://github.com/danielmiessler/SecLists/tree/master/Passwords/Leaked-Databases}} have length at most 15. If desired one could easily adjust the TypTop system to support longer passwords. However, the scheme would either become less efficient or we would need to leak a single bit of information about the user password i.e., indicating whether or not the length of the password is more than 32 characters.  

In our empirical analysis we register a user password of length $\leq 32$ and then generate a sequence of 1000 correct and 1000 incorrect login requests. Incorrect login requests are randomly selected as (1) CAPSLOCK error, (2) Hamming Distance $\leq 2$, (3) Edit Distance $\leq 1$ or (4) completely different. We analyze the running time for Initialization, Correct Login attempt and for Incorrect login attempts --- see \tableref{CondTypComp}. In our analysis we distinguish between authentication delay and total running time. For example, if authentication fails with an incorrect password $pw_u'$ then we will want to run our conditional encryption algorithm $\Cond\Enc_{pk_u}(c_u, pw_u', pw_u')$ so that, if $pw_u'$ is close enough to the real password, we can recover it at a later point in time. However, we can immediately inform the user of the outcome of authentication before performing this somewhat expensive computation. Similarly, if a user logs in with the correct password $pw_u$ then we can recover the symmetric key $K_u$ and decrypt $sk_u$. At this point we will want to decrypt all of the conditional ciphertexts in our vault, but again we can immediately inform the user that the authentication attempt was successful before decrypting these conditional ciphertexts.

\begin{table*} \fontsize{8.5}{7}\selectfont
	\caption{TypTop: Computation and Storage Overhead with/without Conditional Encryption. Here we considered 80-bit level of security when $n =32$.}\tablelab{CondTypComp}
	\makebox[\textwidth]{
	\begin{tabular}{|p{1.3 in}|p{.5 in}|p{.5 in}p{.5 in}|p{.5 in}p{.5 in}|p{.5 in}|}
		\hline
		&  &    \multicolumn{2}{c|}{ Correct login}                   &         \multicolumn{2}{c|}{ Incorrect login}   & \\ \hline
		& Init& \multicolumn{1}{l|}{Auth Delay } & Total Running Time & \multicolumn{1}{l|}{Auth Delay  } & Total Running Time & Storage (KB)  \\ \hline
		
	Typtop \cite{CCS:CWPCR17} 	& 171.95 (ms) & \multicolumn{1}{l|}{26.41 (ms)} & 53.384 (ms) & \multicolumn{1}{l|}{156.556 (ms) } & 158.71 (ms)  & 1  \\ \hline
	
	CondTyptop & 6.771 (s) & \multicolumn{1}{l|}{25.7 (ms)} & 11.203 (s) &  \multicolumn{1}{l|}{160.3 (ms)} & 0.617 (s)  & 246  \\ \hline
	
	CondTyptop (Optimized)	&  7.13 (s)& \multicolumn{1}{l|}{24.12 (ms)} & 8.690 (s) & \multicolumn{1}{l|}{160.33  (ms)} & 0.629 (s) & 246 \\ \hline
	
	Typtop (mhf) \cite{CCS:CWPCR17}	& 5.738 (s) & \multicolumn{1}{l|}{0.943 (s)} & 0.784 (s)  & \multicolumn{1}{l|}{5.644 (s)} & 5.856(s)  & 1\\ \hline
	
	CondTyptop(mhf)	& 19.672 (s) & \multicolumn{1}{l|}{0.933 (s)} &  16.558 (s)& \multicolumn{1}{l|}{5.446 (s)} & 6.162 (s) & 246  \\ \hline
	
	CondTyptop(mhf/opt)	& 14.456 & \multicolumn{1}{l|}{0.729 (s)} & 10.995 & \multicolumn{1}{l|}{4.376(s)} & 5.87 (s) & 246 \\ \hline
	
	\multicolumn{7}{l}{* CondTyptop is the modified typtop scheme when conditional encryption is used.}   \\
	\end{tabular}}
\end{table*}

\subsubsection{Discussion} We can use conditional encryption to strengthen the security guarantees of TypTop without increasing authentication delay for users. The usage of conditional encryption does increase offline computation and storage requirements, but the overhead is still manageable. The reason why the offline computation is higher after a correct login attempt is because this allows us to recover the secret decryption key and then decrypt all of the conditional ciphertexts in our waitlist so that we can consider adding them to our cache of acceptable typos. It is worth noting that the TypTop system maintains the invariant that there are always $10$ (conditional) ciphertexts in the waitlist. The invariant, which is maintained by seeding the waitlist with dummy ciphertexts, ensures that an attacker cannot infer the number of incorrect login attempts. If we don't maintain this invariant then an attacker who breaches the authentication server would learn the total number of incorrect login attempts that were submitted since the last correct login. Our modified implementation of TypTop maintains the same invariant. However, such leakage should arguably not be viewed as problematic since it does not reveal anything about incorrect login attempts or the user's password. In this case we could reduce offline computation (and storage) of TypTop by only placing conditional ciphertexts in the waitlist when there is an incorrect login attempt i.e., if there were no incorrect login attempts since the last correct login attempt then there would be no offline work to decrypt the conditional ciphertexts in the waitlist because the waitlist would be empty!    

\subsection*{Acknowledgements}
This work was supported in part by the National Science Foundation under CAREER Award CNS 2047272. Any views expressed in this paper are those of the authors and do not necessarily represent the position of the National Science foundation. The authors wish to thank anonymous CCS reviewers for constructive feedback and for the suggestion to consider circuit private FHE.


\bibliographystyle{alpha}
\bibliography{extra,cryptobib/abbrev0,cryptobib/crypto}

\appendix
\fullversion{
\section{Details on Paillier Cryptosystem}\applab{App:PailDetails}
In this part we will provided a formal definition of Paillier cryptosystem with more details.  We should highlight that the Paillier cryptosystem \cite{EC:Paillier99} consists of three main algorithms $\Pi_{\Pail}= (\Pail.\KG,$ $ \Pail.\Enc, \Pail.\Dec) $, and two main operations $\Pail.\Add, \Pail.\PtoCMul$ which are described as follows.

\begin{itemize}
	\item $ (pk, sk)\gets \Pail.\KG\left(1^\lambda;\left(p, q\right)\right) $: This algorithm takes as input the security parameter $ \lambda $ and two uniformly at random sampled large $ \mathsf{poly}(\lambda) $\footnote{We should highlight that based on the security parameter we desire, there exists a polynomial function like $ \mathsf{poly} $ which determined the bit length of the prime numbers.} bit prime numbers $ p, q \leq 2^{\mathsf{poly}(\lambda)}  $, and generates the secret-public key pair $ (pk, sk) $ as follows. Then the algorithm sets $ N =pq $ and computes $ \beta = \lcm(p-1, q-1) $. Let we define the function $ L(u) = \frac{u-1}{N} $ in which $ u \in [N^2] $ is a variable. So using the defined function to compute $ \mu = (L(g^\beta \mod N^2))^{-1}  \mod N$ in which $ g\in \mathbb{Z}^*_{N^2} $ is sampled uniformly at random. Finally, the public key and its corresponding secret key is set as follows: $ pk= (N, g) $  and $ sk = (\beta, \mu) $. In this paper we considered $g = N+1$. In this case, $\beta = \lcm{\left( p-1, q-1 \right)}$ and $\mu = \phi (N)^{-1} \mod N$. 
	
	\item $ c\gets \Pail.\Enc_{pk}(m; r) $: The encryption algorithm takes as input the message $ 0\leq m <N $ and random coin $ r\in_R\Mbb{Z}^*_N $, and computes the ciphertext as follows: $ c:= (1+N)^m r^N \mod (N^2) $. 
	
	\item $ m := \Pail.\Dec_{sk} (c)$: The decryption algorithm takes as input the ciphertext $ c\in \Mbb{Z}^*_{N^2} $ and decrypts it as follows: $ m := L(c^{\beta} \mod N^2) . \mu  \mod N $.
	
	\item $ c = \Pail.\Add(c_1, c_2) $: Given $ c_1 = \Pail.\Enc_{pk}(m_1) $ and $ c_2= \Pail.\Enc_{pk}(m_2)  $ for two messages $ m_1, m_2 $, this algorithm computes a ciphertext of $ m_1 + m_2 \mod N $ under the same public keys as follows. $ c = c_1 . c_2 \mod N^2 $. Intuitively we have: 
	\begin{align}
	 c  &= (g^{m_1} r_1^N) . (g^{m_2}r_2^N)  \mod N^2 \nonumber\\
	    &= g^{m_1 + m_2} (r_1r_2)^N \mod N^2 \nonumber\\
	   &=  g^{m} r^{N} \mod N^2
	\end{align}	in which $  m = m_1+m_2 $ and the resulting randomness is $ r = r_1r_2 $. 
	
	For sake of simplicity, we use symbol $ \boxplus $  for this algorithm and we have $ c_1\boxplus c_2 = \Pail.\Add(c_1,c_2) $. Similarly, for the subtraction, we can define $ \boxminus $ and we have $ c_1 \boxminus c_2  = \Pail.\Add(c_1, c_1^{-1}) $. We can also define $ \uplus $ which add $ k $ ciphertexts $ c_1, \ldots, c_k $ and we have $ \uplus_{i = 1}^{k} c_i = c_1 \boxplus \ldots \boxplus c_k  $. 
	
	\item $ c = \Pail.\PtoCMul(m_1,  c_2) $: This algorithm takes as input the plaintext message $ m_1 $ and $ c_2 = \Pail.\Enc_{pk}(m_2) $, and computes ciphertext of $ m_1.m_2 $ as follows: 
	 \begin{align}
	 c  &=  (c_2)^{m_1}  \mod N^2 
	  = (g^{m_2}r_2^N)^{m_1}  
	 = g^{m_1m_2} (r_2^{m_1})^N \mod N^2  \nonumber\\
	 & =  g^{m} r^{N} \mod N^2
	 \end{align} in which $ m = m_1m_2 $ and the resulting randomness is $ r =r_2^{m_1} \mod N^2 $.
	 
	 For sake of simplicity, we use symbol $ \boxtimes $  for this algorithm and we have $ m_1\boxtimes c_2 = \Pail.\PtoCMul(m_1,c_2) $.  
\end{itemize}

\section{Details on Secret Sharing (SS)}\applab{Sec:SecretSharing}
Let $ \Mbb{F} $ be a field of size $ |\Mbb{F}|\geq n $. We define interpolation algorithm $ \InterPol $ over $ t $-degree polynomial $ f: \Mbb{F} \to \Mbb{F} $ which takes as input $ t $ tuples $ (x_i, y_i =f(i)) $ for $ 1\leq i \leq t $, and outputs $ f(0) $. So we have $f(0) = \InterPol((x_1, f(x_1)), \ldots, (x_t, f(x_t)) ) $. The $ (n, t) $- secret sharing scheme for the secret message $ s \in \Mbb{F}$ is defined based on two algorithms $ \Share$, $\recover $ which are described in what follows. 

\begin{itemize}
	\item $ (\ldb s \rdb_1, \ldots, \ldb s \rdb_n)  \gets \Share(n, t, s)$. This algorithm takes as input the secret $ s\in \Mbb{F} $, the threshold value $ t $ and the number of shares $ n $. Then, it randomly samples the $ t $-degree polynomial $ \phi: \Mbb{F}\to \Mbb{F} $ such that $ s = \phi(0) $ and sets the shares as $ \ldb s \rdb_i  = \phi(i)$, for all $ 1\leq i \leq n $.

	\item $ s\gets \recover(\shrs = (\ldb s \rdb_1, \ldots, \ldb s \rdb_{t'}) ) $. This algorithm takes the set of share $\mathsf{shares} $ of size $ t' $ as input, and recover the message $ s $ if there exists set of valid shares $ \ValidShr \subset \shrs  $  s.t.  $ |\ValidShr | \geq t $. To recover $ s $ the algorithm runs the interpolation algorithm $ \hat{s} =  \InterPol ((1,$ $\ldb s \rdb'_1), \ldots, (t, \ldb s \rdb'_{t}) ) $ such that  $ \ldb s \rdb'_i, \in \ValidShr, \forall 1\leq i\leq t$. Finally the algorithm outputs $ s = \ToOrig (\hat{s})$.  
 
\end{itemize}

\begin{definition}[Correctness]
	Given the $ (t,n) $-secret sharing scheme $ \Pi = (\Share, \recover)  $, $\forall n\in\Mbb{N} $,  $ \forall s\in\M^n$ and $ \forall S=\{i_1, \ldots, i_t\} \subseteq \{1, \ldots, n\} $ of size $ t $, the correctness of $ \Pi $ enforces that  
	\begin{align}
	\Pr_{ (\ldb s \rdb_1, \ldots, \ldb s \rdb_n)\gets \Share(1^\lambda, n, t, s) } [\recover(\ldb s\rdb_{i_1}, \ldots, \ldb s\rdb_{i_t})  = s] = 1
	\end{align}
	
\end{definition}

\begin{definition}[SS Perfect Secrecy]\deflab{SSPerfectSec}
	The $ (t,n) $-secret sharing scheme $ \Pi = (\Share, \recover) $ is perfectly secrecy, if $ \forall n\in \Mbb{N} $, $ \forall s,s'\in\M^n $, $ \forall S\subseteq \{1, \ldots, n\} $ s.t. $ |S|<t $, and for all adversaries $ \A$, 
	\begin{align}
		&\Pr_{ (\ldb s \rdb_1, \ldots, \ldb s \rdb_n)\gets \Share(1^\lambda, n, t, s) } [\A(\ldb s\rdb_i|i\in S) =1 ]\nonumber\\
		 &= \Pr_{\A (\ldb s' \rdb_1, \ldots, \ldb s' \rdb_n)\gets \Share(1^\lambda, n, t, s') } [\A (\ldb s'\rdb_i|i\in S) = 1 ]
	\end{align}

\end{definition}

We will use the Shamir Secret sharing scheme \cite{CACM:Shamir79}. Given any field $\mathbb{F}$ of size $|\mathbb{F}| \geq n$ the construction starts with $n+1$ distinct field elements $x_0, x_1,\ldots, x_n \in \mathbb{F}$. Given a secret $s \in \mathbb{F}$ we generate shares $\ldb s \rdb_1 = (x_1,y_1),\ldots,$ $ \ldb s \rdb_{t-1}=(x_{t-1},y_{t-1})$ where $y_1,\ldots, y_{t-1} \in \mathbb{F}$ are random field elements. We then use polynomial interpolation to find a degree $t-1$ polynomial $p(x)$ such that $p(x_0)=s$ and $p(x_i)=y_i$ for $i \leq t-1$. Finally, for share $i > t-1$ we define $\ldb s \rdb_i = \left(x_i, p\left(x_i\right)\right)$. 

 One crucial property of Shamir Secret sharing that we rely on is that any subset of $t-1$ shares is uniformly random over $\mathbb{F}^{t-1}$. In particular, for all secrets $s \in \mathbb{F}$,  all subsets $S = \{i_1,\ldots, i_{t-1}\} \subseteq [n]$ of size $t-1$ the shares  $\ldb s \rdb_{i_1}, \ldots,  \ldb s \rdb_{i_{t-1}}$ can be viewed as uniformly random independent elements in $\mathbb{F}$ unrelated to the secret $s$ i.e., for all  $y \in \mathbb{F}^{t-1}$ we have \[\Pr\left[ \left( \ldb s \rdb_{i_1}, \ldots, \ldb s \rdb_{i_{t-1}}\right)=y \right] = \left| \mathbb{F}\right|^{-t+1} \ \] where the randomness is taken over $\left( \ldb s \rdb_1, \ldots,  \ldb s \rdb_n\right) \gets \mathtt{ShareGen}(n,t,s)$.

\section{Real or Random Security}
\applab{apdx:RealOrRandomProofs}
Similar to the traditional public key encryption schemes, we require that the encryption scheme is secure against any computationally bounded adversary who does not have the secret key $sk$. In this section, we formally define Real-or-Random security based on the a security exeperiment/game called $\mathtt{CE}^{ROR}_{\A}(1^\lambda)$.

\subsection{RoR Experiment for Conditional Encryption}
In the security game we pick a random $(sk,pk)= \KG(1^{\lambda}; r)$ and the attacker tries to distinguish the encryption oracles $\Enc_{pk}(\cdot)$ and $\Cond\Enc_{pk}(\cdot, \cdot)$ from random encryption oracles. More precisely, we consider the following experiment $\CE^{ROR}_{\A}(1^\lambda)$: (1) The challenger $\C$ picks a random bit $b \in \{0,1\}$ and generates a random public/secret key pair $(sk,pk) \gets \KG(1^{\lambda})$; (2) The challenger sends $pk$ to the attacker $\A(1^\lambda)$; (3) Whenever the attacker submits a message $m$ to the encryption oracle the challenger sets $m_0=m$ and picks a random message $m_1$ and sends back $\Enc_{pk}(m_b)$. (4) Whenever the attacker $\A$ submits a pair $(c,m, m')$ to the conditional encryption oracle the challenger sets $c_0=\Cond\Enc_{pk}(c,m,m')$ and sets $c_1 = \Cond\Enc_{pk}(c,r,r')$ for a random messages $r,r'$ and sends back $c_b$. (5) The game ends when $\A$ outputs a bit $b'$ and the output of the experiment is $\mathtt{CE}^{ROR}_{\A}(1^\lambda)=1$ if and only if $b=b'$. 

\begin{definition} (Real or Random Security) \deflab{CondRoR} We say that a conditional encryption scheme $ \Pi = (\KG, \Enc, \Cond\Enc, \Dec)$ is $\left(t\left(\cdot\right), q\left(\cdot\right), \epsilon\left(\cdot\right)\right)$-secure if for all attackers $\A$ running in time $t(\lambda)$ and making at most $q(\lambda)$ oracle queries in total we have 
$\Pr\left[\mathtt{CE}^{ROR}_{\A}\left(1^\lambda\right)=1\right] \leq \dfrac{1}{2} + \epsilon\left(t\left(\lambda\right), q\left(\lambda\right),\lambda\right)$. 
\end{definition}

\newcommand{\thmROREquality}{Assume that Pallier encryption is $(t(\lambda),q(\lambda),\epsilon(t(\lambda),$ $q(\lambda),\lambda))$-real or random secure. Then \constref{Const:EqualityTest} is $ (t'(\lambda), q'(\lambda), $ $\epsilon'(t'(\lambda),q'(\lambda),\lambda)) $-real or random security with  $t'(\lambda) = t(\lambda)-q'(\lambda)$ $ \poly(\lambda)$, $q'(\lambda)=q(\lambda)$ and $\epsilon'\left(t'\left(\lambda\right),q'\left(\lambda\right),\lambda\right) \leq 3\epsilon\left(t\left(\lambda\right),q\left(\lambda\right), \lambda\right)$. }
\begin{theorem}\thmlab{thm:ROREquality}
\thmROREquality 
\end{theorem}

\begin{proofof} {\thmref{thm:ROREquality}} (Sketch) 
Note that we can assume WLOG that the query to the  $\Cond\Enc$ oracle is of the form $((0,c_1),m_2,m_3)$ where $c_1 \in \mathbb{Z}_{N^2}^*$. Otherwise, if $\mathtt{gcd}(c_1,N) \neq 1$ or if the query has the form $((1,c_1),m_2,m_3))$ then the response will simply be $\bot$ regardless of the secret bit $b$. Let $c_1 \in \mathbb{Z}_{N^2}$ be given (We do {\em not} assume that $c_1 \in \mathbb{Z}_{N^2}^*$) and consider the query $((0,c_1),m_2,m_3)$ to $\Cond\Enc$.

We note that if $b=0$ the query returns $(1,c)=\Cond\Enc_{pk}(c_1,m_2,m_3)$ a random Pallier ciphertext of the message $R(m_1-m_2) + m_3 \mod{N}$ if $b=1$ the query returns $(1,c)=\Cond\Enc_{pk}(c_1,r_2,r_3)$ a random Pallier ciphtext for the the different message $R(m_1-r_2) + r_3 \mod{N}$. We define four hybrids: H0.A, H0.B, H1.A and H1.B. In Hybrid H0.A (resp. H1.A) we play the ROR security game with bit $b=0$ (resp. $b=1$). In Hybrid H0.B (resp. H1.B) we continue to use the bit $b=0$ (resp. $b=1$) for the encryption oracle $\Enc$, but we replace the conditional encryption oracle $\Cond\Enc$ oracle with an oracle that simply returns a random pallier ciphertext i.e., on input $(c,m_2,m_3)$ the oracle simply outputs  $(1,(1+N)^{r_1} r_2^N \mod{N^2})$  for $r_1 \in \mathbb{Z}_N$ and $r_2 \in \mathbb{Z}_{N}^*$. Intuitively, by ROR security for Pallier the 
attacker cannot distinguish between Hybrid H0.B and H1.B except with advantage $\epsilon(t,q,\lambda)$ i.e., is not the regular ROR security game $b=0$ or $b=1$ adding a useless extra oracle which returns random Paillier ciphertexts --- this oracle could easily be simulated. By similar reasoning the attacker cannot distinguish between hybrid H0.A and H0.B (or H0.A and H0.B) except with advantage $\epsilon(t,q,\lambda)$. Intuitively, if the attacker $\mathcal{A}$ distinguishes hybrids H0.A and H0.B  we can construct an attacker $\mathcal{B}$ for the Paillier ROR security game i.e., $\mathcal{B}$ simply runs $\mathcal{A}$ and anytime $\mathcal{A}$ submits a query $(c_1,m_2,m_3)$ we submit the query $m_3$ to the Paillier Encryption oracle to obtain $c_3$ and then return $c=c_3 c_1^R (1+N)^{-m_2R} \mod{N^2}$. If $c_3$ was a random Paillier encryption of $m_3$ then $c$ is distributed exactly like $\Cond\Enc(c_1,m_2,m_3)$ in H0.A (resp. H1.A). On the other hand if $c_3$ is a random Paillier ciphertext of a random message then $c$ is uniformly random ciphertext as in H0.B (resp. H1.B). The running time for the attacker $\B$ is at most $t(\lambda) = t'(\lambda) + q(\lambda) \poly(\lambda)$ where the term  $q(\lambda) \poly(\lambda)$. Thus, attacker $\B$ distinguishes the hybrids H0.B (resp. H1.B) with probability at most $\epsilon(t(\lambda),q(\lambda), \lambda)$. Thus, any attacker running in time $t'(\lambda)$ distinguishes H0.A ($b=0$) from the final hybrid H1.A $(b=1$) with advantage at most $3 \epsilon(t(\lambda),q(\lambda), \lambda)$. 

\end{proofof}

\newcommand{\thmRORHamming}{Assume that Pallier encryption is $(t(\lambda),q(\lambda),\epsilon(t(\lambda),$ $q(\lambda),\lambda))$-real or random secure and that $\Pi_{AE}$ is $(t_{AE}(\lambda),q_{AE}(\lambda),$ $\epsilon_{AE}(t_{AE}(\lambda),q_{AE}(\lambda),\lambda))$ real or random secure. Then \constref{const:ArbHamm} is $ \left(t'\left(\lambda\right), q\left(\lambda\right)', \epsilon'\left(t'\left(\lambda\right),q\left(\lambda\right),\lambda\right)\right) $-real or random secure with  $t'(\lambda) = \min\left\{t\left(\lambda\right), t_{AE}\left(\lambda\right)\right\}-o(n)$, $q'(\lambda)= \min \left\{q\left(\lambda\right)/(4n), \frac{q_{AE}(\lambda)}{2}\right\}$ and $\epsilon'(t'$ $(\lambda),q'(\lambda),\lambda)=\epsilon(t(\lambda),q(\lambda),\lambda) + \epsilon_{AE}(t_{AE}(\lambda),q_{AE}(\lambda)',\lambda)$ 
}

\begin{theorem}\thmlab{thm:RORHamming}
\thmRORHamming
\end{theorem}

\begin{proofof}{\thmref{thm:RORHamming}}(Sketch) Let $c_1 \in \mathbb{Z}_{N^2}$ (we do not assume $c_1 \in \mathbb{Z}_{N^2}^*$) and consider the query $(c_1,m_2,m_3)$ to the encryption oracle. If $b=1$ then we have $(1,\tilde{c}_1,\ldots, \tilde{c}_n, c_{AE}) = \Cond\Enc_{pk}(c_1,m_2,m_3)$ 
where each $\tilde{c}_i = c^{R_i} \left(g^{-m_2\left[i\right]} r_{2,i}^N\right)^{R_i} g^{m_3[i]} r_{3,i}^N \mod{N^2}$ for uniformly random $R_i \in \mathbb{Z}_N$ and $r_{2,i}, r_{3,i} \in \mathbb{Z}_N^*$. By ROR security for Pallier we can apply a Hybrid argument can swap $g^{-m_2[i]} r_{2,i}^N$ with  $g^{r_{1,i}} r_{2,i}^N$ with $r_{1,i} \in \mathbb{Z}_N$ (resp. $g^{r_{4,i}} r_{2,i}^N$ ) with $r_{1,i} \in \mathbb{Z}_N$ (resp.  $r_{4,i} \in \mathbb{Z}_N$)  uniformly random. After $2n$ hybrids we have replaced each $\tilde{c}_i$ with a ciphertext of the form $c^{R_i} \left(g^{r_{i,1}}r_{2,i}^N\right)^R_i g^{r_{4,i}}r_{3,i}^N \mod{N^2}$. In particular, all information about the secret symmetric key $K$ is removed so we can invoke ROR security for Authenticated encryption to replace $c_{AE}$ with a random ciphertext. We can then reverse the hybrids to move to transition from $\Cond\Enc_{pk}(c_1,m_2,m_3)$ (when  $b=0$) all the way to the $b=1$ case $\Cond\Enc_{pk}(c_1,m_2',m_3')$ for random messages $m_2'$ and $m_3'$. Thus, we adjust $q'(\lambda) = q(\lambda)/(4n)$ since we invoke ROR security for Pallier during $4n \times q'(n)$ hybrids. Similarly, we invoke ROR security for authnticated encryption in $2 \times q(\lambda)'$ hybrids.
\end{proofof}

\newcommand{\thmORRealORRAN}{Suppose that we are given $ k $ separate conditional encryption schemes $ \Pi_1, \ldots, \Pi_k $ corresponding to predicates $ P_1, \ldots, P_k $ and that each $\Pi_i$ provides  $(t_i(\lambda), q_i(\lambda), \eps_i(t_i(\lambda), q_i(\lambda), \lambda)) $ real or random security. Then the construction $\Pi_{or}$ provides $(t(\lambda), q(\lambda), \eps(t(\lambda), q(\lambda), \lambda))$ real or random security with  $\eps(\lambda) = \sum_{i=1}^k \eps_i(t_i(\lambda), q_i(\lambda), \lambda))$, $q(\lambda) = \min \{\frac{q_1(\lambda)}{2k}, \cdots, \frac{q_k(\lambda)}{2k} \}$ and $t(\lambda) = \min{\{t_1(\lambda), \cdots, t_k(\lambda)\}} -  O(k \cdot \max(t_{\Cond\Enc_1}, $ $\cdots, t_{\Cond\Enc_k}))$. }

\begin{theorem}{\thmlab{ORComp:ROR}}
\thmORRealORRAN
\end{theorem}

\begin{proofof}{\thmref{ORComp:ROR}} (Sketch) Intuitively, for each conditional encryption query $\left(c_1 = \left(c_{1, \Pi_1}, \cdots, c_{k, \Pi_k}\right), m_2, m_3\right)$, we need to query $\tilde{c}_i = \Pi_i.\Cond\Enc(c_{1, \Pi_i}, m_2, m_3)$ if $b=1$. So we have $ k$ hybrids and in each hybrid we use the real or random security of one predicate. So the adversary's  advantage is bounded by $\sum_{i=1}^k \eps_i\left(t_i\left(\lambda\right), q_i\left(\lambda\right), \lambda\right)$. For each query, the simulator need to compute $\tilde{c}_i$ which takes $t_{\Cond\Enc_i}$. So considering all $k$ hybrids, the loss in time is bounded by $O(k $ $\cdot \max\{t_{\Cond\Enc_1},$ $ \cdots, t_{\Cond\Enc_k}\})$ after these $k$ hybrids. We note that the number of queries for the resulting adversary is bounded by $\min \{q_1(\lambda), $ $ \cdots, q_{k}(\lambda)\}$. We can then reverse the hybrids to move to transition from $\Pi_{or}.\Cond\Enc$ $(c_1,m_2, m_3)$ (when $b=1$) all the way to $b=0$ case $\Pi_{or}.\Cond\Enc(c_1,m'_2, m'_3)$ for random messages $m'_2$ and $m'_3$. 
\end{proofof}


\section{General construction of Conditional Encryption from Circuit-Private FHE}
\applab{CondEncFromFHE}
In this section we show how one can construct conditional encryption for an arbitrary binary predicate $P$ using Circuit-Private FHE assuming that the predicate $P$ can be implemented as a polynomial sized circuit. The basic idea is to define a circuit $C_{P,m_2,m_3}(m_1)$ which outputs $m_3$ if $P(m_1,m_2)=1$ and outputs $0$ if $P(m_1,m_2)=0$. Now the the conditional encryption algorithm $\Cond\Enc_{pk}(c,m_2,m_3)$ simply runs $\FHE.\Eval_{pk}(C_{P,m_2,m_3}, c)$. If the input ciphertext $c$ corresponds to a message $m$ such that $P(m,m_2)=1$ then $\FHE.\Eval_{pk}(C_{P,m_2,m_3}, c)$ will output a ciphertext $c'$ which decrypt to $C_P(m,m_2,m_3)=m_3$. Otherwise, if $P(m,m_2)=0$ we get a ciphertext $c'$ which decrypts to $0$. This construction is formalized in \constref{Const:GenericFHE}.

We show that \constref{Const:GenericFHE} provides conditional encryption secrecy as long as the FHE construction satisfies the notion of {\em circuit privacy} \cite{C:OstPasPas14} --- see \appref{appendix:circuitprivacydef} for the formal definition. Before defining {\em circuit privacy} for FHE we first show that circuit privacy is necessary to prove security of our construction. In particular, if FHE exists then there exists a (non-circuit private) FHE scheme for which \constref{Const:GenericFHE} is not secure --- see \appref{circuitprivacynecessary}. After defining circuit privacy in \appref{appendix:circuitprivacydef} we use circuit privacy to prove that \constref{Const:GenericFHE} satisfies conditional encryption secrecy \thmref{thm:CondEncFromFHE} in \appref{appendix:CondEncSecrecyProofFHECond}.  

 More precisely, let $\FHE=(\Setup,\Enc, \Eval,\Dec) $ be an FHE scheme, then we obtain a conditional encryption scheme $\Pi_{\Cond\Enc}$ as described in \constref{Const:GenericFHE} and \figref{fig:GenericFHE}.

\begin{figure}
\begin{construction}\constlab{Const:GenericFHE}
\end{construction}
\begin{mdframed}
	\noindent ~~ \underline{Key Generation Algorithm}:
	\begin{itemize}
		\item $(pk,sk) \gets \Pi_{\Cond\Enc}.\KG(1^\lambda) = \FHE.\Setup(1^\lambda)$ 
	\end{itemize}
	
	\noindent ~~ \underline{Regular Encryption}:
	\begin{itemize}
		\item $(0,c)\gets \Pi_{\Cond\Enc}.\Enc_{pk}(m_1) = \FHE.\Enc_{pk} (m_1)$ 
	\end{itemize}
	
    \noindent ~~ \underline{Conditional Encryption algorithm}: 
    \begin{itemize}
        \item $ (1,c')\gets \Pi_{\Cond\Enc}.\Cond\Enc_{pk}(c,m_2,m_3)$: 
    
    \begin{enumerate}

	   \item Define the circuit $C_{P,m_2,m_3}(m_1)$ which outputs $m_3$ if $P(m_1, m_2)=1$ and outputs $Z$ if $P(m_1,m_2) =0$.      (Note: $Z$ is an arbitrary fixed constant which is publicly known. If we want the construction to be error detecting then $Z$ needs be outside of the message space for our conditional encryption scheme, but inside the FHE message space.  
        \item Run $c'\gets \FHE.\Eval_{pk}(C_{P,m_2,m_3}, c)$ 
        \item Return $(1,c')$
	
\end{enumerate}
\end{itemize}
    \noindent ~~ \underline{Decryption algorithm}:
    \begin{itemize}
	    \item  $m:= \Pi_{\Cond\Enc}.\Dec_{sk}(b, c)$ 
              \begin{enumerate}
                  \item Compute $m = \FHE.\Dec_{sk}(c)$
                  \item If $m$ is outside the message space return $\bot$; otherwise return $m$
              \end{enumerate}
    \end{itemize}

\end{mdframed}	
\caption{Proposed generic construction of our generic construction of Conditional Encryption from FHE}
	\figlab{fig:GenericFHE}
\end{figure}

\subsection{Circuit Privacy is Necessary} \applab{circuitprivacynecessary}
We demonstrate that \constref{Const:GenericFHE} can be insecure if the underlying FHE scheme does not satisfy circuit privacy. In particular, there exists secure FHE schemes for which the proposed construction is blatantly insecure as a conditional encryption scheme. Security definitions for FHE assume that the attacker does not have the secret decryption key, and there is no requirement that the output $ct=\mathsf{Eval}(C,c_1\ldots, c_n)$ hides information about the underlying message bits $m_1,\ldots,m_n$ when the secret key is known. Given a FHE scheme $\Pi_{\FHE} = \left( \KG, \Enc,\Eval, \Dec \right)$ we can define a new FHE scheme $\Pi_{\FHE}' = \left( \KG, \Enc,\Eval',  \Dec' \right)$ where $\Eval'(C,c_1,\ldots,c_n) = (\Eval(C,c_1,\ldots,c_n), c_1)$ i.e., $\Eval'$ simply appends the first input ciphertext $c_1$ to the final output. Correctness still holds as the decryption algorithm $\Dec'$ can simply ignore $c_1$ and then run the original decryption algorithm $\Dec$. Note that as long as the attacker does not have the secret key that the output of $\Eval'$ does not leak any information about the encrypted message $m_1$ corresponding to $c_1$. Thus, the updated FHE scheme $\Pi_{\FHE}'$ still satisfies the traditional notion of semantic security. However, the message $m_1$ can be directly extracted from the evaluation ciphertext $ct'=\Eval'( C,c_1,\ldots,c_n)$ by any party who has the secret decryption key. Instantiated with $\mathsf{Eval}'$ the proposed conditional encryption scheme would output $\Cond\Enc_{pk}(c_1,m_2,m_3) = \FHE.\Eval'_{pk}(C_{P,m_2,m_3},c_1) = (\FHE.\Eval_{pk}(C_{P,m_2,m_3},c_1),c_1) = $. Supposing that $P(m_1,m_2)=0$ conditional encryption secrecy requires that the conditional ciphertext $ct'=\Cond\Enc_{pk}(c_1,m_2,m_3)$ leaks no information about any of the messages $m_1,m_2$ and $m_3$ except that $P(m_1,m_2)=0$. Yet, the ciphertext $ct'$ still contains an encryption of the secret message $m_1$. In our setting the attacker is given the decryption key allowing the attacker to directly extract the secret message $m_1$ from $ct'$. Thus, \constref{Const:GenericFHE} blatantly violates conditional encryption secrecy when instantiated with $\Pi_{\FHE}'$. If we want to show that \constref{Const:GenericFHE} is secure we will need to rely on additional security properties for FHE (circuit privacy) which rules out FHE constructions like $\Pi_{\FHE}'$.

\subsection{Circuit Privacy Definition} \applab{appendix:circuitprivacydef}

To prove security of \constref{Const:GenericFHE} we rely on the notion of circuit privacy. Gentry proposed a formal definition of (statistical) circuit private FHE in his thesis \cite{Gentrythesis} and suggested how to achieve circuit privacy. Ostrovsky et al. \cite{C:OstPasPas14} introduced the notion maliciously circuit private FHE which requires that privacy holds even when the public key and/or input ciphertexts are generated maliciously. We base our definition of circuit privacy on Gentry's definition of circuit privacy as malicious security is not necessary to show conditional encryption secrecy.

Intuitively, a FHE scheme is circuit private if the ciphertexts $c=\FHE.\Eval_{pk}(C,c_1,\ldots,c_n)$ reveals nothing about the circuit $C$ or the encrypted messages $m_1,\ldots, m_n$ corresponding to the input ciphertexts $c_1,\ldots, c_n$ other than the output $C(m_1,\ldots, m_n)$ of the circuit even if the adversary has the secret key or is computationally unbounded (statistical security). This intuition is formalized by the notion of a simulator who takes as input the public key $pk$ and the circuit output $C(m_1,\ldots, m_n)$ and generates a ciphertext that is statistically indistinguishable from  $\FHE.\Eval_{pk}(C,c_1,\ldots,c_n)$ --- in \cite[Def 2.1.6]{Gentrythesis} the simulator is defined to be $\Sim(pk,C(m_1,\ldots,m_n)) = \FHE.\Enc_{pk}(m_1,\ldots, m_n)$. We slightly rephrase the definition \cite{Gentrythesis} to obtain a concrete security definition instead of an asymptotic definition.

\begin{definition}\deflab{def:Mal-CPrivateFHE}
Let $\Pi_{\FHE} = (\KG, \Enc, \Eval, \Dec)$ denote a homomorphic encryption scheme for a class of circuits $\C$. We say $\Pi_{\FHE}$ is $\epsilon(\lambda)$-circuit private if there is a (possibly unbounded) simulator $\Sim$ such that for all key-pairs $(sk,pk)$ in the support of $\KG(1^\lambda)$, any circuit $C \in \C$, any plaintexts $m_1,\ldots, m_n$ and any fixed ciphertexts $c_1,\ldots, c_n$ such that $c_i$ in the image of $\Enc_{pk}(m_i)$ for each $i \leq n$ (i.e., each ciphertext $c_i$ is a valid encryption of the plaintext $m_i$) and any (unbounded) distinguisher $\D$ has advantage at most $\left|p_{\D,sim} - p_{\D,Eval}\right| \leq \epsilon(\lambda)$ where $ p_{\D,sim}=\Pr[\D(sk, pk, c_1,\ldots, c_n, m_1,\ldots, m_n,$ $\Sim(pk, C(m_1,\ldots,m_n))=1] $ 
and $p_{\D,Eval} = \Pr[\D(sk, pk, c_1,\ldots, c_n, m_1,\ldots, m_n,$ $ \Eval_{pk}(C,c_1,\ldots,c_n) ) = 1] $ where the randomness is taken over the random coins of $\D, \Sim$ and $\Eval$. 

\end{definition}

Since our distinguisher $\D$ is unbounded the requirement that the distinguishing advantage is upper bounded by $\epsilon(\lambda)$ is equivalent to requiring that the statistical distance between $\Eval_{pk}(C,c_1,\ldots,c_n) $ and $\Sim(pk, C(m_1,\ldots,m_n)$ is bounded by $\epsilon(\lambda)$. Giving the distinguisher additional information like $sk, pk$,$c_1,\ldots, c_n$ and $m_1,\ldots, m_n$ does not change the definition.

\subsection{Proving Conditional Encryption Secrecy}
\applab{appendix:CondEncSecrecyProofFHECond}

\thmref{thm:CondEncFromFHE} proves that \constref{Const:GenericFHE} achieves conditional encryption secrecy as long as the underlying FHE scheme satisfies circuit privacy. Intuitively, the conditional encryption simulator will just use the circuit private FHE simulator. See details in the proof of \thmref{thm:CondEncFromFHE}.

\begin{theorem}\thmlab{thm:CondEncFromFHE} Assume that $\Pi_{\FHE}=(\KG, \Enc, \Eval, \Dec)$ is an $\eps(\lambda)$-circuit-private FHE, then $\Pi_{\Cond\Enc}$ provides $(\infty, t_{\Sim}, \eps(\lambda))$-conditional encryption secrecy where $t_{\Sim}$ denotes the running time of the simulator for our circuit private FHE scheme. 
\end{theorem}

\begin{proof}(Sketch) Supposing that $P(m_1,m_2)=0$ we have $C_{P,m_2,m_3}(m_1) = Z$ where $Z$ is a fixed constant (publicly known). Thus, our conditional encryption secrecy simulator $\Sim_{CE}(pk)$ will simply run the FHE circuit privacy simulator $\Sim_{cpFHE}(pk, Z)$. Formally, we define $\Sim_{CE}(pk) \doteq (1,\Sim_{cpFHE}(pk, Z))$. Since $C_{P,m_2,m_3}\left(m_1 \right)=Z$, by circuit privacy of the underlying FHE scheme, the output of $\Sim_{cpFHE}(pk, Z)$ will be statistically indistinguishable from the output of $\Cond\Enc_{pk}(c_1,m_2,m_3)=\left(1,\FHE.\Eval_{pk}(C_{P,m_2,m_3}, c_1)\right)$. In particular, for any distinguisher $\D$ we have
\begin{align}
    &\bigg|\Pr\left[\D\left(sk, pk, c_1, m_1,m_2, m_3,\Sim_{cpFHE}\left(pk, C_{P,m_2,m_3}\left(m_1\right) =Z\right)\right)=1\right] \nonumber\\
    &- \Pr\left[\D\left(sk, pk, c_1, m_1,m_2, m_3, \Eval_{pk}(C_{P,m_2,m_3},c_1) \right) = 1 \right]\bigg| \leq \eps(\lambda)\label{CircuiPrivacyAdv}
\end{align}
By definition the distribution of $\Sim_{CE}(pk)$ its output is identical to $\Sim_{cpFHE}(pk, Z)$. Looking at \constref{Const:GenericFHE}, we observe that $\Cond\Enc_{pk}(c_1,m_2,m_3)=\FHE.\Eval_{pk}(C_{P,m_2,m_3}, c_1)$. Thus we can rewrite Equation \eqref{CircuiPrivacyAdv} as follows:
\begin{align*}
    &\bigg|\Pr\left[\D\left(sk, pk, c_1, m_1,m_2, m_3,\left(1,\Sim_{CE}\left(pk\right)\right)\right)=1\right] \\
    &- \Pr\left[\D\left(sk, pk, c_1, m_1,m_2, m_3, \Cond\Enc_{pk}\left(c_1,m_2,m_3\right) \right) = 1 \right]\bigg| \leq \eps(\lambda)\label{CircuiPrivacyAdv}
\end{align*}
So we have shown the there exists simulator $\Sim_{CE}$ running in time $t_{\Sim}(\lambda) = t_{\Sim_{cpFHE}}(\lambda)$ such that any unbounded adversary can distinguish between $\Cond\Enc_{pk}(c_1, m_2,m_3)$ and $\Sim_{CE}(pk)$ with advantage of at most $\eps(\lambda)$ and \constref{Const:GenericFHE} provides $(\infty,t_{\Sim}(\lambda), \eps(\lambda))$ conditional encryption secrecy.

\end{proof}

\subsection{Correctness and Error Detection}

\newcommand{\ThmGeneralFHECorrectness}{	Assuming that the underlying fully homomorphic scheme $\FHE$ satisfies perfect correctness 
 \constref{Const:GenericFHE} satisfies perfect correctness and perfect error detection i.e., the construction is $1-\epsilon(\lambda)$-error detecting and $1-\epsilon(\lambda)$-correct with $\epsilon(\lambda)=0$. }

\begin{theorem} \thmlab{thm:GenFHECorrectness}
\ThmGeneralFHECorrectness
\end{theorem}
\begin{proofof}{\thmref{thm:GenFHECorrectness}}

Correctness of \constref{Const:GenericFHE} follows immediately from the correctness of the underlying FHE scheme. Observe that for any message $m$ in the message space and any secret key pair $(sk,pk)\leftarrow  \Pi_{\FHE}.\KG(1^\lambda)$ we have  
\begin{eqnarray*}
  \Pi_{\Cond\Enc}.\Dec_{sk} \left(  \Pi_{\Cond\Enc}.\Enc_{pk}(m) \right) =&   \Pi_{\Cond\Enc}.\Dec_{sk} \left(0,  \Pi_{\FHE}.\Enc_{pk}(m) \right) \\
   =&   \Pi_{\FHE}.\Dec_{sk} \left(  \Pi_{\FHE}.\Enc_{pk}(m) \right) \\
   =& m \ .
\end{eqnarray*}
The first  equality holds by the definition of $\Pi_{\Cond\Enc}.\Enc$. The second equality follows by correctness of the underlying \FHE scheme and by the definition of $  \Pi_{\Cond\Enc}.\Dec_{sk}$  since we assumed the message $m$ is in message space for our conditional encryption scheme. The final equality follows by the correctness of the underlying fully homomorphic encryption scheme. 

Similarly, consider any messages $m_1,m_2,m_3$ for which $P(m_1,m_2)=1$ and $m_3$ is an arbitrary message in our message space. Let $(0,c_1) \gets \Pi_{\Cond\Enc}.\Enc_{pk}(m_1) = \left(0, \Pi_{\FHE}.\Enc_{pk}\left(m_1\right)\right)$ be any encryption of $m_1$. By construction we have \[\Pi_{\Cond\Enc}.\Cond\Enc_{pk}\left((0,c_1),m_2,m_3 \right) = \left(1, \Pi_{\FHE}.\Eval_{pk}\left(C_{P,m_2,m_3},c_1\right)\right) \ ,  \]
and by correctness of $\Pi_{\FHE}$ we have 
\[\Pi_{\FHE}.\Dec_{sk}\left(\Pi_{\FHE}.\Eval_{pk}\left(C_{P,m_2,m_3},c_1\right)\right) = C_{P,m_2,m_3}(m_1) = m_3 \ , \]
where the first equality follows by $\FHE$ correctness --- note that $c_1 = \Pi_{\FHE}.\Enc_{pk}(m_1)$. The second equality follows by definition of $C_{P,m_2,m_3}$ since, by assumption, we have $P(m_1,m_2)=1$. Since we  assume that $m_3$ is in our message space by definition of $\Pi_{\Cond\Enc}.\Dec$ it follows that 
\[\Pi_{\Cond\Enc}.\Dec_{sk}\left(\Pi_{\Cond\Enc}.\Cond\Enc_{pk}\left(\left(0,c_1\right),m_2,m_3\right)\right) = m_3 \ . \]

Our argument that \constref{Const:GenericFHE} satisfies perfect error detection also follows from the correctness of the underlying FHE scheme. Let $m_1$ and $m_2$ be any messages in the message space for which the predicate does not hold i.e., $P(m_1,m_2)=0$. Let $(0,c_1) \gets \Pi_{\Cond\Enc}.\Enc_{pk}(m_1) = \left(0, \Pi_{\FHE}.\Enc_{pk}(m_1)\right)$ be any encryption of $m_1$. By construction we have \[\Pi_{\Cond\Enc}.\Cond\Enc_{pk}\left((0,c_1),m_2,m_3 \right) = \Pi_{\FHE}.\Eval_{pk}(C_{P,m_2,m_3},c_1) \ ,  \]
and by correctness of $\Pi_{\FHE}$ we have 
\[\Pi_{\FHE}.\Dec_{sk}\left(\Pi_{\FHE}.\Eval_{pk}\left(C_{P,m_2,m_3},c_1\right)\right) = C_{P,m_2,m_3}(m_1) = Z \ , \]
where $Z$ is a message explicitly chosen to be outside the message space of our conditional encryption scheme. The first equality above follows by $\FHE$ correctness since $c_1 = \Pi_{\FHE}.\Enc_{pk}(m_1)$ and the second equality follows by definition of $C_{P,m_2,m_3}$ since, by assumption, we have $P(m_1,m_2)=0$. Thus, by definition of $\Pi_{\Cond\Enc}.\Dec$ it follows that \[\Pi_{\Cond\Enc}.\Dec_{sk}\left(\Pi_{\Cond\Enc}.\Cond\Enc_{pk}\left((0,c_1),m_2,m_3 \right)    \right) = \bot \ . \]

\end{proofof}

\section{Missing Proofs}
\applab{apdx:MissingProofs}

\begin{remindertheorem} {\thmref{thm:EqCor}}
\ThmEqCorrectness
\end{remindertheorem}
\begin{proofof}
{\thmref{thm:EqCor}}
 Since $\Enc_{pk}(m)$ simply runs regular Pallier encryption perfect correctness of Pallier immediately implies that 
 $\Dec_{sk}\left( \Enc_{pk}(m)\right) =  \Dec_{sk}\left(P.\Enc_{pk}(\ToInt(m) \right)= \ToOrig$ $(\ToInt(m)) = m$ with probability $1$ for all messages $m \in \Sigma^{\leq n}$ and all public/private key pairs in the support of $\KG$. Similarly, if $c_1=\Enc_{pk}(m)$ and $P_{=}(m_1,m_2) = 1$ then $\Cond\Enc_{pk}(c_1,m_2,m_3)$ will output $(1,c')$ where $c' = g^{\ToInt{(m_3)}} r^n \mod{N^2}$ for some $r \in \mathbb{Z}_N^*$. Thus, $c'$ is a valid pallier ciphertext for $\ToInt{(m_3)}$ and, by correctness of Pallier, $\Dec_{sk}(1,c')$ will return $m_3$. 

 On the other hand if $P_{=}(m_1,m_2) = 0$ then by \thmref{thm:EqualityTestSecrecy} the ciphertext $c'$ is a valid Pallier Ciphertext for some uniformly random integer $y \in \mathbb{Z}_N$ and we will have $\Dec_{sk}(1,c') = \bot$ as long as $y > |\Sigma|^{n+1}$. Thus, the construction is $1-\epsilon(\lambda)$-error detecting conditional encryption scheme with $\eps(\lambda) = \frac{|\Sigma|^{n+1}}{N} \leq \frac{1}{\max\{p,q\}}$.
\end{proofof}

\begin{remindertheorem}{\thmref{thm:EqualityTestSecrecy}}  
\ThmEqTestSecrecy
\end{remindertheorem}
\begin{proofof}{\thmref{thm:EqualityTestSecrecy}}
We define the simulator $ \Sim(pk) $ as follows. The simulator $ \Sim(pk) $ takes as input the Paillier public key $ pk $ and then selects $ R_s\in_R \mathbb{Z}_{N} $ and $r_s\in_R \mathbb{Z}^*_N$ uniformly at random and then encrypts $ R_s $ as $ C_{\Sim} = \Pail.\Enc_{pk}(R_s; r_s) = g^{R_s} r_s^{N} \mod N^2$ and outputs it. We now argue that for any $m_1, m_2 \in \Sigma^{n}$ with $P_{=}(m_1,m_2) = 0$ any payload message $m_3$ and any Pallier key $\left(pk=\left(N=pq,g\right),sk\right)$  which satisfies our condition that $|\Sigma|^{n+1} \leq \min\{p,q\}$ and any encryption $c_1 = g^{m_1} r_1^{N} \mod{N^2}$ of $m_1$ under $pk$ that the distributions $ \left(pk, sk, m_1, m_2, m_3, c_1, C_{m_3} = \Cond\Enc_{pk}\left(c_{m_1}, m_2, m_3\right) \right)$ and $ (pk, sk,$ $ m_1, m_2, m_3,c_1, C_{\Sim}= \Sim(pk)) $ are identical. In particular, it suffices to argue that $C_{\Sim}= \Sim(pk)$ and $\Cond\Enc_{pk}(c_{m_1}, m_2, m_3)$ are distributed identically. 

To see this consider the generation of $\Cond\Enc_{pk}(c_{m_1}, m_2, m_3)$. First, we pick a random $R \in \mathbb{Z}_N$ and generate an encryption of $(-R \cdot m_2 \mod N)$ as $c_2 = g^{-R \cdot m_2} r_2^N \mod{N^2}$ where $r_2 \in \mathbb{Z}_N^*$ is picked randomly. We then compute $c_1^R = g^{m_1 \cdot R} r_1^{RN}$. Finally, we output
\begin{eqnarray*}&&c_1^Rc_2 \cdot g^{m_3} \mod{N^2} =  g^{m_3+R(m_1-m_2)} r_1^{RN} r_2^N \\ 
&&= g^{m_3+R(m_1-m_2) \mod N} \left(r_1^Rr_2 \mod{N}\right)^{N} \mod{N^2} \ . \end{eqnarray*}
where the values $R \in \mathbb{Z}_N$, $r_2 \in \mathbb{Z}_N^*$ are fresh random values. In the last step we implicitly used the fact that if $r_1^Rr_2 = aN+b$ where $b = \left[r_1^Rr_2 \mod{N}\right]$ then \[ (aN+b)^N = \sum_{i=0}^N {N \choose i} (aN)^ib^{N-i} = b^{N} \mod{N^2} \ . \]

Let us first focus on the term $m_3+R(m_1-m_2)$ in the exponent of $g$. We observe that $[m_1-m_2 \mod N] \in \mathbb{Z}_N^*$ since $1 \leq |m_1-m_2| \leq \left| \Sigma\right|^{n} \leq \min\{p,q\}$. It follows that for any $m_3$ that $R(m_1-m_2) + m_3$ is distributed uniformly at random in $\mathbb{Z}_N$ when $R \in \mathbb{Z}_N$ is picked randomly. We next consider the term $\left(r_1^Rr_2 \mod{N}\right)$ and argue for any fixed $r_1 \in \mathbb{Z}_N^*$ and $R \in \mathbb{Z}_N$ that $\left(r_1^Rr_2 \mod{N}\right)$ is distributed uniformly at random in $\mathbb{Z}_N^*$ when $r_2 \in \mathbb{Z}_N^*$ is picked randomly. It follows that for any $r_1 \in \mathbb{Z}_N^*, m_1,m_2,m_3$ such that $1 \leq |m_1-m_2| \leq \min\{p,q\}$ that the simulated ciphertext ($C_{\Sim}= \Sim(pk) = g^{R_s} r_s^N \mod{N^2}$ for random $r_s \in \mathbb{Z}_N^*$ and $R_s \in \mathbb{Z}_N$ ) is identically distributed to $\Cond\Enc_{pk}(c_{m_1}, m_2, m_3)$.
\end{proofof}

\begin{remindertheorem}{\thmref{thm:StatDistUak}}  
\thmStatDistTwo
\end{remindertheorem}
\begin{proofof} {\thmref{thm:StatDistUak}}
    ‌Based on the definition of statistical distance we have 
    \begin{align}
        \mathtt{SD}(\D_{ak}, \U_b) &= \frac{1}{2} \sum_{i = 0}^{b-1} |\Pr_{y\in_R \D_{ak}}[y = i] - \Pr_{y\in_R \U_b}[y = i]| \nonumber\\ 
        & = \frac{1}{2} (ak) (| \frac{1}{b} - \frac{1}{a} \cdot\frac{1}{k}| + (b-ka) (|\frac{1}{b} - 0|)\nonumber\\ 
        & = \frac{1}{2}(ak) (\frac{b-ak}{abk}) + \frac{1}{2}(r)(\frac{1}{b}) = \frac{r}{b} \leq \frac{1}{k+1}
    \end{align}
    
\end{proofof}

\begin{theorem}{\thmlab{thm:ArbHamm}}
      \ThmHammingDistCorrectness
\end{theorem} 
\begin{proofof}{\thmref{thm:ArbHamm}}
We first note Authenticated Encryption security implies that the term $\eps_{AE}(\lambda)$ is negligible. Otherwise, an AE attacker could simply pick a random key $K'$ and use $c=\Enc_{K'}(m)$ as an attempted forgery for the unknown secret key $K$! 

 There are two conditions in the \defref{CondCorr} which need to be proved. The first condition is regular encryption correctness and the other one is the conditional encryption correctness. 

The observation that $\Dec_{sk}\left( \Enc_{pk}(m;r)\right) = m $ for all messages $m \in \Sigma^n$, random coins $r$ and all $(sk,pk)$ in the support of the key generation algorithm follows immediately from the correctness of Pallier encryption.

It remains to to show that for all messages $ m_1, m_2 \in \Sigma^n$ such that $P_{\ell, \Ham}(m_1,m_2) $, all payload messages $m_3$, all $\{sk,pk\}$ in the support of our Key Generation algorithm and all random strings $ r_1, r_2 \in_R (\Mbb{Z}_{N}^*)^n$
we have

\begin{align}\eqnlab{EQ:EncCor2}	
\Pr \Biggl[\Dec_{sk}(\Cond\Enc_{pk}(c_1, m_2, m_3; r_2)) = m_3 \Biggl| \begin{matrix}
	(sk, pk)\gets \KG (1^\lambda) \\ 
	c_1 = \Enc_{pk}(m_1; r_1) \\
	P_{t,\Ham}  (m_1, m_2) = 1
	\end{matrix} \Biggl] \geq 1 - \eps \ .
  	\end{align}  
   
Let $K$ denote the authenticated encryption key and let  $ \ldb s \rdb_1,\ldots,   \ldb s \rdb_n$ denote the shares of $K$ that were generated by the conditional encryption algorithm. Let $\tilde{c} = (b,\tilde{c}_1,\ldots, \tilde{c}_n, C_{AE})$ denote the output of $\Cond\Enc_{pk}(c_1, m_2, m_3; r_2))$, and let  $\ldb s' \rdb_i = \RanDec\left( P.\Dec_{sk}(\tilde{c}_i) \right)$ denote the shares that are recovered. Finally, let $S^* = \{ i \in [n]~: ~m_2[i]=m_1[i]\}$ denote indices of the characters where $m_2$ and $m_1$ match. By correctness of Pallier we have $\ldb s' \rdb_i = \ldb s \rdb_i$  for {\em all} $i \in S^*$. For $i \not \in S^*$ the distribution over $\ldb s' \rdb_i$ is as follows: sample a uniformly random item $y_i$ from $\mathbb{Z}_N^*$ and output $y_i \mod{2^{\lambda}}$.

If $P_{Hamm,\ell}(m_1,m_2)=1$ we have $|S^*| \geq n-\ell$ and there is some subset $S \subseteq S^*$ of size $|S|=n-\ell$ such that \[ K=K_{S} = \recover\left( \left\{\left(i, \ldb s' \rdb_i\right)_{i \in S} \right\} \right)   \ . \]
From the correctness of the authenticated encryption scheme it follows that $\Auth.\Dec_{K_{S}}(c_{AE}) = m_3$.  

Thus, the only possible to output an incorrect message $m'$ is if for some $S \subseteq n$ of size $n-\ell$ we have $K \neq K_S = \recover\left( \left\{\left(i, \ldb s' \rdb_i\right)_{i \in S}\right\}\right)$ and $\Auth.\Dec_{K_{S^*}}(c_{AE}) \neq \bot$. However, if $K_S \neq K$ then $S \not \subseteq S^*$ and we can find some $i \in S \setminus S^* $. For now assume that for all $i \not\in S^*$ the value of $\ldb s' \rdb_i$ is uniformly random we can view $K_S$ as a uniformly random key. If we view each $K_S$ as random then we have $\Pr[\Auth.\Dec_{K_S}(c_{AE}) \neq \bot] \leq \epsilon_{AE}$ and $\Pr[\exists S \subseteq [n]~.  \Auth.\Dec_{K_S}(c_{AE}) \not\in \{m_3, \bot \} ] \leq {n \choose \ell} \eps_{AE}$. 

In the previous paragraph we assumed that the value $\ldb s' \rdb_i$ is uniformly random for each  $i \not\in S^*$ the value. This is close, but it is not quite true. In reality the distribution of $\ldb s' \rdb_i$ is described by sampling a uniformly random $y_i \in \mathbb{Z}_N^*$ and then outputting $y_i \mod{2^{\lambda}}$.However, by \thmref{thm:StatDistUak} the statistical distance between original/modified distribution of our recovered shares $ \ldb s' \rdb_1,\ldots,   \ldb s' \rdb_n$ is upper bounded by $2^{-\lambda}$. This follows since we are guaranteed that $N > n 2^{2\lambda}$ by definition of the key generation algorithm. Thus, we have 
\[ \Pr \Biggl[\Dec_{sk}(\Cond\Enc_{pk}(c_1, m_2, m_3; r_2)) \neq m_3 \Biggl| \begin{matrix}
	(sk, pk)\gets \KG (1^\lambda) \\ 
	c_1 = \Enc_{pk}(m_1; r_1) \\
	P_{t,\Ham}  (m_1, m_2) = 1
	\end{matrix} \Biggl] \leq  {n \choose \ell} \epsilon_{AE} + 2^{-\lambda} \ . \]

Similarly, if $P_{\ell,Hamm}(m_1,m_2)=0$ then for all $S \subseteq [n]$ of size $|S|=n-\ell$ we can (essentially) view $K_S$ as random since there is some $i \in S \setminus S^*$. It follows that 

\[ \Pr \Biggl[\Dec_{sk}(\Cond\Enc_{pk}(c_1, m_2, m_3; r_2)) \neq \bot \Biggl| \begin{matrix}
	(sk, pk)\gets \KG (1^\lambda) \\ 
	c_1 = \Enc_{pk}(m_1; r_1) \\
	P_{t,\Ham}  (m_1, m_2) = 0
	\end{matrix} \Biggl] \leq {n \choose \ell} \epsilon_{AE} + 2^{-\lambda} \ . \]

\end{proofof}

\begin{remindertheorem}{    \thmref{thm:CondSecArbHamm}}
 \thmSemiHonest
\end{remindertheorem} 
\begin{proofof}{\thmref{thm:CondSecArbHamm}}
	To prove this theorem we use a hybrid argument. In the first hybrid (Hybrid 0, real world) the distinguisher is given the actual ciphertext output conditional encryption and in the last hybrid contains the adversary is given a ciphertext output by our simulator --- described in \figref{fig:SimArbHamm}. As the hybrids are indistinguishable, we can conclude that the first and last hybrid are indistinguishable as well which implies that the our suggested construction is secure and provides conditional encryption secrecy in the semi-honest model. Then we concretely compute the distinguishing advantage of the defined hybrids. In what follows, we describe the hybrids with more details.

	 	\begin{itemize}
	 	\item \textbf{Hybrid 0}: In this hybrid the distinguisher $ \D $ is given    $(sk, pk,  m_1, ,m_2,$ $ m_3, c_{m_1}, (1, \tilde{c}))$  in which $ \tilde{c} = (\tilde{c}_1, \dots, \tilde{c}_n, c_{AE})  \leftarrow \Cond\Enc_{pk}(c_1,m_2,m_3)$. 
	 	
	 	\item \textbf{Hybrid 1}: Let $T = \{i : m_2[i] \neq m_1[i]\}$ be the set of the indexes that $ m_1 $ and $ m_2 $ have different characters. We define \textbf{Hybrid 1} similar to \textbf{Hybird 0}, except for all $j\in T $ we replace $$\tilde{c}_{j}= P.\Enc_{pk}\Big(R_i(m_2[j]-m_1[j])+ \RanEnc(\ldb s \rdb_i)\Big)$$ with $P.\Enc(R_j’)$ where $ R_j' \in_R \Mbb{Z}_N$ are fresh and uniform random values chosen from $\Mbb{Z}_N$.

	 	\item \textbf{Hybrid 2}: This hybrid is exactly the same as the previous hybrid except we replace all the remaining ciphertexts $ j\in [1:n]/ T  $ with $ \tilde{c}_{j}= P.\Enc_{pk}\Big(R_j(m_2[j]-m_1[j])+ \RanEnc(\ldb s_r \rdb_j)\Big) $ where $ \ldb s_r \rdb_j \in_R \{0,1\}^{\lambda} $ are fresh uniformly random elements (chosen independently from the secret $K$)  chosen from the field $\Mbb{F}_{2^\lambda}$.

           \item \textbf{Hybrid 3}: This hybrid is exactly the same as the previous hybrid except we replace all the ciphertexts $ j\in [1:n]/ T  $ with $ \tilde{c}_{j}= P.\Enc_{pk}\Big(R_j(m_2[j]-m_1[j])+ \hat{R}_j)\Big) $ where $ \hat{R}_j\in_R \Mbb{Z}_N $ are chosen from $\Mbb{Z}_{N}$ uniformly at random.

       \item \textbf{Hybrid 4}: This hybrid is exactly the same as the previous hybrid except we replace ciphertexts $ \tilde{c}_{j}$ for all $ j\in [1:n]/ T  $, with $\Pail.\Enc_{pk}(R'_j)$ in which $ R'_j\in_R \Mbb{Z}_N $ are chosen from $\Mbb{Z}_{N}$ uniformly at random.
           
	 	\item \textbf{Hybrid 5}: This hybrid is exactly the same as the previous hybrid unless we replace $ c_{AE} $ with $ c'_{AE} \in_R \{0,1\}^{l(\lambda)}$ a $\lambda$-bit string chosen uniformly at random. We note that $l(\lambda)$ is a polynomila over the security parameter $\lambda$ which represents the ciphertext size of authenticated encryption. 
	 
	 	\item \textbf{Hybrid 6}: We replace the ciphertext of the conditional encryption with the output of the simulator $ \Sim $ described in \figref{fig:SimArbHamm}.

	 	\end{itemize}

	Now we are proving that the defined hybrids are equivalent.

\subsubsection{\textbf{Hybrid 0} $\equiv$ \textbf{Hybrid 1} } These hybrids are  equivalent i.e., we have 
	\begin{align}
& \Pr[\D^{H_{0}} =1] = \Pr[\D^{H_1} =1] \ . 
\end{align} 
Where $\D^{H_i}=1$ denotes the event that the distinguisher outputs $1$ in hybrid $i$. 
The argument is essentially the same as what we had for the security of \textit{Equality test} predicate --- see the proof of \thmref{thm:EqualityTestSecrecy}. In particular, for each $j \in T$ we have $m_1[j] \neq m_2[j]$ and $\left|m_1\left[j\right]-m_2\left[j\right]\right| \leq \min\{p,q\}$ which implies that $\left(m_1\left[j\right]-m_2\left[j\right]\right) \in \mathbb{Z}_N^*$. It follows that $R_j \times \left(m_1\left[j\right]-m_2\left[j\right]\right)$ is uniformly random in $\mathbb{Z}_N$.

\subsubsection{\textbf{Hybrid 1} $ \equiv $ \textbf{Hybrid 2}} We have information theoretically eliminated all information about shares $\shrs{}$ with $j \in T$. Since $P_{\ell,\Ham}(m_1,m_2)=0$ we have $|T| > \ell$ and $|\overline{T}| < n-\ell$. Let $T=\{i_1,\ldots, i_t\}$ with $t < n-\ell$. Shamir Secret Sharing guarantees that $(s_{i_1} , s_{i_2} , \ldots , s_{i_t})$ is uniformly random in $\mathbb{F}_{2^{\lambda}}^t$. Thus, we can simply replace the shares with uniformly random values. We have \begin{align}
& \Pr[\D^{H_{1}} ] =   \Pr[\D^{H_2} ] \ . 
\end{align}

\subsubsection{Statistically indistinguishability of \textbf{Hybrid 2} $ \equiv $ \textbf{Hybrid 3}} We apply \thmref{thm:StatDistUak} with $a=2^{\lambda}$, $k = \lfloor \frac{N}{2^\lambda} \rfloor$ and $b=N$. We first observe that when $i \in \overline{T}$ the value of $s_i \in \mathbb{F}_{2^{\lambda}}$ is uniformly random so that $\RanEnc(s_i)$ is equivalent to $\mathcal{D}_{ak}$. It follows that the statistical distance between $\RanEnc(s_i)$ and the uniform ditribution $\Mbb{Z}_N$ is at most $\frac{1}{k} = \lfloor \frac{N}{2^\lambda} \rfloor^{-1}$. Since we are replacing the random value in $|\overline{T}|$ ciphertexts the overall statistical distance is upper bounded by $\frac{|\overline{T}|}{k} \leq \frac{n}{k}$  we have: 

\begin{align}
	& |\Pr[\D^{H_{2}} ] -   \Pr[\D^{H_3}]| \leq \frac{2^\lambda n}{N-2^{\lambda}} \leq 2^{-\lambda} \ . 
\end{align}
The last inequality follows since  we pick $N \geq 2n 2^{2\lambda}$ so that $\frac{2^\lambda n}{N-2^{\lambda}} \leq 2^{-\lambda}$.

\subsubsection{\textbf{Hybrid 3} $\equiv$ \textbf{Hybrid 4}} These hybrids are statistically indistinguishable as $R_j (m_2[j]-m_1[j]) + \hat{R}_j$ is already uniformly random in $\mathbb{Z}_N$. We have 
\begin{align}
	& \Pr[\D^{H_{3}} ] = \Pr[\D^{H_4} ] 
\end{align}

\subsubsection{Indistinguishability of  \textbf{Hybrid 4} and \textbf{Hybrid 5}}  By Hybrid 4 we have information theoretically elimated any information about the secret key $K$ for our authentication encryption scheme from $(\tilde{c}_1,\ldots, \tilde{c}_n)$. Thus, by AE security any adversary running in time at most $t_{AE} = t_{AE}(\lambda)$ can distinguish between $c_{AE}$ and $c_{AE}'$ with the advantage of at most $\eps_{AE}(t_{AE}, \lambda)$. So we have 

	\begin{align}
	& |\Pr[\D^{H_{4}} ] -   \Pr[\D^{H_1}=1]|  \leq \eps_{AE}(t_{AE}, \lambda)
	\end{align}

\subsubsection{\textbf{Hybrid 5}  $\equiv $ \textbf{Hybrid 6}} Looking at the definition of our our simulator in \figref{fig:SimArbHamm}, we can see that the conditionally encrypted ciphertext in Hybrids 5 and 6 are generated in exactly the same way. It follows that the hybrids are 
information-theoretically equivalent and we have 
\begin{align}
	& \Pr[\D^{H_{5}}] = \Pr[\D^{H_6} ] 
\end{align}

Putting everything together we have 
\begin{align}
&\Big| \Pr\left[\D \left(sk, pk, m_1, m_2, m_3. \Cond\Enc_{pk}\left(\Enc_{pk}\left(m_1\right), m_2\right)\right)=1\right] \nonumber\\
&-  \Pr\left[\D \left(sk, pk, m_1, m_2, m'_3,\Sim\left(pk\right)\right)=1\right]\Big| \nonumber \\
&= \left|\Pr\left[\D^{H_0}\right]-\Pr\left[\D^{H_6}\right]\right| \nonumber \\
&\leq \sum_{i=0}^5 \left|\Pr\left[\D^{H_i}\right]-\Pr\left[\D^{H_{i+1}}\right]\right| \nonumber \\
&< \eps_{AE}(t',\lambda) + \frac{n2^\lambda}{N-2^{\lambda}} \leq \eps_{AE}(t',\lambda) + 2^{-\lambda}  \nonumber \ .
\end{align}

\begin{figure*}
		\begin{itemize}
			\item [] \underline{Design of simulator $ \Sim(pk) $}
			\begin{itemize}
				\item [1.] Sample, $ r''_1, \ldots, r''_n \in_R\Mbb{Z}^*_N, R''_1, \ldots, R''_n \in_R{\Mbb{Z}_N}^{n} $ uniformly at random  
    
				\item [2.] For all $ 1\leq i \leq n $ compute $ \tilde{c}'_i = \Pail.\Enc_{pk}(R''_i; r''_i ) $
				
				\item [3.] Pick $ R_K\in_R \set{0,1}^{l(\lambda)} $ uniformly at random and set $ c'_{AE} = R_K $.  //{\color{blue} $l(\lambda)$ represents the ciphertext size of our authenticated encryption.}

				\item [4.] Output $ \tilde{c}' = (1, \tilde{c}'_1, \cdots, \tilde{c}'_n,  c'_{AE}) $. 
			\end{itemize}
		\end{itemize}\caption{Steps of designing the simulator $ \Sim $ for the conditional encryption secrecy when the predicate is $P_{\ell, \Ham}$}
		\figlab{fig:SimArbHamm}
\end{figure*}

\end{proofof}

\begin{remindertheorem} {\thmref{thm:EDCor}}
\thmEDCorrect
\end{remindertheorem}

\begin{proofof}{\thmref{thm:EDCor}}
Note that $\Enc_{pk}(m)$ includes $c[0] = P.\Enc_{pk}(\ToInt(m))$ and that therefore by correctness of Pallier we have  $\Dec_{sk}\left( \Enc_{pk}(m)\right) = \Dec_{sk}\left( P.\Enc_{pk}(\ToInt(m))\right) = \ToOrig$ $(\ToInt(m)) = m$ with probability $1$ for all messages $m \in \Sigma^{\leq n}$ and all public/private key pairs in the support of $\KG$. 

Recall that if $c = (0,c[0],\ldots, c[n])=\Enc_{pk}(m)$ then $\Cond\Enc_{pk}(c,m',m'')$ will output a ciphertext of the form $(1,\tilde{c}_0,\ldots,\tilde{c}_{2n})$. If $P_{1,\ED}(m,m') = 0$ then we have $m_{-j} \neq m'$ and $m \neq m'_{-j}$ for all $0 \leq j \leq n$. Thus, by \thmref{thm:EqualityTestSecrecy} each $\tilde{c}_j = g^{y_j} r_j^n \mod{N^2}$ for random values $r_j \in \mathbb{Z}_N^*$ and $y_j \in \mathbb{Z}_N$. Thus, we have \[\Pr[\Dec_{sk}(1,\tilde{c}_0,\ldots,\tilde{c}_{2n}) \neq \bot] \leq \Pr[\exists j. y_j < |\Sigma|^{n+1}] \leq \frac{(2n+1) |\Sigma|^{n+1}}{N} \ .\] This implies that the construction is $1-\epsilon(\lambda)$-error detecting conditional encryption scheme with $\eps(\lambda) = \frac{|\Sigma|^{n+1}}{N} \leq \frac{1}{\max\{p,q\}}$.

Finally, if $P_{1,\ED}(m,m') = 0$ then by perfect correctness of our conditional encryption scheme for equality predicate there exists some $j$ such that $\tilde{c}_j = g^{y_j} r_j^N \mod{N^2}$ is a valid Pallier encryption of  $y_j=\ToInt(m'')$. Furthermore, we have already shown that $\Pr[\exists j. y_j < |\Sigma|^{n+1} \wedge y_j \neq \ToInt(m'')] \leq \frac{(2n+1) |\Sigma|^{n+1}}{N}$. It follows that, except with probability $\frac{(2n+1) |\Sigma|^{n+1}}{N}$ that we will have $\Dec_{sk}(1,\tilde{c}_0,\ldots,\tilde{c}_{2n}) = \ToOrig(\ToInt(m''))$. 
    
\end{proofof}

\begin{remindertheorem}{\thmref{ORComp:Security}}\thmORPrivacy
\end{remindertheorem}
\begin{proofof}{\thmref{ORComp:Security}}
    The simulator $\Sim_{OR}(pk)$ for $\Pi_{OR}$ will run the simulator $\Sim_i(pk_i)$ for each conditional encryption scheme \footnote{In the malicious security setting the simulator $\Sim_{OR}(b, pk)$ is also given a bit $b=1$ if and only if $\Cond\Enc_{pk}(c,m',m'')=\bot$ i.e., if and only if $\Pi_i.\Cond\Enc_{pk}(c,m',m'')=\bot$ for some $i \leq k$. If $b=1$ then $\Sim_{OR}(b,pk)$ outputs $\bot$. Otherwise we simply run $\Sim_i(0,pk_i)$ for each $i \leq k$.} and concatenate all of the ciphertexts. Clearly, the running time of the simulator is $t_{\Sim}'(\lambda) \approx \sum_{i=1}^k t_{\Sim,i}(\lambda)$. We can now define a sequence of $k+1$ hybrids Hybrid $0$ to Hybrid $k$. Intuitively, in hybrid $i$ we set $c_j = \Sim_i(pk)$ for $j\leq i$ and $c_j = \Pi_j.\Cond\Enc(c,m',m'')$ for $j > i$. Note that in Hybrid $0$ we have $c_j = \Pi_j.\Cond\Enc(c,m',m'')$ for all $j$ and thus the final output is $\Cond\Enc(c,m',m'')$. By contrast, in Hybrid $k$ we have   $c_j = \Sim_j(pk_j)$ for all $j \leq k$ and thus the final output is $\Sim_{OR}(pk)$. 
        
    By assumption any attacker running in time $t'(\lambda)=t(\lambda) - o(t(\lambda))$ can distinguish hybrids $i-1$ and $i$ with probability at most $\epsilon_i(t(\lambda), \lambda)$. It follows that any attacker running in time $t'(\lambda)=t(\lambda)  - o(t(\lambda))$ can  distinguish hybrid $0$ from hybrid $k$ with probability at most  $\epsilon'(\lambda,t'(\lambda)) = \sum_{i=1}^k  \epsilon_i(t(\lambda), \lambda)$. 
\end{proofof}

\begin{remindertheorem}{\thmref{ORComp:Correct}}
    \thmORCorrect
\end{remindertheorem}
\begin{proofof}{\thmref{ORComp:Correct}}
Let $T= \{j: P_j(m_1,m_2)=1\}$ and $\overline{T} = \{j: P_j(m_1,m_2)=0\} = [k] \setminus T$.  We first suppose that $P_{OR}(m_1,m_2)=0$ which implies that $P_i(m_1,m_2)=0$ for all $i \leq k$ i.e., $\overline{T} = [k]$. 

Now let $(pk,sk)$ be any honestly generated public/secret key and let $c = (0,c_1,\ldots,c_k) = \Cond\Enc_{pk}(m_1)$ with $c_i \doteq \Pi_i.\Enc_{pk}(m_1)$. The probability that $\Pi_i.\Dec_{sk}\left(\Pi_i.\Cond\Enc_{pk}(c_i,m_2,m_3)\right) \neq \bot$ is at most $\epsilon'_i(\lambda)$. Union bounding over all $i \leq k$ the probability that there exists $i$ such that $\Pi_i.\Dec_{sk}\left(\Pi_i.\Cond\Enc_{pk}(c_i)\right) \neq \bot$ is at most $\sum_{i=1}^k \epsilon'_i(\lambda) = \epsilon'(\lambda)$.

On the other hand suppose that $P_{OR}(m_1,m_2)=1$ which means that $P_j(m_1,m_2)=1$ for some $j\leq k$. Clearly, if $|T|\geq 1$ and $\Pi_i.\Dec_{sk}\left(\Pi_i.\Cond\Enc_{pk}(c_i)\right) = m_3$ for all $i \in T$ and $\Pi_i.\Dec_{sk}\left(\Pi_i.\Cond\Enc_{pk}(c_i)\right) = \bot$ for all $i \not \in T$ then $\Dec_{sk}$ will output the correct message $m_3$. As before the probability that there exists $i \in \overline{T}$ such that $\Pi_i.\Dec_{sk}\left(\Pi_i.\Cond\Enc_{pk}(c_i)\right) \neq \bot$ is at most $\sum_{i=1}^k \epsilon'_i(\lambda) = \epsilon'(\lambda)$. Similarly, the probability that there exists $j \in T$ such that $\Pi_i.\Dec_{sk}\left(\Pi_i.\Cond\Enc_{pk}(c_i)\right) \neq m_3$ is at most $\sum_{j=1}^k \epsilon_i(\lambda)$. 

Thus, we have $\Pr\left[\Dec_{sk}\left(\Cond\Enc_{pk}\left( \Enc_{pk}(m_1) ,m_2,m_3 \right)\right) \neq m_3 \right] \leq \sum_{i=1}^k \left( \epsilon_i'(\lambda)+\epsilon_i(\lambda)\right) = \epsilon(\lambda)$.

\end{proofof}

\section{System Model of Personalized Typo Tolerance}
In this section, we will concentrate on the application of the conditional encryption scheme in designing a secure mechanism to supporting password typos in a secure way. We first start with describing the syntax and API of a typical password-based authentication server, and then we use our introduced conditional encryption schemes for handling the typos when the user logs in with a wrong password close the original one. Indeed we expect the if the miss-typed password distance from the original one is \textit{small}, then this typo can be used by the user for future logins. 

 A password-based authentication scheme $ \Pi  $ is a stateful  mechanism that includes three main algorithms: $ \Init $, $\Sgin$ and $ \Lgin$ which are described as follows.

\begin{itemize}
 	\item $ (\sigma_0)\gets \Init (1^\lambda) $: The initialization algorithm (potentially randomized) $ \Init $ takes as input the security parameter $ \lambda $ and setups the system and outputs the initial state $ \sigma_0 $.
	
	\item $ (\sigma', d\in \{\acpt, \rjct\}) \gets \Sgin (\sigma, \id, \pwd)$: This is potentially a randomized algorithm which takes as input the current state of the system $ \sigma $, the user identity $ \id  $ and its corresponding password $ \pwd $, and updates the state of the system to $ \sigma' $ and output $ \acpt $ (for the successful registration when the user id $ \id $ is new and previously is not registered) or $ \rjct $ (for unsuccessful registration). 
	
	\item $ (\sigma', d\in \{\acpt, \rjct \} )\gets \Lgin (\sigma', \id, \pwd)$: Let $ \sigma $ be the current state. This is also potentially is a randomized algorithm which takes as input the state $ \sigma $, the identity $ \id $ and password $ \pwd $, and outputs the updated state $ \sigma'$ and the login result  $ d $. It either outputs $ d= \acpt $ (for successful login) or $ d = \rjct $ (unsuccessful login attempt e.g., using wrong password).
\end{itemize}

 In what follows, we will formally define the required properties of  $ \Pi $. Before that, we will start by introducing some predicates which will be used in our definitions.
 
\begin{itemize}
	\item $ \{0,1 \} := \isReg(\id, \sigma) $:  This predicate takes as input the identity-password pair $ (\id, \pwd) $, the current state of the system and outputs $ 1 $ if the user was previously registered with the corresponding identity $ \id $; otherwise, it outputs $ 0 $. 
	
	\item $ \{\pwd, \bot \} \gets \PullPwd(\id, \sigma)$: This algorithm takes as input the user identity $ \id $ and current state $ \sigma $, and outputs  $ \pwd $ the associated password to $ \id $ if $ \isReg(\id, \sigma) = 1$; otherwise, it outputs $ \bot $. 
\end{itemize}
 
\begin{definition}[Correctness]
	Let $ \Pi = (\Sgin, \Lgin) $ be a password-based authentication mechanism, $ P $ a binary predicate \footnote{The predicate determines if two passwords' distance satisfies the target distance metric or not. For example, if the distance metric is Hamming 2, for two password $ \pwd_1, \pwd_2 $, we have $ P(\pwd_1, \pwd_2) = 1 $ if $ \Ham(\pwd_1, \pwd_2)  \leq 2 $}, and $ \EvType = \{\Register, \Login \} $ be the set of registration/login actions.  Then, for all sequence of registration/login events $ U_1, \ldots, U_i \in \mathcal{ID}\times \Pwd \times \EvType$ resulting the state $ \sigma_i $, the correctness of $ \Pi $ enforces two following conditions:

\begin{itemize}
	\item \textbf{Login Correctness:} For all events \[ U_{i+1} = (\id, \pwd, \Login) \in \Id\times \Pwd \times \EvType\] we have $ (\sigma_{i+1}, \acpt)\gets \Lgin(\sigma_i, \id, \pwd)  $ if \[ i =\isReg(\id, \pwd, \sigma_i) \] Intuitively, this basically means that the user with identity $ \id $ has already successfully registered with the corresponding password $ \pwd $ and for all future login events using the the id-password pair $ (\id, \pwd) $ the login algorithm $ \Lgin $ always outputs $ \acpt $.  
	
	\item \textbf{$ (P,\theta) $-Typo Resilience:} 
	Let $ P $ be binary predicate,  $ 0\leq \theta \leq 1 $ be an arbitrary real number, $ U_{i+1}  = (\id_{i+1}, \pwd_{i+1},\Login) \in \Id\times \Pwd \times \EvType$ be a login event s.t. $ 1 := \isReg $ $(\id_{i+1}, \pwd_{j}, \sigma_{i})   \wedge  P(\pwd_{i+1}, \pwd_j)$ for some $ 1\leq j \leq i $, and $ (\sigma_{i+1}, \rjct) \gets \Lgin(\id_{i+1}, \pwd_{i+1}, \sigma_i) $. We say $ \Pi $ is $(P, \theta)$-Typo resistant if after login event $ U_k = (\id_{i+1}, \pwd_j, \Login)  $ for some $ k > i+1 $, for all $ k'> k $ we have $ (\sigma_{k'+1}, \acpt )\gets \Login (\id_{i+1}, \pwd_{i+1}, \sigma_{k'})  $ with probability at least $ \theta $. Intuitively, this implies that after an unsuccessful login with a password which has small distance to the original password, if we have a login with the original password, then, we have the chance of successful login (at least with probability $ \theta $) with the miss-typed password and the that typo will be added to the cache of password which will grant successful login.

\end{itemize} 

\end{definition}

\section{Typo Vault: Security Definitions}
\applab{apdx:TypoPrivacySecDef}
In this section, we formally describe the security definitions based on a game between an adversary $ \A $ and a challenger $ \C $. In the game we use a predicate which checks if the received login query is valid or not.  

\begin{definition}[Valid Login Query]\deflab{Def:EqPwdD}
	Let $ \pwd_0, \pwd_1, \pwd \in \Pwd$ be three passwords from password space $ \Pwd $. We say that the login query $ (\pwd_0, \pwd_1, \pwd)$ is a valid query under predicate $ P $ if we either have 
	\begin{enumerate}
		\item $ \pwd_0 = \pwd_1 $ if $ 1 = P(\pwd_0, \pwd) $ or $ 1 = P(\pwd_1, \pwd) $. 
		\item Or $ P(\pwd_0, \pwd) = 0 \wedge P(\pwd_1, \pwd) = 0 $. 
	\end{enumerate}
	
	We denote $ \ValidLginQ(\pwd_0, \pwd_1, \pwd) = 1 $ when the query is valid with regard to the mentioned two conditions, otherwise we have $\ValidLginQ(\pwd_0, \pwd_1,\pwd)  = 0$.
\end{definition}

\begin{definition}[Typo Privacy]\deflab{def:TypoPrivacy} We say that a password-based authentication scheme $ \Pi $ is $ \left(t\left(\lambda\right), q\left(\lambda\right),\eps\left(t,q,\lambda\right)\right) $-typo private under the predicate $ P $ if for all adversaries \A running in time $ t $ and making at most $ q $ queries to \Login/\Register we have 
	\begin{align}
	\Pr\left[ \Experiment_{\Pi, \mathsf{Typo-Privacy}} \left(\A, \lambda, q\right) = 1\right] \leq \eps (t,q,\lambda)
	\end{align}
\end{definition}

\begin{figure*}
	
		\def\arraystretch{0}
\fontsize{8}{10}\selectfont
\begin{experiment}
	\explab{Exp:TypoPrivacy}
\end{experiment}
		\makebox[\textwidth]{
		\begin{tabular}{|p{2 in}|p{2.6in}|}
	\hline		
   \underline{$ \Experiment_{\mathsf{\Pi, Typo-Privacy}} [\A, \lambda, q]$}:

			\begin{itemize}
				\item \textbf{Init}
				\begin{enumerate}
					\item \C runs $ \sigma_0 \gets \Init(1^\lambda) $. 
					\item \C randomly selects $ b\in_R\{0,1\} $.  
					\item \C sets $ t  = 0 $ as the starting time. 
				
				\end{enumerate}
 				
				\item \textbf{Query phase}: \A makes at most $ q $ queries to \QSgin/\QLgin oracles: 
	
				\begin{itemize}
			
					\item $ \QSgin(\id_i, \pwd_i) $

					\item $  \QLgin(\id_i, \pwd_{i,0}, \pwd_{i,1}) $				
				\end{itemize}

				\item \textbf{Guess}: Let $\view  = (\lambda, \sigma_0, \sigma_1, d_1, \ldots, \sigma_q, d_q ) $. 
				\begin{itemize}
					\item $ b' = \A ( \view ) $.
				\end{itemize}
			
				\item \textbf{Experiment Output}:
				 \item [] $\Experiment_{\mathsf{\Pi, Typo-Privacy}} [\A, \lambda,q] = 1 \iff  b' == b$

			\end{itemize}

			&

			\underline{$ \QSgin(\id, \pwd) $:}
			\begin{enumerate}
				\item If $ t>q  $ return $ \bot $. {\color{green} // at most $ q $ queries is allowed.}
				\item set $ \sigma = \sigma_t $
				\item $ (\sigma_{t+1}, d_{t+1})\gets \Sgin(\sigma, \id, \pwd) $
				\item set $ t = t +1 $. 
			\end{enumerate}

				\underline{$ \QLgin(\id, \pwd_0, \pwd_1) $:}
				\begin{enumerate}
					\item If $ t> q $ return $ \bot $ { \color{green} // at most $ q $ queries are allowed.}
					\item set $ \sigma = \sigma_t $
					
				\item if $ 1 = \isReg(\id, \sigma) $
			
				\begin{itemize}
					\item $ \pwd_{\id} = \PullPwd(\id, \sigma) $
				
					\item if $ 1 = \ValidLginQ(\pwd_0, \pwd_1,  \pwd_{\id})  $
					\begin{itemize}
						\item [-] $ (\sigma_{t+1}, d_{t+1}) \gets \Lgin (\id, \pwd_b, \sigma)  $.
						\item [-] set $ t= t+1 $. 
					\end{itemize}
					\item else 
				\begin{itemize}
					\item [-] set $ d_{t+1}  = \rjct$, $ \sigma_{t+1} = \sigma $, $ t = t+1 $
				\end{itemize}
				\end{itemize}
			
			\item else 
			
			\begin{itemize}
				\item set $  \sigma_{t+1} = \sigma,  d_{t+1} =  \rjct $
				\item  set $  t = t+1 $
			\end{itemize}

				\end{enumerate}
	\\ \hline
		\end{tabular}}

\caption{Formal description ``$ \Experiment_{\mathsf{\Pi, Typo-Privacy}} $'' which the experiment defining the typo privacy. The experiment is defined based on the interaction between unlimited adversary \A and the challenger \C. }
\figlab{fig:Exp:TypoPrivacy}
\end{figure*}

In \expref{Exp:TypoPrivacy} we defined two main oracles that the adversary can make at most $ q $ queries to them which are $ \QSgin $ and $ \QLgin $. \QSgin receives an ID and password, and assuming that it knows the current state of the system, first checks if the ID is registered previously or not. If not, it registers the user by running $ \Sgin $ as described in \expref{Exp:TypoPrivacy} and updates the state. $ \QLgin $ receives the and ID and a pair of two PWDs $ \pwd_0 $ and $ \pwd_1 $ as  a request for login attempt. Similarly, assuming that oracle knows the current state of the system, the oracle uses $ \pwd_b $ to run the login algorithm $ \Lgin $ as described in \expref{Exp:TypoPrivacy}. We also defined the time variable $ t $ to track the number of queries and the state updates when the adversary access to the oracles.

We remark that there are other security definitions like \textit{offline distinguishing}, \textit{offline guessing} and \textit{Random-or-Real} which are defined for a  typo tolerable password-based authentication scheme and introduced and discussed in \cite{CCS:CWPCR17}. However, due to the page limitation we ignore to formally review and prove them. However, we should highlight that our construction also provide these security properties. Basically, the source of difference between our proposal and the TypTop mechanism is the underlying the public key encryption scheme. Moreover, it is worth mentioning that their Cheetarjee et al's construction dose not support the security definition we just introduced, i.e.,  \defref{def:TypoPrivacy}. 
  
 
 \subsubsection{No Typo Privacy for Original TypTop}\seclab{NoTypoPrivacyFortyptop}
 In what fallows we will show that TypTop \cite{CCS:CWPCR17} does not satisfy Typo Privacy we described in \expref{Exp:TypoPrivacy}. Before that, we just briefly review TypTop mechanism and then we will show why their proposal does not support our suggested security property.
 
Let $ \pwd $ be the user's password. In TypTop, the server uses a password-based  key derivation function to extract the secret key $ k = \PKDF(\pwd) $ and also uses public key cryptography and assigns a pairs of secret-public key $ (pk, sk) $ to each user in time of registration. The user assigned public key will be stored in the sever along with his/her other credentials. Moreover, let $ \AE = (\AE.\Enc, \AE.\Dec) $ be an authenticated encryption scheme. Then, the server encrypts $ sk $ using $ k $ to store the ciphertext $ C_{sk} = \AE.\Enc_k(sk) $ in the server side. After an unsuccessful login attempt using a miss typed password $ \pwd' $,  the server will encrypt the typo $ \pwd' $ using $ pk $, and adds it to the waiting list. In the future login using the original password $ \pwd $, the server extracts $ k = \PKDF (\pwd )$, and decrypts $ C_{sk}  $ to extract the secret key $ sk $. Then the server uses the secret key $ sk $ to decrypt all the ciphertexts appeared in the waiting list.  Finally, those that have small edit distance to the original password $ \pwd $ have the chance to be added to the cache of valid passwords/typos which will grant successful login for the future attempts.
 
Looking at the adversary's ability/view, we can see that the designed TypTop dose not satisfy Typo-privacy. Based on the experiment description, we can see that the adversary knows the passwords and therefore she can decrypt the challenge ciphertext to determine which password is chosen by the challenger. This was a high level intuition of the security issue and in what follows you may find more formal technical details of the mentioned issue.  For simplicity we assume that number of queries is at most $ q  =2$ the first one is the registration query of a user with identity $ \id  $ and password $ \pwd $. And the second query is t a login query with password pairs $ \pwd_0, \pwd_1 $ of the adversary choice and we have $ P(\pwd_0, \pwd) \neq 	1$ (so $ \pwd_0 $ and $ \pwd_1 $  are not necessarily equal while based on the condition of the experiment if $ P(\pwd, \pwd_0) =1 $, then they have to be equal i.e., $ \pwd_0 = \pwd_1 $, i.e., $ \ValidLginQ(\pwd_0, \pwd_1, \pwd) =1 $). Regarding the description of TypTop, the challenger will add the ciphertext $ {C_{\pwd_b} }= \Pi_{\mathsf{PubKey}}.\Enc(pk, \pwd_b)  $ to the updated state as the encryption of a potential typo. Based on the description of the game, the adversary chooses the registration password $ \pwd $ and knows. So simply can extract the authenticated decryption key $ k  = \PKDF(\pwd) $ and computes the secret key $ sk  = \Pi_{\mathsf{PubKey}}.\Dec(k, C_{sk}) $. Finally, the adversary computes $ \pwd_b = \Pi_{\mathsf{PubKey}}.\Dec(sk,  C_{\pwd_b}) $ and simply outputs $ b $ ($ \A $ knows both $ \pwd_0,  \pwd_1 $). 
 
 The above attack represents the simplified version of the actual Typo privacy and actually is a weaker definition. However, as the TypTop scheme does not provide this weaker requirement, it the does not satisfy Typo privacy as well.


\section{Personalized Typo Tolerance Mechanism from Conditional Encryption} \applab{CondTyoTopCompelete}

In this part we will describe our generic construction of a password typo vaults which safely caching incorrect login attempts with conditional encryption. In the rest of the paper we call it `` CondTypTop'' to imply construction of TypTop which is designed using conditional encryption. The main building blocks of our proposed construction are conditional encryption $ \Pi_{\CE}  = (\KG, \Enc, \Cond\Enc, \Dec)$, authenticated encryption $ \Pi_{\AE} = (\Enc, \Dec)  $ and password-based key derivation function $ \PKDF $\footnote{which takes as input the password and deterministically extracts a key for symmetric key encryption scheme and we have $ k_{\pwd} = \PKDF (\pwd) $}. Let the underlying conditional encryption be over the binary predicate $ P $. Considering the mentioned building blocks, our construction of $ \Pi_{P} = (\Init, \Sgin, $$\Lgin)$ is described in \figref{fig:TyptopFromCond}. 

Intuitively, we should highlight that construction is similar to TypTop, however the main and basic idea and difference is replacing the traditional public key encryption part with our proposed conditional encryption scheme. So after an incorrect attempt using the typo included password $ \pwd' $ we encrypt it using conditional encryption using the encryption of original password ciphertext. In fact, in this case $ \pwd' $ will be considered as the payload of $ \Pi_{\CE}.\Cond\Enc $.   For this aim, we have to encrypt the original password $ \pwd $ using $ C_{\pwd} = \Pi_{\CE}.\Enc_{pk} (\pwd; r)  $ and store it in the server along with user's public key $ pk $. Then, in future login with the miss typed password $ \pwd' $, we add the ciphertext $C_{pwd'} = \Pi_{\CE}.\Cond\Enc_{pk}(C_{\pwd}, \pwd', \pwd'; r') $ to our waiting list. So, in the future login with the original password $ \pwd $ we extract $ k_\pwd = \PKDF(\pwd) $ and decrypt $ sk =\Pi_{\AE}.\Dec(k_\pwd, C_{sk}) $. Finally, if $ P(\pwd, \pwd') = 1 $, the conditional decryption returns the underlying payload (here is the password with small typo $ \pwd' $) $\pwd' =  \Pi_{\CE}.\Dec_{sk}(C_{\pwd'}) $; otherwise, the decryption algorithm returns uniformly random element chosen from the password space $ \Pwd $. 

In what follows, we will provide the detailed description of algorithms $ \Init $, $ \Sgin $ and $ \Lgin $. 

\begin{itemize}
	\item $ \sigma_0 \gets \Init (1^\lambda) $: This algorithm takes as input the security parameter $ \lambda $, and sets the initial state $ \sigma_0 = (\lambda,S_W, S_T, W,T) $ in which $ W $ is allocated waiting list, $ T $ the cache of valid typos granting successful login. $ S_W $ and $ S_T $ are respectively the size of waiting list and cache size allocated for each user. Initially, as we don't have any registered user, $ W $ and $ T $ are empty. 
	
	\item $ (\sigma', d = \{\acpt, \rjct\})\gets \Sgin(\sigma, \id, \pwd)$: Let $ \sigma $ be the current state of the system, and $ (\id, \pwd) $ be the pair of password and user identity. If $ 1 = \isReg(\sigma, \pwd)  $, the algorithm outputs $  (\sigma, d =\rjct) $; otherwise, the algorithm computes symmetric key $ k_\pwd =\PKDF (\pwd || \salt_\id )$ in which $ \salt_\id\in_R\{0,1\}^{\lambda} $ is the assigned user's slat, and $ (pk_\id, sk_\id) \gets \KG(1^\lambda)  $. Then, algorithm computes $ C_{\id, \pwd}  = \Pi_{\CE}.\Enc_{pk_\id}(\pwd; r) $ under random coin $ r\in_R\{0,1\}^\lambda $. Then the algorithm will assign a waiting list $ W_{\id} $ of size $ S_W = \sigma[1] $ and will fill $ W_{\id} $ with conditional encryption of garbage messages, i.e., $ W_{\id} = \Pi_{\CE}.\Cond\Enc_{pk_\id} $ $(C_{\id, \pwd}, r',r';r") $ s.t. $ r'\in_R \Pwd $\footnote{We remind you that $ \Pwd $ is the space of all possible passwords.} and $ r'' \in_R\{0,1\}^\lambda $. Finally, the array $ T_\id $ of size $ S_T = \sigma[2] $ as a cache of passwords which granting successful logins will be assigned to the user with identity $\id $ such that  $T_\id[0] = \Pi_{\AE}.\Enc(k_\pwd, sk_\id)$ . To track the number of logins, the algorithm sets $ c_\id $ as the counter which will be increased by one after each attempt for login. We set  $\sigma_\id = (\id, \salt_\id, pk_\id, W_\id, T_\id, C_{\id, \pwd},c_\id)$    The updated state will be $ \sigma' = (\sigma, \sigma_\id) $ and the algorithm outputs $ (\sigma', d = \acpt) $.

	\item $ (\sigma', d  = \{ \acpt, \rjct  \}) \gets \Lgin (\sigma, \id, \pwd) $: The algorithm first checks if the user is registered and output $ (\sigma'=\sigma, d = \rjct) $ if $ 0 = \isReg(\sigma, \id) $; otherwise, it extracts the state $ \sigma_\id \in \sigma $ which was previously assigned to the user $ \id  $. Then it obtains the salt $ \salt_\id = \sigma_\id [1] $ and computes $ k_\pwd = \PKDF(\pwd || \salt_\id ) $. Then considering the cache $ T_\id  = \sigma_\id[4] $, the algorithm search for the ciphertext $ T_\id [i]  $ for all $ 1\leq i \leq S_T $ such that $(1, sk_\id) =  \Pi_\AE.\Dec(k_\pwd, T_\id[i]) $, this is a successful login and the algorithm sets $ d= \acpt $; otherwise,  $ d = \rjct$. Now the counter is updated i.e., $\sigma_\id[6] = \sigma_\id[6]+1$, we have two cases: 
	
	\begin{enumerate}
		\item \textbf{Successful login ($ d= \acpt $)}: Now the algorithm will process waiting list using the extracted secret key $ sk_\id $. For this aim, the algorithm will decrypt all the ciphertexts in $ W_\id = \sigma_\id [3] $ as $ m_j = \Pi_\CE.\Dec_{sk_\id}(W_\id [j]) $ for all $ 1\leq j \leq S_W $. For those $ m_j $ that have $ P(\pwd, m_i) = 1 $  we add them to the cache $ T_\id $ if we have empty space for new typos in $ T_\id $. We can assign a weight to $ m_i $ based on its appearance in the sequence $ m_1, \ldots, m_{S_W} $ and randomly selects them based on their weight until the  cache become full. For the selected $ m_k $ the algorithm will use $ k_{\pwd, m_k} = \PKDF(m_k || \salt ) $ to compute \[C_{\pwd, m_k} = \Pi_\AE.\Enc_{k_{\pwd, m_k}}(sk_\id) \ .\] Now, a random shuffling will be applied on the elements of cache with the constrain that the first element is always associated to the original password. After processing waiting list, similar to the $ \Sgin $, the algorithm fill the waiting list with garbage ciphertexts and updates $ W_\id  $. In this, we should highlight that the updated $ \sigma' $ is the same as $ \sigma $ and the only different is the updated cache/waiting list  $ T_\id $/$ W_\id $ and.  This is the successful login and the algorithm outputs and the output is $ \sigma' = \sigma $ 
				
		\item \textbf{Unsuccessful login ($ d = \rjct $)}: In this case we need to conditionally encrypt $ \pwd $ and add it to the waiting list. So, first we extract $ C_{\id, \pwd'} =  \sigma_\id[5]  $ and compute \[ C_{\pwd, \pwd'} \gets \Pi_{\CE_{pk_\id}}.\Cond\Enc _{pk_\id}(C_{\id, \pwd'}, \pwd, \pwd; r)\] in which $ r\in_R \{0,1\}^\lambda $ (remind that $ \lambda  = \sigma[0] $). Assuming the updated value of counter $ c_\id = \sigma_\id[6] $, the algorithm computes $ i = c_\id \mod S_W $ and replace the $ i $-th element of the waiting list. Then it will randomly shuffle the waiting list and update the state as $ \sigma' $. We highlight that, $ \sigma' $ is different with $ \sigma $ related to the counter and the changes applied to the waiting list. 

	\end{enumerate}
	Now the login attempt is done and the algorithm outputs $ (\sigma', d) $. 
\end{itemize}



\subsection{Security proof of CondTypTop}
In this part we will prove the security of the CondTypTop. We should highlight that we just concentrate on the typo-privacy that we defined in this paper. The other security properties defined for TypTop are still preserving as our construction is similar to the original TypTop. So, due to page limitation, we leave discussing them here and with similar reasoning all the remaining security properties can be proved in the same way. 

To prove that CondTypTop provides Typo-privacy, we use hybrid argument and will define three hybrids. Then, using the security of conditional encryption, we show that all these hybrids are computational indistinguishable. Finally, by indistinguishablity of the first and the last hybrid, we conclude that the our CondTypTop has typo privacy. Intuitively, the first hybrid is the original typo privacy game. In the second hybrid, we take the current state of the system and replace all the ciphertexts that are the output of the $ \Cond\Enc $ algorithm with the output of $ \Sim(pk) $ of conditional encryption. As we assumed that the underlying conditional encryption is secure, inherently such simulator exists. Based on the construction of CondTypTop, the ciphertexts we replacing are the conditional encryption of $ \pwd_{i,b} $ for all $ i $ such that the $ i $-th query is $ \QLgin $ query. Finally, in the last hybrid, we just replace $ \Sim(pk) $ with conditional encryption  of $ \pwd_{i,1-b} $. In what follows we formally prove the typo privacy of CondTypTop.

\begin{theorem}\thmlab{thm:CondTypTopSec}
 Given the $ (t, q, \eps) $-secure conditional encryption $ \Pi_{\CE} = (\KG, \Enc, \Cond\Enc, \Dec) $, then CondTyoTop $ \Pi $	descried in \constref{const:ConTypTop}  is $ (t', q', \eps') $-typo private.
\end{theorem}

\begin{proofof}{\thmref{thm:CondTypTopSec}}
	To prove the security of our CondTyoTop, we should highlight that for the ciphertextsof conditional encryption which presented in the final state, two main cased can be considered. Without loss of generality, let the $ i $-th query be a $ \QLgin $ query. Then state of the system contains the conditional encryption $ C_{ib} = \Cond\Enc_{pk}(C_{\pwd_{\id}}, \pwd_{i, b}) $ in which $ C_{\pwd_{\id}} =\Enc_{pk}(\pwd_{\id}) $ for some previously registered user with identity $ \id $. So the two cases are as follows: 
	\begin{itemize}
		\item $ \pwd_{i,0} = \pwd_{i,1}  $. In this case, both ciphertexts are statistically indistinguishable and their input plaintexts are exactly the same. So no adversary can distinguish between their ciphertexts. 
		\item $ P(\pwd_{i,0}, \pwd_{\id}) = 0 \wedge P(\pwd_{i,0}, \pwd_{\id}) = 0  $. For this case we need to prove adversary's advantage to win in the experiment is polynomially close to the advantage of adversary who breaks the security of conditional encryption. We will show it by defining three following hybrids \textbf{Hybrid 0}, \textbf{Hybrid 1} and \textbf{Hybrid 2}. 
	\end{itemize}
As we mentioned before, to prove the security of \constref{const:ConTypTop}, we defined three hybrids and then prove all these hybrids are indistinguishable.  Indistinguishability of these hybrids implies that the advantage of the adversary to win in $ \Experiment_{\Pi, \mathsf{Typo-Privacy}}  $ is negligible.

\begin{itemize}
	\item[\textbf{Hybrid 0}:]This is the original experiment $ \Experiment_{\Pi, \mathsf{Typo-Privacy}}  $. 
	
	\item [\textbf{Hybrid 1}:] Let $ \sigma $ be the final state of the typo privacy experiment defined in \textbf{Hybrid 0}. Suppose $ Q = \{ C_{i,b} \in \sigma  \} $ be set of all conditional encryption of $ \pwd_{i,b} $ for all the login queries $ 1\leq i \leq q' $.  Due to the security of conditional encryption, we know that there exists simulator $ \Sim (pk) $ who simulates the ciphertext of conditional encryption scheme. In this hybrid, we replace all $ C_{i,b} \in Q $ with $ C'_{i,b } \gets  \Sim(pk) $. 
	
	\item [\textbf{Hybrid 2}: ] In this hybrid we simply replace all $ C'_{i,b} $ with $$ C_{i,1-b}  = \Cond\Enc_{pk}(C_{\pwd_id},\pwd_{i,1-b}) $$.  
	
\end{itemize}

\subsubsection{Indistinguishability of the hybrids}. Due to the security of conditional encryption, the advantage of adversary $ \A $ to distinguish between \textbf{Hybrid 0} and \textbf{Hybrid 1} is $ \eps $. More formally, we should highlight that as $ P(\pwd_{i, b}, \pwd_{\id})= 0  $, the simulator can simply select a random number and encrypt it using the public key $ pk $. As the security of conditional encryption guarantees that if $ P(m_1, m_2) =1 $, then the resulting ciphertext will be the encryption of a message chosen uniformly at random. And this is basically what our simulator $\Sim (pk) $ does. With similar argument, the adversary advantage in distinguishing between the hybrids \textbf{Hybrid 1} and \textbf{Hybrid 2} is also $ \eps $. As a result, the advantage of adversary $ \A $ spending time $ t' =t $ to distinguish between \textbf{Hybrid 0} and \textbf{Hybrid 1} is $ \eps' = \eps  $.

\end{proofof}



\section{Constructions}
\applab{AppConstructions}
\begin{figure}
\begin{construction}\constlab{Const:EqualityTest}
\end{construction}
\begin{mdframed}
	
	\noindent ~~ \underline{$ (pk, sk) = \KG( 1^\lambda) $}:
	\begin{enumerate}
		\item Pick random $ r\in_R \{0,1\}^\lambda $
		\item Compute $ \left(sk =\left(\beta, \mu\right), pk = \left(N, g\right)\right) \gets \Pail.\KG(1^\lambda; r) $\footnote{We use the simple variant of Pallier key generation which picks primes $p$ and $q$ of equivalent length and sets  $ N = pq $, $g=N+1$, $\beta  =\lcm (p-1, q-1) $ and $ \mu = \varphi(N)^{-1} \mod{N}$ where $\varphi(N) = (p-1)(q-1)$. See details in \appref{App:PailDetails}. } 
		
		\item If $ |\Sigma|^{n+1}\geq \min(p,q) $ or $\mathtt{gcd}(N, (p-1)(q-1)) \neq 1$, repeat step (1); else, output $ (sk, pk) $
	\end{enumerate}
	
		\noindent ~~ \underline{$ c_{m_1}  \gets  \Enc_{pk}(m_1) $}:
	\begin{enumerate}
		\item Pick random $ r\in_R \{0,1\}^\lambda $ 
		\item Compute $ \hat{m}_1 = \ToInt(m) $. 
		\item Return $ \left(0, c_{m_1}  :=  \Pail.\Enc\left(\hat{m}_1; r\right) \right)$
	\end{enumerate}
	
\noindent ~~ \underline{$ \tilde{c} = \Cond\Enc_{pk}( c_{m_1},m_2,m_3) $}: 
\begin{enumerate}

	\item Parse $ (b,c) \gets c_{m_1}$ and $(N,g) \gets pk$.
	\item If $b=1$ or $\mathtt{gcd}(c,N) \neq 1$ return $\bot$. 			
	\item Compute $\hat{m}_3= \ToInt(m_3)$,  $\hat{m}_2 = \ToInt(m_2)$ and select $R\in_R \Mbb{Z}_N$, $r_2\in_R \mathbb{Z}^*_N$

    \item $ c^{=}  = g^{\hat{m_3}} \cdot \left[\left(R \boxtimes c\right) \boxplus \Pail.\Enc_{pk} \left(\left(-R\right)\cdot m_2; r_2\right)  \right] $ 
	
	\item Return $\tilde{c} = \left(1, c^{=}\right)$
\end{enumerate}

\noindent ~~ \underline{$ m := \Dec_{sk}( \tilde{c}) $}:
\begin{enumerate}
	\item Parse $ (b, c) \gets \tilde{c} $
	\item Compute $ x \gets \Pail.\Dec_{sk}(c) $.
    \item compute $m \gets \ToOrig(x)$
    \item If $b=1$ and $m > |\Sigma|^{n+1}$ return $\bot$
    \item Otherwise return $m$
\end{enumerate}

\end{mdframed}	

\end{figure}

\begin{figure}
     \begin{construction}
			\constlab{const:ArbHamm}
	\end{construction}
    \begin{mdframed} \fontsize{8}{8}\selectfont
        \underline{$ (sk, pk)\gets\KG(1^\lambda; r) $}:
		\begin{itemize}
			\item[1.]  $\run \left(\Pail.sk, \Pail.pk = \left(N= pq,g=N+1\right)\right) \gets \Pail.\KG(1^\lambda; r) $. 
			\begin{itemize}
				\item [-]  if $ \min \{p,q\}  < |\Sigma| $ or $\mathtt{gcd}(N,(p-1)(q-1))\neq 1$ repeat step 1. {\color{green} //run $ P.\KG $ again.}  
			\end{itemize}
            \item [2.] $\sset  \SS = (\Share, \recover) $ over field $ \Mbb{F} _{2^\lambda}$
			\item [3.] $\sset sk = \Pail.sk $ and $ pk = (\Pail.pk, 2^\lambda)$
			
		\end{itemize}

        \underline{$ c \gets  \Enc_{pk}(m; r) $:}
	        \begin{itemize}
		        \item [1.]  $\pparse m = \left(m\left[1\right], \dots, m\left[n\right]\right),  r = \left(r_1, \ldots, r_n\right)\in \left(\Mbb{Z}^*_N\right)^n $. 

		        \item [2.] $ \forall 1\leq i \leq n $, $\ccompute \hat{m}[i] = \ToInt(m[i]), c_i = \Pail.\Enc_{pk}\left(\hat{m}\left[i\right]; r_i\right) $
		        \item [3.] $\ooutput c = (b =0, c_1, \ldots, c_n) $.
	        \end{itemize}
	
		\underline{$ c\gets\Cond\Enc_{pk}(c_m, m', m''; r) $:} 
	        \begin{itemize}
		          \item [1.] $\pparse  m' = \left(m'\left[1\right], \ldots, m'\left[n\right]\right)\in \Sigma^n $,  $r = \left(r_1, \ldots, r_n\right)\in \left(\Mbb{Z}^*_N\right)^n $, $c_m = \left(b, c_1, \cdots, c_n\right)  $ 
		
		          \item [2.]  $\ccompute \hat{m}'[i] = \ToInt\left(m'\left[i\right]\right) $ to map each letter to integer.
		
		          \item [3.] $ \iif b = 1 $, $ \ooutput \bot $. 
		
		          \item [4.] $ \forall 1\leq i \leq n $, $\ccompute \tilde{c}'_i =\Pail.\Enc_{pk}\left(\hat{m}'\left[i\right]; r_i\right) \boxminus c_i  $.

		          \item [5.] Pick $ K\in_R \set{0,1}^{\lambda} $, compute $ c_{AE}  = \Auth.\Enc_K( m'' )$. 
		          \item [6.] $ \left(\left(i_1, \ldb s \rdb_1\right), \ldots,  \left(i_n, \ldb s \rdb_n\right) \right) \gets \SS.\Share(n, n-l,K )$. 
		
		          \item [7.] $\forall 1\leq i \leq n: \ssample a_i\in_R \left[1, \left\lfloor \frac{N-1}{2^\lambda} -1\right\rfloor\right] $, $\tilde{r}_i \in_R \Mbb{Z}^*_N$

                \item [8.]  $\forall 1\leq i \leq n:\ccompute $ $ y_i = \ldb s \rdb_i + a_i 2^\lambda, \tilde{c}_{i}  =  \Pail.\Enc_{pk}(y_i; \tilde{r}_i ) \boxplus (R_i \boxtimes  \tilde{c}'_i)$
                
		          \item [9.] $ \ooutput  \tilde{c} = (b = 1, \tilde{c}_1, \cdots, \tilde{c}_n,c_{AE}) $
        	\end{itemize}
        
        \underline{$ \{m, \bot\} = \Cond\Dec_{sk}\left(\left(b,\tilde{c}\right)\right) $:}
		\begin{itemize}
			\item [2.1] $\iif b = 1 $ {\color{green} // The ciphertext is extracted from $ \Cond\Enc(\cdot) $}
			\begin{itemize}
                \item $ \pparse\tilde{c}=  (\tilde{c}_1, \cdots, \tilde{c}_n, c_{AE}) $, $ \aand \sset m =\bot$
				\item $ \ffor 1 \leq i \leq n: \hat{y}_i = \Pail.\Dec_{sk} (\tilde{c}_i) $, $ \ldb s' \rdb_i = \RanDec(\hat{y}_i) =  \hat{y}_i \mod 2^\lambda$. 
				\item For all ${ n\choose {n-\ell}}$  possible  $S \subseteq [n] $ with $|S|=n-\ell$, 
				\begin{itemize}
					\item[-] $\ccompute K_S = \recover\left( \left\{\left(i, \ldb s'\rdb_{i}\right)\right\}_{i \in S}  \right)   $ 
                    \item[-]$ (v', \hat{m}'') = \Auth.\Dec_{K_S}(c_{AE})$,  $ \iif v' =1, \rreturn m  =\hat{m}'' $\footnote{Here $m = \bot$ as for all possible $K_S$ the $\Auth\Dec_{K_S}$ algorithm failed. }
				\end{itemize}
			\end{itemize}  
			\item[2.2] $\iif b = 0 $ {\color{green} // The ciphertext is extracted from $ \Enc(\cdot) $}
			\begin{itemize}
                \item[-] $ \pparse\tilde{c}=  ( \tilde{c}_1, \cdots, \tilde{c}_n) $, $ \aand \sset m =\bot$
				\item[-] $ \ffor 1 \leq i \leq n  $, $\ccompute\hat{m}_i = \Pail.\Dec_{sk} (\tilde{c}_i) $
				\item[-] $ \rreturn m = \ToOrig(\hat{m}_1)\| \ldots \| \ToOrig(\hat{m}_n) $
			\end{itemize}
		\end{itemize}
    \end{mdframed}
    \caption{ Concrete construction of conditional encryption over binary predicate $ P_{\ell,\Ham} $ using secret sharing.}
\end{figure}

\begin{figure}\fontsize{10}{10}\selectfont

			\begin{construction}
			\constlab{const:ConTypTop}
		\end{construction}
		
		\begin{mdframed}
		\underline{$ \sigma_{0}\gets \Init(1^\lambda)$:}
		
		Given the security parameter $ \lambda $ as input, this algorithm acts as follows. 
		
		\begin{itemize}
			\item [1. ] For all $ 1\leq i \leq L $ sample $ w_i \in_R \Pwd $ uniformly at random, and Set $ W = [\Pi_{\condit}.\Enc(w_1), \ldots,\Pi_{\condit}.\Enc(w_L)] $. 
			\item [2. ] Set $ T = [] $ as the list for collecting legitimate typos
			\item [3. ] Set $ \sigma_0 = (W, T) $
		\end{itemize}
		\underline{$ (\sigma_{i}, d = \{\acpt, \rjct\})\gets \Sgin(\sigma_{i-1}, \id, \pwd)$:}
		\begin{enumerate}
			\item [1. ] For all $ 1\leq i \leq L $ sample $ w_i \in_R \C_{\Pi_{\condit}}$\footnote{We note that $ w_i $s are random element sampled from the ciphertext space $ \C_{\Pi_{\condit}} $ associated to $ \Pi_{\condit} $. } , and Set $ W = [w_1, \ldots, w_L] $, and set $ W_{\id} = W $. 
			\item [2. ] Set $ T_{\id} = [] $  //As the list for collecting legitimate typos
			\item[3.] Compute $ b = \mathsf{IsRegistered}(\sigma_{i-1}, \id) $. 
			\begin{itemize}
				\item []If  $ b =0 $, set $ d =\acpt $ and go to step 4; 
				\item []If $ b = 1 $, set $ d= \rjct$, break and return $ (\sigma_i, d) $
 			\end{itemize} 
			\item[4. ] Run $ (pk_{\id}, sk_{\id})\gets \Pi_{\condit} .\KG(1^\lambda) $. 
			\item[5.] Set $ k_{\id, \pwd} = PKDF(\pwd) $, compute $ C_{\id,\pwd}  = AE.\Enc_{k_{\id, \pwd} } (sk)$, and $ c_{\pwd}  = \Pi_{\condit}.\Enc_{pk_{\id}}(\pwd)$
			\item[6.] Set $ T_{\id}[0] = C_{\id,\pwd}$, and $ W_{\id} = W $. 
			\item[7.] Update $ \sigma_{i} =  \apnd(\sigma_{i-1},  (c_{\pwd} , pk_{\id}, W_{\id}, T_{\id}))$. 
			\item[8.] $ \rreturn (\sigma_{i}, d)$
		\end{enumerate}
	\underline{$ (\sigma_i, d = \{\acpt, \rjct\})\gets \Lgin(\sigma_{i-1}, \id, \pwd')$:}
	\begin{enumerate}
		\item [1. ] Compute $ s_{\id}= \mathsf{Extract}(\id, \sigma_{i-1})$. 
		\begin{itemize}
			\item [] If $ s_{\id} = \bot$, return $(\sigma_{i+1}  =\sigma_i, d= \rjct) $ and break; 
			\item[] Else, parse $ s_{\id} = c_{\pwd} , pk_{\id}, W_{\id}, T_{\id} $
		\end{itemize}
		\item[2.] Compute $ k = PKDF(\pwd) $
		\item[3.] If $\forall x\in T_{\id} $, $ \bot = AE.\Dec_{k}(x)  $ 
		\begin{itemize}
			\item[2.1] Set $ d =\rjct $, compute $ c_{\pwd'} = \Pi_{\condit}.\Cond\Enc(c_{\pwd}, \pwd') $, update $ W_{\id} =\apnd(W_{\id}, c_{\pwd'} )  $. //Adding typo's ciphertext to waiting list $ W_{\id} $. 
			\item[2.2] Set $ s_{\id} = (c_{\pwd} , pk_{\id}, W_{\id}, T_{\id} ) $, compute $ \sigma_{i}= \mathsf{Replace} (\id, s_{\id}, \sigma_{i-1}) $, return $ (\sigma_i, d) $, and break.
		\end{itemize}
		\item [3.] Find $ x\in T_{\id} $ s.t. $\bot \neq  sk_{\id}  = AE.\Dec_{k} (x)$, and  $\forall i\in [|W_{\id}|] $  compute $ pw_i = \Pi_{\condit}.\Dec_{sk_{\id}}(W_{\id}[i]), $
		\begin{itemize}
			\item if $ 1= P(pw_i, \pwd)$: compute $ k_{\id, pw_i} = PKDF(pw_i) $, $ C_{\id, pw_i} = AE.\Enc_{k_{\id, pw}}(sk_{\id}) $, and update $T_{\id} = \apnd (T_{\id},  C_{\id, pw}) $. 
		\end{itemize}
	\item[4. ] For all $ 1\leq i \leq L $ sample $ w_i \in_R \C_{\Pi_{\condit}}$, and set $ W = [w_1, \ldots, w_L] $, and set $ W_{\id} = W $. 
	\item [5. ]  Set $ s_{\id} =  (c_{\pwd} , pk_{\id}, W_{\id}, T_{\id} )$,  and compute $ \sigma_{i} = \mathsf{Replace}(\id, \sigma_i, s_{\id} ) $. 
	\item [6. ] Return $ (\sigma_{i}, d= \acpt) $
		
	\end{enumerate}
	
	\end{mdframed}
	\caption{Proposed generic construction of our password typo vaults for secure caching incorrect login attempts with conditional encryption }
	\figlab{fig:TyptopFromCond}
\end{figure}

}{}

\end{document}